%% file: qcomp_notes_v3.tex
\documentclass[11pt]{article}
\pdfoutput=1
\usepackage{jhepmod}
\usepackage{booktabs}
\usepackage{amssymb} 
\usepackage[english]{babel}
\usepackage{amsmath,amssymb,amsbsy,amstext, amsthm, simplewick}
\usepackage{graphicx}
\usepackage{amsfonts}
\usepackage{mathtools}
\usepackage{amssymb}
\usepackage{upgreek}
\usepackage{listings}
\usepackage{exscale,relsize}
\usepackage[makeroom]{cancel}
\usepackage{soul}
\usepackage[utf8]{inputenc}
\RequirePackage{color}
\usepackage{amsopn, amstext, wasysym}
\usepackage{colonequals}
\usepackage{rotating}
\usepackage{multirow}
\usepackage{xspace}
\usepackage{datetime}
\usepackage[framemethod=tikz]{mdframed}

\usepackage[T1]{fontenc}
\usepackage{graphicx,scalerel}
\usepackage{amsfonts} 
\newmdenv[innerlinewidth=0.5pt, roundcorner=4pt,linecolor=mycolor,innerleftmargin=6pt,
innerrightmargin=6pt,innertopmargin=6pt,innerbottommargin=6pt]{mybox}

\newmdenv[innerlinewidth=0.5pt, roundcorner=4pt,linecolor=mauve,innerleftmargin=6pt,
innerrightmargin=6pt,innertopmargin=6pt,innerbottommargin=6pt]{mybox2}

\usetikzlibrary{backgrounds,fit,decorations.pathreplacing,calc}

\definecolor{navyblue}{rgb}{0.0, 0.0, 0.9}
\definecolor{rindou1}{rgb}{0.4431,0.2862,0.7960}
\definecolor{rindou2}{rgb}{0.078,0.1215,0.4392}
\definecolor{BrickRed}{rgb}{0.0, 0.53, 0.74}
\definecolor{BlueNCS}{rgb}{0.0, 0.53, 0.74}
\definecolor{mycolor}{rgb}{0.122, 0.435, 0.698}
\definecolor{mycolor2}{rgb}{0.02, 0.435, 0.698}
\definecolor{dkgreen}{rgb}{0,0.6,0}
\definecolor{gray}{rgb}{0.5,0.5,0.5}
\definecolor{mauve}{rgb}{0.58,0.33,0.82}
\definecolor{green2}{cmyk}{0, 1, 0.5, 0}
\definecolor{lightgreen}{cmyk}{0.2, 0, 0.2, 0.2}
\definecolor{lightgray}{cmyk}{0.1,0.2,0,0.1}
\definecolor{lightgray2}{cmyk}{0.4,0.4,0,0.8}
\definecolor{black}{cmyk}{1.0,1.0,1.0,1.0}
\definecolor{celadon}{rgb}{0.67,0.88,0.69}

\usepackage{amsthm}
\usepackage{upgreek}
\usepackage{slashed}
\usepackage{verbatim} 
\usepackage{scalerel}
\usepackage{lmodern}
\usepackage{framed}
\usepackage{lipsum}
\usepackage{xcolor}
\usepackage{mathrsfs}
\usetikzlibrary{arrows}
\usepackage{colortbl}
\usepackage[framemethod=tikz]{mdframed}

{\endMakeFramed}

\newenvironment{frshaded*}{%
	\MakeFramed {\advance\hsize-\width \FrameRestore}}%
{\endMakeFramed}

\usepackage{pgf,tikz}
\usepackage{mathrsfs}
\usetikzlibrary{arrows}
\usepackage{placeins}
\usepackage{pgf,tikz}
\usepackage{mathtools}
\usetikzlibrary{arrows.meta}
\usepackage[utf8]{inputenc}
\usepackage[english]{babel}
\usepackage{graphicx,tipa}
\usepackage{csquotes}
\usepackage{amsmath}
\usepackage{amssymb}
\usepackage{xcolor}
\usepackage{color,wasysym}
\usepackage{bbold}
\usepackage[abs]{overpic}
\usepackage[T1]{fontenc}
\usepackage{hyperref}
\usepackage{cancel}
\usepackage[braket, qm]{qcircuit}

\definecolor{redred}{HTML}{D53E4F}

\definecolor{blueblue}{HTML}{1B57B6}

\newcommand{\kp}{\ket{\psi}}
\newcommand{\bp}{\bra{\psi}}

\newcommand{\MAT}{\textsc{Mathematica}}
\newcommand{\QIS}{\textsc{Qiskit}}

\usepackage{tikz,xcolor,hyperref}
\definecolor{lime}{HTML}{A6CE39}
\DeclareRobustCommand{\orcidicon}{%
	\begin{tikzpicture}
	\draw[lime, fill=lime] (0,0) 
	circle [radius=0.16] 
	node[white] {{\fontfamily{qag}\selectfont \tiny ID}};	\draw[white, fill=white] (-0.0625,0.095) 
	circle [radius=0.007];	\end{tikzpicture}
	\hspace{-2mm}}
\foreach \x in {A,...,Z}{%
	\expandafter\xdef\csname orcid\x\endcsname{\noexpand\href{https://orcid.org/\csname orcidauthor\x\endcsname}{\noexpand\orcidicon}}
}

\usepackage{environ}
\usepackage{varwidth}
\newlength{\TheoremWidthTweak}%

\NewEnviron{short-shaded}[1][]{%
	\setlength{\TheoremWidthTweak}{\dimexpr%
		+\mdflength{innerleftmargin}
		+\mdflength{innerrightmargin}
		+\mdflength{leftmargin}
		+\mdflength{rightmargin}
	}%
	\savebox0{%
		\begin{varwidth}{\dimexpr\linewidth-\TheoremWidthTweak\relax}%
			\BODY
		\end{varwidth}%
	}%
	\begin{mdframed}[backgroundcolor=lightgray,userdefinedwidth=\dimexpr\wd0+\TheoremWidthTweak\relax, #1]
		\usebox0
	\end{mdframed}
}

\begin{document}
	\begin{titlepage}
		\setcounter{page}{1} \baselineskip=15.5pt \thispagestyle{empty}
		\bigskip\
		\vspace{1cm}
		\begin{center}
			{\fontsize{19}{38}\textsc{Notes on Quantum Computation and information}}  
		\end{center}
		\vspace{0.2cm}
		\begin{center}
			{\fontsize{12}{30}\selectfont Raghav G.~Jha\orcidA{}\,}
		\end{center}

		\begin{center}
			\vskip 7pt
			\textsl{Thomas Jefferson National Accelerator Facility, Newport News, VA 23606, USA\\
			Perimeter Institute for Theoretical Physics, Waterloo, Ontario N2L 2Y5\\
				}
			\vskip 6pt
		\end{center}
		
		\vspace{2.6cm}
		\hrule \vspace{0.2cm}
		\noindent {\sffamily \bfseries Abstract} \\[0.1cm]
We discuss basic fundamentals of quantum computing and information - quantum gates, circuits, algorithms, theorems, error correction, 
methods for unitary time-evolution and provide a collection of \QIS\footnote{It is an open source software 
development kit (SDK) for working with OpenQASM and the IBM Q quantum processors. Various IBM quantum simulators can be found   
\href{https://quantum-computing.ibm.com/composer/docs/iqx/manage/simulator/index}{here}.} programs
and exercises for the interested reader. 
\vspace{1.0cm}
\hrule
\vspace{1.0cm}
\noindent \emph{Note:} Some of the material is based on lectures given at Rensselaer Polytechnic Institute (RPI) 
Summer School in June 2022, Hampton University Graduate Studies (HUGS) program and Quantum Computing Bootcamp 
at Jefferson Lab in June 2023. I thank the organisers of those programs for inviting me to teach and learn about this research area. 
Please send suggestions/corrections to \texttt{raghav.govind.jha@gmail.com} 

\vspace{0.5cm}		
		\tableofcontents
		\vspace{0.6cm}
	\end{titlepage}

	\newpage
	\section{Introduction}
	
One of the quotes that profoundly expresses the present state of the development in quantum computation and information is by Wheeler. 
In his colorful language, he said - `I think of my lifetime in Physics as divided into three periods. In the first period, I was in the
the grip of the idea that everything
is particles. I call my second
period, everything is
fields, and now I am in the grip of
a new vision that everything is information'. 
In fact, we can safely add `quantum' to this since it appears that everything 
is `quantum information'. There are several physicists who have explored this line of thought, see for example Refs.~\cite{Landauer_1991, Landauer_1996, Preskill:1999he} and references therein to start a trail.\footnote{The quintessential example of how crucial information is can be understood from a well-known paradox due to Maxwell (1867). This is known as `Maxwell's demon'. This thought experiment is an apparent violation of the second law of thermodynamics. Szilard wrote an interesting paper in 1929 which took a big step towards the resolution, but it did not 'totally' explain the paradox. It was Landauer's work \cite{Landauer_1961} and Bennett's follow-up \cite{Bennett_1982} that was resolved after 115 years and was one of the major problems that raged physicists throughout the century. The solution lies in the fact that information is physical, just like the entropy or energy in thermodynamics. Additionally, it is interesting to note that von Neumann in his letter to Teller on April 8, 1947 wrote - ``I have some definite ideas concerning automata, memory, clearing, and irreversibility and entropy, somewhat connected to Szilard's treatment of Maxwell's demon and my old treatment of entropy and observability in quantum mechanics'' [p. 245, John von Neumann Selected Letters, American Mathematical Society (October 1, 2005)] but we believe it was never written up by either of the Martians. The statement appears to the author to be closely related to the idea of erasure and non-violation of the second law as established by Landauer et al.} 
	
	The goal of theoretical physics is to understand the principles that govern 
	the universe at different length scales. It is a challenging problem since the associated scales 
	vary over a wide range (about thirty orders of magnitude) from inside the nucleus ($10^{-15}$ m) to the 
	size of typical galaxies ($10^{15}$ m). We have two extremely accurate theories that govern each of them separately.
	The incredible progress made in the last 125 years has given us the basic principles of quantum 
	mechanics and the special theory of relativity which when combined into the framework of 
	`quantum field theory' (QFT) holds the key to explaining a major portion of all physical phenomena 
	occurring in nature at sub-nuclear scales and culminating in the Standard Model of particle physics. 
	Some jewels of this framework are - calculating the magnetic moment of the electron to about ten decimal places, and the discovery of the Higgs boson at the Large Hadron Collider (LHC) at CERN.
	On the other hand, at the other extreme of the spectrum, we have the highly successful theory of 
	general relativity which describes classical gravity, predicts the existence of 
	black holes and has been tested to great accuracy culminating with the discovery of gravitational waves. 
	
\begin{figure}
		\centering 
		\includegraphics[width=0.8\textwidth]{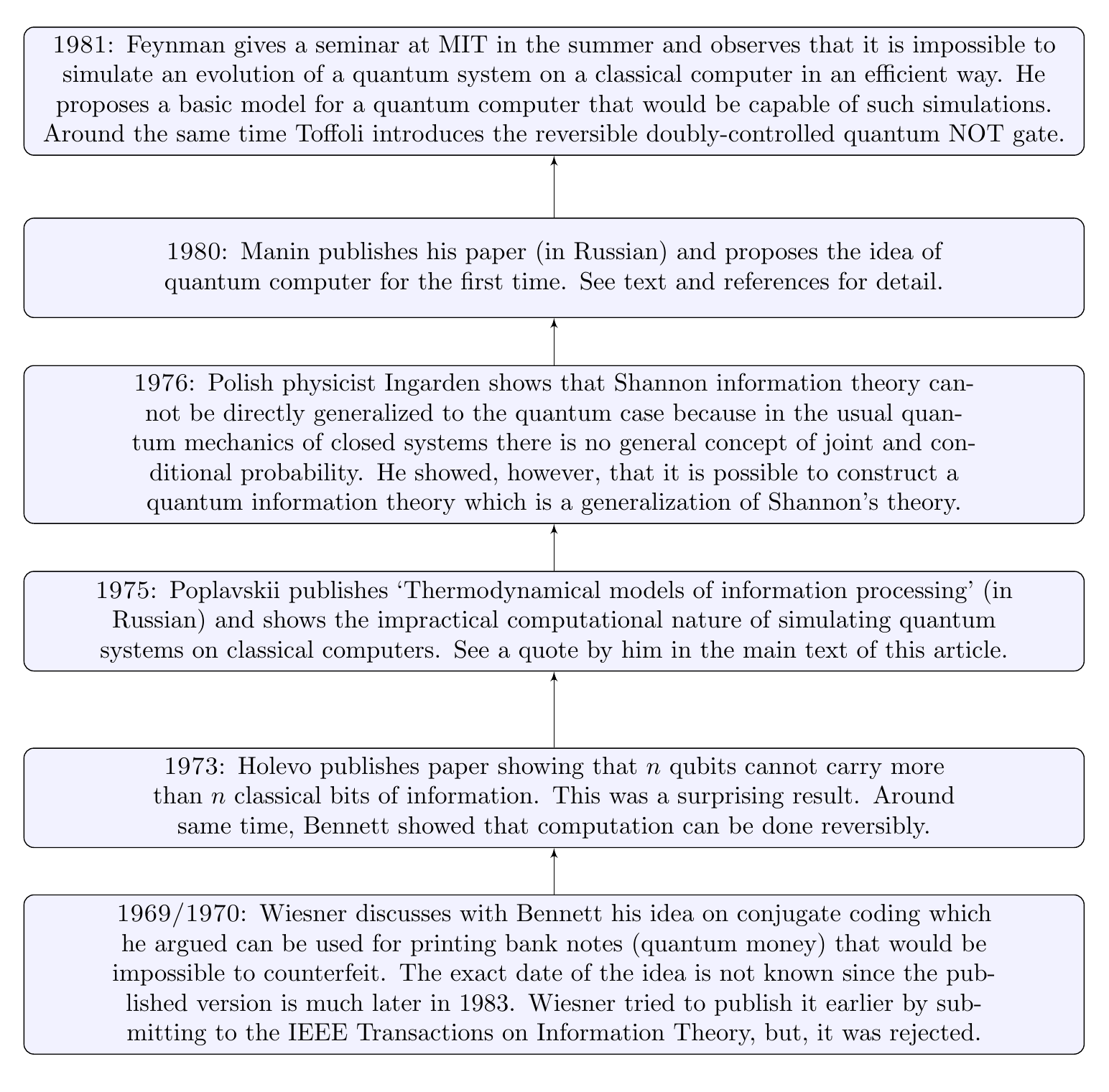}
		\caption{\label{fig:historyQC}The brief history of quantum computation.}
	\end{figure}

	Another topic close to physicists is that of computing and since computing is a physical process 
	it is governed by the laws of Physics. Computers\footnote{Computer is a machine that maps an array of bits (0 and 1) from one configuration to another. If it 
	manipulates quantum bits, it is called a quantum computer.} have changed and affected every facet of life around us. 
	From handling complex machines like airplanes to the RSA (Rivest–Shamir–Adleman) key protocol, which forms 
	the basis of all Internet transactions. 
	But there are certain classes of problems where even the fastest supercomputers of today's era fail. 
	They fail because we have to wait for thousands of years to solve a problem using classical ideas of bits, on which they are based. Though classical computation has rapidly progressed to the extent that we can now have a billion transistors in a single laptop, it is probably not possible that the size of these transistors can be made smaller than the size of say hydrogen atom which is about $10^{-10}$ m or the frequency of the clock speed can be increased more than 1 PHz ($10^{15}$ Hz) which is roughly the frequency of atomic transitions. So, it is clear that supercomputers will soon exhaust their capability, and problems with exponential complexity won't be solved using them. This leads to a paradigm shift in our idea of computing. 
	Along this direction, already in the 1970s, Benioff started thinking about the theoretical feasibility of quantum computing and his paper \cite{Benioff_1980, 			Benioff_1982} which described a quantum mechanical model of Turing Machines was one of the first papers in the field of quantum computing. This work was 			based on a classical description in 1973 of reversible Turing machines from Ref.~\cite{Bennett_1973}. In the same year, there was 
	another work \cite{Manin:1980abc} by Russian mathematician Manin who argued that the superposition and entanglement features of quantum mechanics would 		make for a substantial speedup compared to classical 
	methods. Soon after, Feynman \cite{Feynman1982} argued that quantum computers will be better suited to these tasks than any classical computer we can ever build because they can store much more information compared to classical computers. He pointed out that - `Let the computer itself be built of quantum mechanical elements which obey quantum mechanical laws'. In the early days of quantum computing, Poplavskii argued \cite{Qinfobook} - ``The quantum-mechanical computation of one molecule of methane requires about $10^{42}$ grid points. Assuming that at each point we have to perform only 10 elementary operations, and assume that those calculations are performed at temperatures of 0.003 Kelvin, we would still have to use all the energy produced on Earth during the last century''. This makes it clear that a full-fledged computation of quantum dynamics is beyond the reach of any classical computer. Nature is quantum mechanical and hence we have to make use of it. These problems and ideas gave rise to the field of quantum computation. In summary, quantum computation addresses one of the two issues that limit any classical computation: space (memory) and time. In quantum computers, we can store exponentially large information, so it is memory-wise way more efficient than its classical counterpart. We provide the first twelve years of the timeline of quantum computing in Fig.~\ref{fig:historyQC}.

	The usefulness of classical computers has been a monumental 
	achievement of human mankind in the past century. The basic fundamental
	idea of classical computers (or computing) is the binary two-state signal: 0 and 1. 
	It is practically the only thing a computer understands at its most fundamental level. 
	A computer practically does two things: 1) it stores information as a string of 0s and 1s
	(also known as classical bits), and 2) Transform them based on the instructions given. 
	It achieves the second task by using gates which perform simple logic operations to transform bits. Some examples of such gates are AND, NOT, OR gates. However, we know that at the microscopic level, the laws of Physics and hence nature
	is purely quantum-mechanical. It marks a paradigm shift in our understanding compared to the classical world, which is deterministic. In simple terms, this means that while a 
	classical bit can at a given time be either 0 or 1, a quantum bit (known as `qubit')
	\footnote{If the vector space dimensions is $d > 2$, then it is called qudits. If we have qudits with $d \to \infty$, then 
	we refer to it as `qumodes' or a continuous variable (CV) system. A quantum harmonic oscillator is one example.} can be 
	in a superposition of these two options and hence we can never conclude exactly which 
	one until we measure it. The existing quantum devices contain $\mathcal{O}(100)$ qubits
	but they are noisy, so actual efficiency might be a small fraction of it. They are sometimes referred to as 
	`Noisy Intermediate-Scale Quantum (NISQ)' devices are also sometimes referred to as `NISQ era'.  
	
To understand the power of principles of quantum mechanics, consider the amount of digital data in the world. It is estimated that the total digital data in the world by 2025 will be close to 175 ZB (1 ZB = $10^{21}$ bytes). This data can be contained in total if we have about 78 fully efficient qubits. This is the power of qubits (growth of `power of 2').\footnote{
Once a king offered any reward to a man who had done good for the kingdom. 
The man asked that a single grain of rice be placed on the first square of the chessboard. 
Then two grains on the second square, four grains on the third, and so on, and doubling each time.
The king was taken aback by such a small request and agreed. He ordered the treasurer to pay the sum agreed upon. 
After a few days, the treasurer explained that the sum could not be paid since by the time he got halfway through the board, 
the amount of grain required was more than the entire kingdom possessed. This is the exponential 
growth known to kings several thousand years before. I first heard this story from my grandfather when I was 8, however, 
this is now easily found online and also appears in some popular science books. 
}
However, such power often comes with its own set of troubles and limitations. For a review of quantum algorithms and other details, see Ref.~\cite{Shor2000}. 
As of today, there exists only a handful of useful\footnote{Polynomial or exponential speedup compared to best-known classical algorithms} quantum algorithms, and it is not yet known how many classes of 
problems will eventually benefit by exploiting the quantum features. One of the major applications of quantum computing is in the 
field of quantum many-body systems. The mightiest of computers today in the world cannot deal fully (without approximation and problems) with 
even a quantum system of highly 
interacting 100 electrons! Therefore, solving this is one of the major challenges in theoretical and computational Physics. Even though 
we only have a relatively small (< 100) number of non-error corrected qubits and the implementation faces several problems, 
it is expected that in the coming 2-3 decades the situation will change drastically. 
Even 100-200 error-corrected qubits in the future can completely 
revolutionize the nuclear computations and the whole of quantum chemistry
and several areas of Physics. 

Though we will mostly restrict to the theoretical side of quantum computing in these notes, it is good to point out some basic ideas about the experimental side of this story. 
The act of fabricating a proper logical qubit is a challenging task. In order to actually realize a quantum computer, we need to have qubits that can retain 
their properties and those which can be made to evolve/change as computationally desired. We must also prepare qubits in some initial state and measure them conveniently. In order to construct a computer as envisaged by Manin and others with these properties, DiVincenzo in 2000 \cite{DiVincenzo_2000} listed some conditions necessary to achieve this based on his attempts to construct a quantum computer (a machine that can efficiently simulate quantum systems such as solving the quantum many-body problems). These conditions are now known as `DiVincenzo criteria' and are mentioned below: 
	
	\begin{itemize}
		\item A scalable physical system with well-characterized qubits. 
		\item The ability to initialize to some fiducial state such as $\ket{000000\cdots 00}$. 
		\item Long decoherence time compared to gate operation time ($\tau_{\text{deco.}} \gg \tau_{\text{op.}})$. 
		\item A universal set of quantum gates. 
		\item A qubit-specific measurement capability. 
	\end{itemize}
	These conditions are for quantum computation, such as superconducting quantum computing or trapped ion approaches. There are two more conditions that apply to quantum communication. In order to understand these two conditions, we have to 
	explain stationary and flying qubits. Stationary qubits are those which are in the computer and on which we apply gates while flying qubits are ones that move (say, like photons). With these definitions, the last two conditions are: 
	\begin{itemize}
		\item The ability to interchange between stationary and flying qubits\footnote{Flying qubits have restricted use compared to stationary qubits. They are only used to pass the information over some distance, 
		whereas stationary qubits have a two-fold task: store information and perform calculations.}.  
		\item The ability to faithfully transmit flying qubits between specified locations. 
	\end{itemize}
	The experimental challenge is that these conditions can often only be partially met. Single photons have two polarization states and are almost ideal qubits, but the optical material required to make them interact is difficult in practice without loss of coherence. One can represent spin-1/2 particles as qubits as is often done theoretically and experimentally for the trapped ion method to quantum computing, but the phonons that are used to mediate the interactions have short coherence time. Another approach is to use electric charge as a qubit representation and use excellent techniques from modern electronics, but even here lies the problem of decoherence. One advanced approach is to use charge carriers in not a usual metal but in a superconductor. The superconductor qubit representation is favored due to the robustness of the Cooper pairs involved. In fact, there are now some advancements achieved over this as well. As of today's technology, we have some major categories for these constructions. The \emph{first} is the 
	`superconducting qubits' (SC). It was first observed that Josephson junction (JJ)\footnote{A Josephson junction is made of a non-superconducting material between two layers of superconducting electrodes. A Cooper pair of electrons can tunnel through the non-superconducting barrier from one superconductor to another.}
is the element that provides the condition to turn a superconducting circuit into a qubit. One example is Google's Sycamore 53-qubit chip, which belongs to this category. 
	In fact, these quantum processors use an advanced version of this which has more than one JJ and is known as `transmons' which is the shortened name for `transmission line shunted plasma oscillation qubit'. Both IBM and Google quantum processors use this modified qubit approach. It is closely related to the charge qubit (cooper pair box) but has better effectiveness due to improvement in decoherence time. There is \emph{another} approach known as `topological qubits' (Microsoft, Bell Labs) which underlies the application of Majorana fermions to create qubit states. There are other technologies based on trapped ions, silicon dots (Intel), and diamond vacancies. 
We will not discuss these here, since it would really digress us from the goal of this article. 
	
In these notes, the emphasis is to introduce the idea of quantum computing inspired by the connection to classical computing and then use some SDKs to perform some simple computations on IBM's backend machines. These are not meant to be exhaustive, and the purpose is not to solve some model in Physics which cannot be solved in a reasonable time using computers we have today, but to introduce the elements which might in coming decades be part of advanced quantum algorithms and computation that can achieve those dreams. However, we want to make it clear to the reader that even with quantum computing and reliable error correction some physical
problems might still not be solved in polynomial time. In order to understand this issue, one must talk in terms of complexity classes i.e., the problem of how the computational resources scale with increasing system size. This is one of the main problems in computer science. We say that $\textbf{P}$ is the set of problems where time scales as polynomial in system size on a deterministic Turing machine. 
This is also sometimes referred to as a problem that can be solved\footnote{It is interesting to note that G\"{o}del's letter to von Neumann in 1956 
(when the latter was in the hospital) discussed a function $\phi$ which has some relation to complexity classes. It is probably the first written instance of 
$\textbf{P}$ versus $\textbf{NP}$ issue. However, since von Neumann died early 1957, his answer to this letter was never known.} efficiently.
The classical complexity classes introduced for Turing machines of 
both kinds (deterministic or probabilistic) have since then been extended to admit quantum complexity classes. For example, 
$\textbf{BQP}$ stands for bounded-error quantum polynomial time and is referred to as the class of decision problems solvable by a quantum computer in polynomial time. It is the quantum analog of the complexity class $\textbf{BPP}$\footnote{$\textbf{BPP}$
is class of decision problems or languages that can be solved in polynomial time by probabilistic Turing machines with error probability bounded by 1/3}. But since a quantum circuit can simulate a classical one, both $\textbf{P}$, and its probabilistic analog, $\textbf{BPP}$, are contained in $\textbf{BQP}$. From a perspective of a physicist, it is important to know which problems belong to this complexity class since that would potentially benefit the most from the quantum computers in the future. For instance, the preparation of ground state of Hubbard model is not in $\textbf{BQP}$ but the time evolution is. It is also known that 
the scattering problem in interacting QFT lies in $\textbf{BQP}$. In fact, not all physical problems lie in $\textbf{BPP}$ or $\textbf{BQP}$, some of them belong to $\textbf{NP}$ which stands for \emph{non-deterministic polynomial time} requiring more than polynomial resources assuming that $\textbf{P} \neq \textbf{NP}$\footnote{$\textbf{P} \neq \textbf{NP}$ has enormous practical and economic importance because modern cryptography is based on assumption that they are not equal. This means that there exist problems that are impossible for computers to solve, but for which the solutions are easily checked (polynomial). A failure of this could create havoc in modern cryptography resulting in breakdown of secure transactions all over the world.} but whose solution can be verified in polynomial time. The quantum analog of $\textbf{NP}$ with polynomial intractable problems is denoted by $\textbf{QMA}$\footnote{QMA
is an acronym for Quantum Merlin Arthur}, and 
problems that belong to this class are \emph{unlikely} to be solved efficiently even 
with a quantum computer assuming $\textbf{BQP} \neq \textbf{QMA}$. 
$\textbf{QMA}$-hard (defined similarly to $\textbf{NP}$-hard), 
is a class of problems if every problem in $\textbf{QMA}$ 
can be reduced to it. A problem is said to be $\textbf{QMA}$-complete 
if it is $\textbf{QMA}$-hard and in $\textbf{QMA}$.
One of the motivations for quantum computers is the potential to solve interesting problems which are $\textbf{NP}$ hard like the model with 
sign problems. These cannot be efficiently solved on classical computers. However, some of these might still not belong to $\textbf{BQP}$ and therefore might still be 
inaccessible with quantum resources. For example, it was shown in Ref.~\cite{Jordan:2017lea} 
that scattering in scalar QFT is $\textbf{BQP}$-complete, indicating that all problems in $\textbf{BQP}$ 
can be mapped to scattering in scalar QFT with polynomial scaling time to solution, and further that the scattering problem itself is in 
$\textbf{BQP}$ and thus can be efficiently simulated quantum mechanically. We say a problem is $\textbf{BQP}$-complete when it is both 
$\textbf{BQP}$ and $\textbf{BQP}$-hard. Similarly, for the classical version, this means that a problem is said to be $\textbf{NP}$-hard if 
everything in $\textbf{NP}$ can be transformed in polynomial time into it even though it may not be in $\textbf{NP}$. 
Conversely, a problem is $\textbf{NP}$-complete if it is both in $\textbf{NP}$ and $\textbf{NP}$-hard. 

We now give an outline of the notes. In Sec.~\ref{sec:2}, we explain the notation and basic fundamentals. 
In the next section, we introduce the idea of quantum states, quantum logic gates, density matrices, and the important 
quantum Fourier transform method. In Sec.~\ref{sec:TQA}, we mention several quantum algorithms such as phase kickback, Deutsch algorithm, 
Grover's search algorithm, Kitaev's phase estimation, and Shor's algorithm. In Sec.~\ref{sec:VQE}, we discuss the variational quantum eigensolver (VQE) 
method to solve some simple problems. 
In Sec.~\ref{sec:QEC}, we mention an important feature of quantum computing i.e., quantum error correction 
and explain bit and phase flip and universal code to correct arbitrary (phase or flip) single-qubit errors. 
In the Appendix, i.e., Sec.~\ref{sec:last}, we provide codes and instructions on how the user
can run some simple simulations on a web browser (Google Collaboratory).

\begin{figure}
\centering 
\includegraphics[width=0.65\textwidth]{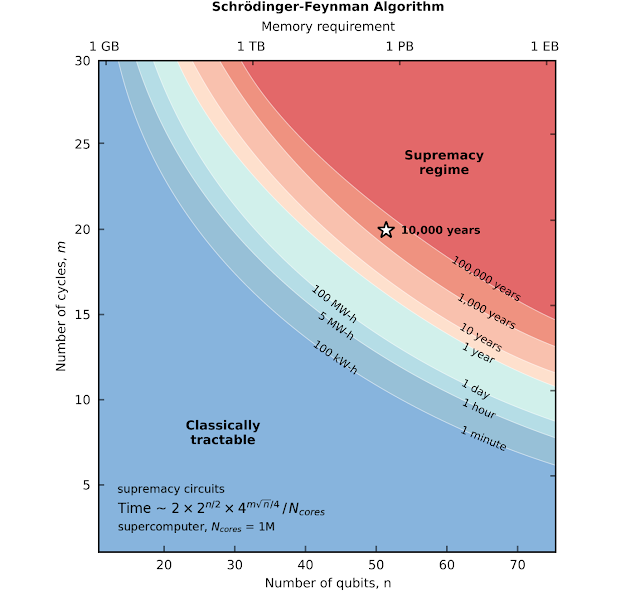}
\caption{\label{fig:flow}Estimate of the equivalent classical computation time assuming 1M CPU cores for quantum supremacy circuits as a function of the number of qubits and a number of cycles for the Schrödinger-Feynman algorithm. The star shows the estimated computation time for the largest experimental circuits. To know more about Schrödinger-Feynman (SF) algorithm, please see Ref.~\cite{SA2016} and references therein. 
In short, Schrödinger's method takes about $m^{2n}$ time and about $2n$ memory, while Feynman's method takes about $4^{m}$ time and about $m+n$ memory. Feynman's algorithm saves memory because it calculates an amplitude instead of keeping track of the state vector throughout like in the Schrödinger approach. However, the fact that more intermediate steps are to be done means more time. There are some methods that combine features of both, and those belonging to the class of SF algorithms. This figure is taken from Ref.~\cite{Google2019}. This diagram is only meant to provide a basic idea and should not be taken at face value.}
\end{figure}

\section{\label{sec:2}Notation and all that}

Before we delve into describing the notations and discussing basic ideas, we emphasize and point out the differences between digital classical and quantum
computing in Table \ref{tab:1}. 
\begin{table}[ht]
\centering
\begin{tabular}{|c||c|c|c| } 
\hline
Elements & Classical & Quantum \\ 
\hline \hline 
Algebra & Boolean & Linear \\  \hline 
Basic Unit & Bit & Qubit  \\ \hline
Gates & Logic gates & Unitary gates \\  \hline 
Reversibility of gates & Sometimes & Always \\ \hline
Example of universal gate set & $\{\rm{NAND}\}$ &  $\{\rm{H, T, CNOT}\}$ \\ \hline
Correction of errors & Repetition & Shor-like code \\ \hline
\hline
  \end{tabular}
  \caption{Comparison between classical and quantum digital computing.}
  \label{tab:1}
\end{table}
We will restrict to the well-known Dirac's\footnote{Probably inspired by Grassmann's similar notation used several decades before} bra-ket notation\cite{Dirac_1939} which deals with states in Hilbert space based on von Neumann abstract and mathematical precise formulation of quantum mechanics written down around 1932\footnote{It might be surprising that 
just about three years later in 1935, he wrote to Birkhoff from 1935 -- `I would like to make a confession which may seem immoral: I do not believe in Hilbert space anymore'. We encourage the interested reader to explore this direction if interested.}. We will briefly review this for interested readers. Suppose we have a column and row vector as given below:
\begin{equation}
\ket{B}  = \begin{pmatrix} B_{1} \\ \ddots \\ \ddots \\ B_{N} \end{pmatrix}, ~~~~
\bra{A}  = \begin{pmatrix} A_{1}^{*}  \ddots  \ddots ~A_{N}^{*}. \end{pmatrix}
\end{equation}
We can do two operations using them by computing the inner product, i.e., $\langle A \vert B \rangle$ or the outer product denoted by $\ket{A} \ket{B}$. The first returns a number, while the latter gives a matrix. The row vector is called `bra' while the column vector is called `ket' and they are both dual to each other (in the vector space sense). 
This state denoted as `ket' is postulated to contain all the information about the physical state
which we wish to know. We have the following relations: $\bra{A}^{\dagger} = \ket{A}$, and $\ket{A}^{\dagger} = \bra{A}$. In quantum mechanics, the state of a system is described by a ray $\ket{\psi}$ in Hilbert space, $\mathcal{H}$
where the ray has unit norm. If we have, $\lambda = e^{i \phi}$ then we see that $\ket{\psi}$ and $\lambda \ket{\psi}$ both have the same norm because $\lambda$ is just a phase, and they represent the same physical state. The phase of the ray is not an observable. The rays of Hilbert space $\mathcal{H}$ are the equivalence class of unit vector that only differ by a phase. 
Two states can be added together as: $  \ket{\alpha} + \ket{\beta}  = \ket{\gamma}$. In fact, if 
$\ket{\psi_0}$ and  $\ket{\psi_1}$ are two orthogonal states, then their 
linear superposition ($c_1 \ket{\psi_0} + c_2 \ket{\psi_1})$ is also a well-defined state but must satisfy 
$|c_1|^2 + |c_2|^2 = 1$. We refer to $c_1$ and $c_2$ as `amplitudes' corresponding to $\ket{\psi_0}$ and $\ket{\psi_1}$
respectively. We can multiply a ket by a complex number $c$ as: $ \ket{\alpha} = c \kp$  or $\kp c $, 
however, physically speaking there is no difference between $\ket{\alpha}$ and $\kp$ as discussed above. 
When we want to measure an observable, we apply the corresponding operator from the left on the ket
as: $ A \cdot \kp$ or  $\hat{A} \kp $. Two kets denoted by $\alpha$ and $\beta$ are said to be orthogonal if: $ \bra{\alpha} \ket{\beta} = \langle \alpha \vert \beta \rangle =  \langle \beta \vert \alpha \rangle^{*} = 0$. 
There are also some operations that are not allowed. The operator must be either on the left of ket 
or on the right of a bra, i.e., $ \kp A, A \bp$ are both incorrect for an operator $A$. Some products like $ \ket{\alpha} \ket{\beta}$ are not allowed if the ket vectors belong to the same vector space. Note that when we write $\ket{01} = \ket{0} \ket{1}$ 
later on, we mean vector spaces of two distinct qubits so it is allowed. The expectation value of any observable corresponding
to the Hermitian operator $A$ is given by: $ \langle A \rangle = \langle \alpha \vert \hat{A} \vert \alpha \rangle$. In fact, we can 
also write this as:

\[ \langle A \rangle  = \sum_{a^{\prime}, a^{\prime\prime}} \langle \alpha \vert a^{\prime\prime} \rangle  \langle a^{\prime\prime} \vert A \vert a^{\prime} \rangle \langle a^{\prime}  \vert \alpha \rangle  = \sum_{a^{\prime}} a^{\prime} ~  |\langle a^{\prime} \vert \alpha \rangle|^{2} \]

Consider a spin-1/2 particle with spin-up $\ket{\uparrow}$ or down $\ket{\downarrow}$. The Hilbert space in this case is two-dimensional with two orthonormal basis which we denote as $\ket{0}$ and $\ket{1}$. This is an example of a quantum bit (`qubit'). One often refers to the orthonormal basis $\{\ket{0}, \ket{1}\}$ as computational basis. Any state $\ket{\alpha}$ can be decomposed as a convex combination of projectors onto pure
states \footnote{Often mentioned as `inserting an identity'} i.e., $\ket{\alpha} = \sum_{i} p_{i} \ket{\psi_i} \langle \psi_i \vert \alpha \rangle$. In quantum computation, we have the following standard definitions:
\begin{equation}
\ket{0} = \begin{pmatrix} 1 \\ 0  \end{pmatrix}, ~~~ \ket{1} = \begin{pmatrix} 0 \\ 1  \end{pmatrix}. 
\end{equation}
These are one-qubit state. We can have two-qubit states as well, and then we use shorthand like $ \ket{00} = \ket{0} \ket{0} = \ket{0} \otimes \ket{0}$,  $ \ket{10} = \ket{1} \ket{0}$ and so on. Therefore we have, $\ket{10} = \begin{pmatrix} 0 \\ 0 \\ 1 \\ 0  \end{pmatrix}$, 
$\ket{11} = \begin{pmatrix} 0 \\ 0 \\ 0 \\ 1  \end{pmatrix}$ and so on. We refer to a state as `normalized' if $\langle \psi \vert \psi \rangle = 1$. This is called the inner product of a `bra' and `ket'. 
For readers comfortable with this notation, nothing remains to be said, but for those who are new, we give a small example of `bra-ket' gymnastics now. Suppose we have a system in state $\ket{i}$ and another in $\ket{j}$, then the composite state is written as $\ket{ij} = \ket{i} \ket{j}$. 
For example, $ \bra{j} \bra{i} \ket{k} \ket{l} = \ket{ij}^{\dagger} \ket{kl} = \bra{j} \langle i \vert k \rangle \ket{l} = 
\langle i \vert k \rangle  \langle j \vert l \rangle $. One can also define an `outer product' as, $ \kp \bra{\phi} $ which is a matrix. Suppose we take 
$\ket{0}, \ket{1}$ and construct $\rho = 1/2(\ket{0} \bra{0} - \ket{0} \bra{1} - \ket{1} \bra{0} + \ket{1} \bra{1})$. 
The matrix $\rho$ is given by: 

\[  \rho = \frac{1}{2}
\begin{pmatrix}
1& -1 & \\ 
-1 & 1 & \\ 
\end{pmatrix}. \] 
More systematically, we define the outer product as:
\begin{equation}
\kp \bra{\phi} = \begin{pmatrix}
\alpha & \\ 
\beta & \\ 
\end{pmatrix} \begin{pmatrix}
\gamma^{*}  & \delta^{*} & \\
\end{pmatrix} = \begin{pmatrix}
\alpha \gamma^{*} & \alpha \delta^{*} & \\ 
\beta \gamma^{*}  &  \beta \delta^{*} & \\ 
\end{pmatrix}.
\end{equation}
In fact, any square matrix can be written as a linear combination of outer products. The simplest example is 
to write any 2 $\times$ 2 matrix in this form given by:
\begin{equation}
A = A_{00} \ket{0} \bra{0} + A_{01} \ket{0} \bra{1} + A_{10} \ket{1} \bra{0} + A_{11} \ket{1} \bra{1}
\end{equation} 
This representation of a matrix in terms of `ket-bra' also provides a simple interpretation 
of some quantum gates. For example, $  \ket{00} \bra{00} +  \ket{01} \bra{01} +  \ket{10} \bra{11} + \ket{11} \bra{10}$
gives a representation of CNOT gate\footnote{We will soon learn about this gate in the next sections} 
and implies the corresponding transformation.

\begin{mybox2}
\textsc{Example 1:} Consider $\kp = \ket{a} \otimes \ket{b}$, $A \ket{a} = a\ket{a}$, and 
$B \ket{b} = b\ket{b}$. Compute $ A \otimes B \kp$. 
\end{mybox2}
\begin{proof}
We can write $ A \otimes B \kp$ as $ A \otimes B \ket{ab} = (A \otimes B) \ket{a} \otimes \ket{b} $
and then by distributing the operators write this as $ A \ket{a} \otimes B \ket{b} = ab \kp$. 
\end{proof}

\section{\label{sec:3}States, quantum gates, density matrices, and rules} 

In classical logic gates, the input signal can be a string of either 1 or 0 and there is a single output signal. 
The classical gates can be reversible or non-reversible. The NOT gate is an example of a reversible gate. 
For example, to see this clearly, consider the truth table representing the action of a 
XOR gate for two input classical bits, A and B. The output is given by $ A \oplus B$. 
We mention this along with other two logic gates in the table below. 

\begin{center}
\begin{tabular}{ |c|c|c|c|c|c| } 
\hline
$A$ & $B$ & AND ($A \cdot B$) & OR ($A + B$) & XOR($A \oplus B$) \\
\hline
0 & 0 & 0 & 0 & 0 \\  \hline 
0 & 1 & 0 & 1 & 1  \\  \hline 
1 & 0 & 0 & 1 & 1  \\  \hline 
1 & 1 & 1 & 1 & 0  \\
\hline
\end{tabular}
\end{center}
This can be extended easily to have three input lines, the output is given by $A \oplus B \oplus C$ and will be high (1) if the number of highs in the input is an odd number. The fact that there is a single output line clearly implies that we can never run this operation backward. Given an output, we can never know what was the actual input. 
So XOR seems like a non-reversible gate. However, if we allow a garbage output (say $A$), then XOR can be
made reversible. The AND gate is not reversible (even if we allow an extra output bit). NAND and NOR are the 
universal gate set for classical computers (which means that any gates can be constructed from them). The representation of a NAND gate made up of two transistors is shown below.

\begin{figure}
\centering 
\includegraphics[width=0.35\textwidth]{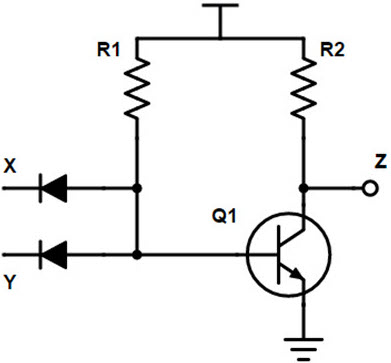}
\caption{\label{fig:NAND}The construction of NAND gate using transistor and diodes. The Z is the output and the T-shaped sign signals the input $+V$ voltage. We say the input is high (1) if it is $+V$ while it is 0 if grounded.}
\end{figure}

\subsection{States on the Bloch sphere and gates}

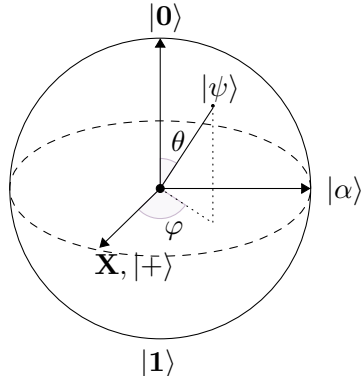
\begin{figure} 
\begin{center}
\begin{tikzpicture}[line cap=round, line join=round, >=Triangle]
\clip(-2.19,-2.49) rectangle (3.26,3.28);
\draw [shift={(0,0)}, lightgray, fill, fill opacity=0.1] (0,0) -- (56.7:0.4) arc (56.7:90.:0.4) -- cycle;
\draw [shift={(0,0)}, lightgray, fill, fill opacity=0.1] (0,0) -- (-135.7:0.4) arc (-135.7:-33.2:0.4) -- cycle;
\draw(0,0) circle (2cm);
\draw [rotate around={0.:(0.,0.)},dash pattern=on 3pt off 3pt] (0,0) ellipse (2cm and 0.9cm);
\draw (0,0)-- (0.70,1.07);
\draw [->] (0,0) -- (0,2);
\draw [->] (0,0) -- (-0.81,-0.79);
\draw [->] (0,0) -- (2,0);
\draw [dotted] (0.7,1)-- (0.7,-0.46);
\draw [dotted] (0,0)-- (0.7,-0.46);
\draw (-0.08,-0.3) node[anchor=north west] {$\varphi$};
\draw (0.01,0.9) node[anchor=north west] {$\theta$};
\draw (-1.01,-0.72) node[anchor=north west] {$\mathbf {X}, \ket{+}$};
\draw (2.07,0.3) node[anchor=north west] {$\mathbf {\ket{\alpha}}$};
\draw (-0.3,2.6) node[anchor=north west] {$\mathbf {\ket{0}}$};
\draw (-0.4,-2) node[anchor=north west] {$\mathbf {\ket{1}}$};
\draw (0.4,1.65) node[anchor=north west] {$|\psi\rangle$};
\scriptsize
\draw [fill] (0,0) circle (1.5pt);
\draw [fill] (0.7,1.1) circle (0.5pt);
\end{tikzpicture}
\caption{\label{fig:BS}The representation of the possible states of single qubit on the Bloch sphere. We have  $\ket{\alpha} =  \frac{1}{\sqrt{2}}(\ket{0} + i \ket{1})$ and $\ket{\psi} = \cos(\theta/2) \ket{0} + e^{i \varphi} \sin(\theta/2) \ket{1}$ which is just $ P(\phi) R_{y}(\theta) \ket{0}$. The $Z$-basis is the `computational basis'. }  
\end{center}
\end{figure}

The inputs for the quantum case are known as `states'. They can either be $\ket{0}$, $\ket{1}$ or in a superposition of these two states or in some entangled state of these qubits. We will see this in more detail later on. The classical logic gate XOR we considered above, or any other logic gate behaves very differently compared to a quantum logic gate. In quantum circuits, the data (or states) are represented by `qubits', the operations are described by unitary quantum logic gates and the results are obtained by doing measurements. The first point of difference is that in this case, the number of inputs is the same as the number of outputs. The circuit can be run backward, and we can reproduce input if we want. One striking property of quantum logic gates is that they can change the inputs in ways that cannot be done classically. One example of this is to look at Hadamard gate (H). 
If we take an input qubit to be $\ket{0}$ and then act on it with $H$ gate, we have:
\[ H  \ket{0} \equiv \ket{+} = \frac{1}{\sqrt{2}} \Big( \ket{0} + \ket{1} \Big), \] 
\[ H  \ket{1} \equiv \ket{-}  = \frac{1}{\sqrt{2}} \Big( \ket{0} - \ket{1} \Big), \] 
which is now in a superposition state. The Hadamard gate can also be expressed as a 
$\pi/2$ rotation around the Y-axis, followed by a $180^{\circ}$ rotation around 
the X-axis i.e. $H = \sqrt{XY}$. We can represent this gate by the following matrix: 
\begin{equation} 
H = \frac{1}{\sqrt{2}} 
\begin{pmatrix} 1 & 1 \\ 1 & -1 
\end{pmatrix}
= \frac{\vert 0 \rangle + \vert 1 \rangle}{\sqrt{2}} \langle 0 \vert  + \frac{\vert 0 \rangle - \vert 1 \rangle}{\sqrt{2}} \langle 1 \vert 
\end{equation}
A natural way of representing the qubit state is by using the Bloch sphere\footnote{Note that this is mathematically
referred to as `Riemann sphere' and is denoted as $\mathbb{CP}^1$ which is $\equiv SU(2)/U(1)$} representation as shown in Fig.~\ref{fig:BS}. 
The Hadamard transformation (also known as `Walsh-Hadamard') is a special case of the more general Fourier transform (FT) which plays a major role in quantum algorithms. One special property of this gate is that $H = H^{-1}$ and therefore: 
\[ H \Big(\frac{1}{\sqrt{2}} \Big( \ket{0} + \ket{1} \Big)\Big)  = \ket{0}. \]
One of the frequently used quantum logic gate is the quantum analog of XOR gate and is known as 
CNOT (controlled NOT) or CX gate. The matrix representation of CNOT is given by: 
\begin{equation} 
\begin{pmatrix}
1 & 0 & 0 & 0 \\ 
0 & 1 & 0 & 0 \\ 
0 & 0 & 0 & 1 \\ 
0 & 0 & 1 & 0 \\ 
\end{pmatrix}
\end{equation}
and the truth table in this case can be written as:

\begin{center}
\begin{tabular}{ |c|c|c|c| } 
\hline
$\ket{A}$ & $\ket{B}$ & \ket{A} & $ \ket{A \oplus B} $ \\
\hline
\ket{0} & \ket{0} & \ket{0} & \ket{0} \\  \hline 
\ket{0} & \ket{1} & \ket{0} & \ket{1}  \\  \hline 
\ket{1} & \ket{0} & \ket{1} & \ket{1} \\  \hline 
\ket{1} & \ket{1} & \ket{1} & \ket{0} \\  \hline 
\end{tabular}
\end{center} 
This can also be written as:
\begin{equation}
\text{CNOT} = \ket{0} \bra{0} \otimes \mathbb{1} + 
\ket{1} \bra{1} \otimes \sigma_{x},  
\end{equation}
where $\sigma_x$ is one of the Pauli matrices. For this gate, we have two outputs (unlike classical XOR gate) and the first output is sometimes called `garbage output' since it is just the first input. The second input ($\ket{B}$) is called the `target' while the first is called the `control' (or control line) for obvious reasons. The CNOT gate is a two-qubit operation, in which the first qubit 
is usually referred to as the control qubit and the second qubit as the target qubit. Now suppose an input state is given by $ \ket{\psi}  = \begin{pmatrix} \alpha \\ \beta \\ 0 \\0 \end{pmatrix} $, it is easy to check that if we pass it through CNOT gate, it will remain unchanged. We leave this simple exercise for the reader. 
There are many other important quantum gates such as the Pauli-X (simply $X$) which is a rotation through $\pi$ radians around the x-axis and similar for other directions ($y$ and $z$): 
\[ \ X = \sigma_x = \sigma_1 = \begin{pmatrix} 0 & 1 \\ 1 & 0 \end{pmatrix}  \] 
\[ \ Y = \sigma_y = \sigma_2 = \begin{pmatrix} 0 & -i \\ i & 0 \end{pmatrix}  \] 
\[ \ Z = \sigma_z = \sigma_3 = \begin{pmatrix} 1 & 0 \\ 0 & -1 \end{pmatrix}.  \] 
These three matrices also have corresponding rotation matrices, such as 
$R_{X}(\theta) = \exp(-i X \theta/2)$. $X$ basis is also called Hadamard basis, since it can be 
generated from computational basis ($Z$) by acting by $H$. In Fig. \ref{fig:BS}, we denoted the state $\ket{\psi} = \cos(\theta/2) \ket{0} + e^{i \varphi} \sin(\theta/2) \ket{1}$. 
The outer product of this state is closely related to the dot product of the unit vector in spherical coordinates\footnote{Recall that $(n_1, n_2, n_3) = (\cos \phi \sin \theta, \sin \phi \sin \theta, \cos \theta)$}
with Pauli matrices as:

\begin{align}
    \ket{\psi}\bra{\psi}
      & = \begin{pmatrix}\cos \frac{\theta}{2} \\ e^{i\phi}\sin \frac{\theta}{2} \end{pmatrix}
          \begin{pmatrix}\cos \frac{\theta}{2} & e^{-i\phi}\sin \frac{\theta}{2} \end{pmatrix} \nonumber \\
      & = \begin{pmatrix}
            \cos^2(\frac{\theta}{2}) & e^{-i\phi}\cos \frac{\theta}{2} \sin \frac{\theta}{2} \nonumber  \\
            e^{i\phi} \cos \frac{\theta}{2} \sin \frac{\theta}{2} & \sin^2(\frac{\theta}{2})
          \end{pmatrix} \\
      & = \frac{1}{2}
          \begin{pmatrix}
            1 + \cos\theta & e^{-i\phi}\sin\theta \nonumber \\
            e^{i\phi}\sin\theta & 1 - \cos\theta
          \end{pmatrix} \\
      & = \frac{1}{2}
          \begin{pmatrix}
            1 + \cos\theta & \cos\phi\sin\theta - i\sin\phi\sin\theta \nonumber \\
            \cos\phi\sin\theta + i\sin\phi\sin\theta & 1 - \cos\theta
          \end{pmatrix} \\
      & = \frac{1}{2}
          \begin{pmatrix}
            1 + n_3 & n_1 - i n_2 \nonumber \\
            n_1 + i n_2 & 1 - n_3
          \end{pmatrix} \\
      & = \frac{1}{2}(I + \hat{n} \cdot \vec{\sigma}). 
  \end{align}
The matrix $\hat{n} \cdot \vec{\sigma}$ is an involutory matrix (i.e., a matrix which is both unitary and hermitian and whose square is an identity matrix). 
It is straightforward to show that the eigenvalues of $\hat{n} \cdot \vec{\sigma}$ is $\pm1$ and denoting corresponding eigenvectors $\alpha$ and $\beta$, 
we can write: $\hat{n} \cdot \vec{\sigma} = \ket{\alpha} \bra{\alpha}  - \ket{\beta} \bra{\beta}$ and 
$\exp(i \theta \hat{n} \cdot \vec{\sigma}) = \cos(\theta) \mathbb{1} + i \sin \theta \hat{n} \cdot \vec{\sigma}$. 
 \begin{mybox2}
\textsc{Example 2:} Show that $ \hat{n} \cdot \vec{\sigma} = \cos \theta \ket{0} \bra{0} + e^{-i\phi} \sin \theta \ket{0} \bra{1} + e^{i\phi} \sin \theta \ket{1} \bra{0} - \cos \theta \ket{1} \bra{1}$
\end{mybox2}
\begin{proof} 
We begin by noting that we have, 
\begin{equation}
\hat{n} \cdot \vec{\sigma} = \begin{bmatrix}
\cos\theta & \cos\phi\sin\theta - i\sin\phi\sin\theta \nonumber \\
            \cos\phi\sin\theta + i\sin\phi\sin\theta & - \cos\theta
\end{bmatrix}.
\end{equation}
Then using $e^{i \theta} = \cos \theta + i \sin \theta$ and matrix representations of $\ket{00} \bra{00}, \ket{01} \bra{01}, \ket{10} \bra{10}$, and $\ket{11} \bra{11}$, we obtain the required.  
\end{proof}
\vspace{10mm}

\begin{mybox2}
\textsc{Example 3:} Show that any single qubit unitary can be written as:
\begin{equation}
U = \begin{bmatrix}
e^{i(\alpha -\beta/2 - \delta/2)} \cos(\theta) ~~ & e^{i(\alpha -\beta/2 + \delta/2)} \sin(\theta)  \\
e^{i(\alpha +\beta/2 - \delta/2)} \sin(\theta)~~ & e^{i(\alpha +\beta/2 + \delta/2)} \cos(\theta) 
\end{bmatrix}.
\end{equation}
Furthermore, show that $U = e^{i \alpha} R_{z}(\beta) R_{y}(\gamma) R_{z}(\delta)$. 
\end{mybox2}
\begin{proof}
We now give the proof of the above example. 
We can define general $U$ as: 
\[ U=\begin{pmatrix}a&b\\c&d\end{pmatrix}. \]
We have constraints: $\vert a \vert^{2} + \vert c \vert^{2} = 1$, $\vert b \vert^{2} + \vert d \vert^{2} = 1$, 
and $a^{*} b + c^{*} d = 0$. Then we have, 
\[ U=\begin{pmatrix}e^{\alpha_{11}}\cos\theta& e^{\alpha_{12}}\sin\theta\\
e^{\alpha_{21}}\sin\theta & e^{\alpha_{22}}\cos\theta
\end{pmatrix}. \]
The last condition (orthogonal columns) gives: $ e^{i(\alpha_{11}-\alpha_{12})}+e^{i(\alpha_{21}-\alpha_{22})}=0, $ where $\alpha_{11}$ is related to the other three by 
$ \alpha_{11}=\alpha_{12}+\alpha_{21}-\alpha_{22}+\pi $. Now we can set $\theta = \gamma/2$, $ \alpha{12} = \alpha - \beta/2 + \delta/2 + \pi$, $ \alpha{21} = \alpha + \beta/2 - \delta/2 $, and $ \alpha{22} = \alpha + \beta/2 +  \delta/2 $. 
\end{proof}

\vspace{10mm} 

\begin{figure}[h]
\centering
\leavevmode
\large{
\Qcircuit @C=1em @R=1em {
&  \gate{H} & \qw ~~~~~ =
}} ~~~~ $  \frac{1}{\sqrt{2}}   \begin{bmatrix}
1 ~&~ 1 \\
1 ~&~ -1
\end{bmatrix}$~~, ~~~~
\centering
\leavevmode
\large{
\Qcircuit @C=1em @R=1em {
\lstick{\ket{0}}  &  \gate{H} & \qw & \ket{+}   & \\ 
}} ~~~~~~~

\vspace{7mm} 
\centering
\leavevmode
\large{
\Qcircuit @C=1em @R=1em {
&  \gate{X} & \qw ~~~~~=
}} ~~~~~$ \begin{bmatrix}
0 ~&~ 1 \\
1 ~&~ 0   
\end{bmatrix}$~~, ~~~~
\centering
\leavevmode
\large{
\Qcircuit @C=1em @R=1em {
\lstick{\ket{0}}  &  \gate{X} & \qw & \ket{1}   & \\ 
}} ~~~~~~~

\vspace{7mm} 
\centering
\leavevmode
\large{
\Qcircuit @C=1em @R=1em {
&  \gate{Z} & \qw ~~~~~ =
}} ~~~~ $  \begin{bmatrix}
1 & 0 \\
0 & -1
\end{bmatrix}$~~, ~~~~
\centering
\leavevmode
\large{
\Qcircuit @C=1em @R=1em {
\lstick{\ket{1}}  &  \gate{Z} & \qw &  & -\ket{1}   & \\ 
}} ~~~~~~~

\vspace{7mm} 
\centering
\leavevmode
\large{
\Qcircuit @C=1em @R=1em {
&  \gate{Y} & \qw ~~~~~ =
}} ~~~~ $  \begin{bmatrix}
0 ~&~ -i \\
i ~&~ 0
\end{bmatrix}$

\vspace{7mm} 
\centering
\leavevmode
\large{
\Qcircuit @C=1em @R=1em {
&  \gate{P} & \qw ~~~~~ =
}} ~~~~ $  \begin{bmatrix}
1 ~&~ 0 \\
0 ~&~ e^{i \phi} 
\end{bmatrix}$

\vspace{7mm} 
\centering
\leavevmode
\large{
\Qcircuit @C=1em @R=1em {
&  \gate{R_z(\theta)} & \qw ~~~~~ =
}} ~~~~ $  \begin{bmatrix}
e^{- i\theta/2} ~&~ 0 \\
0 ~&~ e^{i \theta/2} 
\end{bmatrix}$

\vspace{7mm} 
\centering
\leavevmode
\large{
\Qcircuit @C=1em @R=1em {
&  \gate{S} & \qw ~~~~~ =
}} ~~~~ $  \begin{bmatrix}
1 ~& 0 \\
0 & i 
\end{bmatrix}$

\vspace{7mm} 
\centering
\leavevmode
\large{
\Qcircuit @C=1em @R=1em {
&  \gate{T} & \qw ~~~~~ =
}} ~~~~ $  \begin{bmatrix}
1 & 0 \\
0 & e^{\frac{i \pi}{4}} 
\end{bmatrix} = e^{\frac{i \pi}{8}}  \begin{bmatrix}
e^{\frac{-i \pi}{8}} & 0 \\
0 & e^{\frac{i \pi}{8}} 
\end{bmatrix}$
\caption{Definition of Hadamard ($H$), Paulis ($X, Z, Y$), Phase ($P$), Rotation around z-axis ($R_z(\theta)$), $\pi/2$ phase ($S$) and $\pi/8$ ($T$) single qubit gates and example of how some of them act on a single qubit.}
\label{fig:gates1}
\end{figure}
Another useful quantum gate which is an extension of CNOT gate is known as 
controlled-controlled NOT (CCNOT) gate or Toffoli gate. This is a three-qubit operation defined by:
\begin{equation} 
\begin{pmatrix}
1 & 0 & 0 & 0 & 0 & 0 & 0 & 0 \\ 
0 & 1 & 0 & 0 & 0 & 0 & 0 & 0 \\ 
0 & 0 & 1 & 0 & 0 & 0 & 0 & 0 \\ 
0 & 0 & 0 & 1 & 0 & 0 & 0 & 0 \\ 
0 & 0 & 0 & 0 & 1 & 0 & 0 & 0 \\ 
0 & 0 & 0 & 0 & 0 & 1 & 0 & 0 \\ 
0 & 0 & 0 & 0 & 0 & 0 & 0 & 1 \\ 
0 & 0 & 0 & 0 & 0 & 0 & 1 & 0 \\ 
\end{pmatrix}
\end{equation} 
resulting in 
\begin{align}
& \ket{000} \to \ket{000}  \nonumber \\ 
& \ket{001} \to \ket{001}  \nonumber \\ 
& \ket{010} \to \ket{010}  \nonumber \\ 
& \ket{011} \to \ket{011}  \nonumber \\ 
& \ket{100} \to \ket{100}  \nonumber \\ 
& \ket{101} \to \ket{101}  \nonumber \\ 
& \ket{110} \to \ket{111} \nonumber  \\ 
& \ket{111} \to \ket{110} 
\end{align}
\begin{mybox}
\textsc{$\blacktriangleright$ Question 1:} Write the $Z$ and $P$ gate in the outer product notation and compute the action of $P$ on a general  
qubit state. 
\end{mybox}
In fact, Toffoli and any single qubit gate which does not preserve the computational basis \
(such as Hadamard) form a universal gate set \cite{shi2002both} for quantum computation. 
But, Toffoli by itself is universal for classical computation. Hence, it appears that 
one only needs to add the Hadamard gate to make a 'classical' set of 
gates quantum universal. And this is true and was proved in Ref.~\cite{aharonov2003simple}. 
Another interesting fact about Toffoli which is often required is how can we efficiently
decompose it into say two-qubit CNOT gates. In Ref.~\cite{2008arXiv0803.2316S}, it was shown that 
one needs at least $2n$ CNOT to implement any $n$-qubit Toffoli. For the 
standard case of $n=3$, one needs at least 6 CNOTs. 
In some of the algorithms we discuss later in the notes, another 
important operation is the swapping of qubits. This is represented by the 
matrix as:
	\[ 
	\rm{SWAP} =
	\begin{pmatrix}
		1 & 0 & 0 & 0 \\
		0 & 0 & 1 & 0 \\
		0 & 1 & 0 & 0 \\
		0 & 0 & 0 & 1
	\end{pmatrix} 
	\] 
	The gate achieves $|a, b\rangle \rightarrow |b, a\rangle$ and can be written in terms of CNOT gates as: \footnote{We can see this as follows: $ \vert a, b \rangle \to \vert a, a \oplus b \rangle$ after application of first CNOT gate. 
		Then the second CNOT (note with control now down and target on upper qubit) 
		transforms $ \vert a, a \oplus b \rangle  \to \vert a \oplus (a \oplus b), a \oplus b \rangle = |b, a \oplus b \rangle$.
		Then the last CNOT transforms this to the desired $ |b, a \rangle$
	}
	\begin{equation}
		\Qcircuit @C=2.2em @R=2.7em @! {
			\lstick{\ket{a}} & \qswap & \qw  \\ 
			\lstick{\ket{b}} & \qswap \qwx  & \qw 
		} 
		\hspace{24mm} 
		\Qcircuit @C=2.2em @R=0.6em @! {
			& \ctrl{2} & \targ  & \ctrl{2} & \qw \\
			= \hspace{25mm} & & & & \\ 
			& \targ & \ctrl{-2}  & \targ & \qw 
		} 
	\end{equation}
A closely related version of this -- which is often used to reorder or swap qubits with a control 
is the controlled SWAP (also known as `Fredkin gate' \footnote{For a photonic based quantum computing implmentation of this gate, see Ref.~\cite{PhysRevLett.62.2124}})is an important gate that can be implemented as: 
$ \texttt{qc.cswap(0,1,2)} $. 
We can see how this gate can be decomposed in terms of 
CNOT gate in \QIS~as: \\
$\texttt{qc.decompose().draw(output='mpl',style='iqx')}$. The representation of the gate is given below: 
\begin{equation} 
\text{CSWAP} =
|0 \rangle \langle 0| \otimes \mathbb{1} \otimes \mathbb{1} +
|1 \rangle \langle 1| \otimes \rm{SWAP} =
\begin{pmatrix}
1 & 0 & 0 & 0 & 0 & 0 & 0 & 0 \\
0 & 1 & 0 & 0 & 0 & 0 & 0 & 0 \\
0 & 0 & 1 & 0 & 0 & 0 & 0 & 0 \\
0 & 0 & 0 & 1 & 0 & 1 & 0 & 0 \\
0 & 0 & 0 & 0 & 1 & 0 & 0 & 0 \\
0 & 0 & 0 & 0 & 0 & 0 & 1 & 0 \\
0 & 0 & 0 & 0 & 0 & 1 & 0 & 0 \\
0 & 0 & 0 & 0 & 0 & 0 & 0 & 1
\end{pmatrix}
\end{equation} 
and is represented by the following symbol:
\[ \Qcircuit @C=2em @R=1.7em {
& \ctrl{2} & \qw \\
& \qswap & \qw \\ 
& \qswap & \qw 
}\]
which means in our notation, $\ket{0,a,b} \to \ket{0,a,b}$, and $\ket{1,a,b} \to \ket{1,b,a}$ respectively. 
The use of control feature can be extended to any unitary gate. For example, consider the controlled RX gate where the transformation is defined by the matrix:
\[ \text{RX}  = \begin{pmatrix} \cos(\frac{\theta}{2}) & -i  \sin(\frac{\theta}{2}) \\  -i  \sin(\frac{\theta}{2}) &  \cos(\frac{\theta}{2}) \end{pmatrix}. \]
If we want to build a controlled-RX gate then the matrix is given by a $ 4 \times 4$ matrix:
\begin{equation}
\text{CRX} = \begin{pmatrix}
1 & 0 & 0 & 0   \\
0 & \cos(\frac{\theta}{2}) & 0 & -i \sin(\frac{\theta}{2})   \\
0 & 0 & 1 & 0 \\
0 & -i \sin(\frac{\theta}{2}) & 0 & \cos(\frac{\theta}{2})   
\end{pmatrix}.
\end{equation}
This gate is included in \QIS~like other gates we have discussed above. It can simply be called
by doing $\texttt{qc.crx($\pi$/2,0,1)}$ but double controlled RX gate is not included in usual gate library in \QIS~and we leave the design of this gate for the interested reader. 
In addition to the RX gate (which corresponds to rotation along $x$-axis), we can have RY or RZ gates. A general rotation gate is 
represented as:
\begin{align}
R_{\vec{n}}(\theta) &= \exp\Bigg[\frac{-i\theta}{2}\Big(n_1 X + n_2 Y + n_3 Z\Big)\Bigg] \\ \nonumber 
& = \begin{pmatrix}
\cos(\theta/2) - i n_3 \sin(\theta/2) & -n_2 \sin(\theta/2) - i n_1 \sin(\theta/2)    \\
n_2 \sin(\theta/2) - i n_1 \sin(\theta/2)   & \cos(\theta/2) + i n_3 \sin(\theta/2)   
\end{pmatrix}.
\end{align}
If we put $n_2 = n_3 = 0$ and $n_1 = 1$, we get back the RX gate\footnote{Note that
the matrices have $\theta$ rather than $\theta/2$. This is due to the
fact that rotations on Bloch sphere by $\theta$ corresponds to $\theta/2$
for the $SU(2)$ unitary matrices due to double cover relation between $SO(3)$ and
$SU(2)$}.

\begin{mybox}
\textsc{$\blacktriangleright$ Question 2:} Construct controlled-controlled-RX gate matrix and then implement it in \QIS. Check that it has the correct behaviour by checking how $\ket{110}$ and $\ket{000}$ transform respectively.
\end{mybox}

\begin{mybox}
\textsc{$\blacktriangleright$ Question 3:} a) Show that the controlled-Hadamard (CH) gate given by: 
\begin{equation}
\text{CH} = \begin{pmatrix}
1 & 0 & 0 & 0   \\
0 & 1& 0 & 0  \\
0 & 0 & \frac{1}{\sqrt{2}}   & \frac{1}{\sqrt{2}}  \\
0 & 0 & \frac{1}{\sqrt{2}}  & -\frac{1}{\sqrt{2}}   
\end{pmatrix}, 
\end{equation}
can be written as $\ket{00} \bra{00} + \ket{01} \bra{01} + \frac{1}{\sqrt{2}} \Big(\ket{10} \bra{10} + \ket{10} \bra{11} + \ket{11} \bra{10} - \ket{11} \bra{11}\Big)$. 
\\ \\ 
b) Suppose we act with $R_{y}(3 \pi/4)$ on starting state $\ket{0}$. What is the probability that we measure the state we started with? Recall that $R_{y}(\theta)$ is defined as: 
\begin{equation}
\begin{pmatrix}
\cos(\theta/2) & -\sin(\theta/2)    \\
\sin(\theta/2)  & \cos(\theta/2)    \\ 
\end{pmatrix}. 
\end{equation}

\end{mybox}



\begin{mybox}
\textsc{$\blacktriangleright$ Question 4:} Show that 
\[ H \cdot S \cdot T \cdot H \ket{0} =  \frac{1}{2} \Bigg[ \Big( 1 + e^{i 3 \pi/4}\Big) \ket{0} + \Big( 1 - e^{i 3 \pi/4}\Big) \ket{1}  \Bigg]\] 
where $H$, $S$, and $T$ are the usual gates defined in the text.
\end{mybox}
	A simple circuit with two CNOT gates is given as: 
	\[ \Qcircuit @C=1em @R=.7em {
		& \ctrl{1} & \targ & \qw \\
		& \targ & \ctrl{-1} & \qw
	}\]
	Though we will use the modern notation for CNOT, the old notation used by Feynman \cite{Feynman1986-FEYQMC} was instead:
	\[ \Qcircuit @C=2em @R=1.7em {
		& \ctrl{1} & \qswap & \qw \\
		& \qswap & \ctrl{-1} & \qw
	}\]
	Furthermore, a slightly more complicated circuit with multiple control lines and an arbitrary gate $U$ can be drawn as: 
	\[ \Qcircuit @C=1em @R=.7em {
		& \ctrl{2} & \targ & \gate{U} & \qw \\
		& \qw & \ctrl{-1} & \qw & \qw \\
		& \targ & \ctrl{-1} & \ctrl{-2} & \qw \\
		& \qw & \ctrl{-1} & \qw & \qw 
	}\]
	It is easy for the reader to see that if we have two qubits and act with $H$ on the first qubit and then apply CNOT gate, it results in an entangled state. 
	For three qubits, this is more interesting. Three qubits can be entangled in two ways that are not related to each other. One of them is called the GHZ state and the other is known as W-state \cite{D_r_2000}. The GHZ state is fully separable, while the W state cannot be separated by any local operations and classical communication (LOCC). These are defined as follows: 
	\[ \text{GHZ}_{3} = \frac{1}{\sqrt{2}} \Big(\vert 000 \rangle + \vert 111 \rangle \Big), \]
	\[ \text{W}_{3} = \frac{1}{\sqrt{3}} \Big(\vert 001 \rangle + \vert 010 \rangle + \vert 100 \rangle \Big). \]
	It is straightforward to extend $W$ state to its $n$-qubit version as:
	\begin{equation}
		\ket{W} = \frac{1}{\sqrt{n}} \Big(  \ket{10 \cdots 00} +  \ket{01 \cdots 00}  + \cdots +  \ket{00 \cdots 01} \Big). 
	\end{equation}
	We can also define something called the `generalized' GHZ state defined as:
	\begin{equation}
	\text{GHZ}_{n} = \cos \theta \ket{000 \cdots 000} +  \sin \theta \ket{111 \cdots 111}.  
	\end{equation}
	It is easy to implement these states in \QIS~and we leave that for the reader. 
Going beyond the discussion of single gates, for several purposes, some collection of quantum gates is often useful. When these gates are collected together, they are referred to as a `set'. One of the examples is the stabilizer set, which is given by a group of three gates: $\{H, \text{CNOT}, P\}$ and the same as Clifford group. These gates are important for error-correcting codes
but do not form a universal set as discussed later in the notes. 

However, it was found that $\{ \rm{CNOT}$, single-qubit unitaries ($SU(2)\}$ form a universal set of gates 
but it is infinite. We would like this to be close to some finite gate set. 
	A finite gate set can only produce a countable set of gates and one would think that 
	every unitary cannot be obtained by this method. However, we do not need to get 
	every unitary and this is where one of the most important theorems in quantum information/computation comes into play. Solovay-Kitaev (SK) theorem argues that if a single-qubit quantum gates generate a dense subset of $SU(2)$ then that set is guaranteed to fill $SU(2)$ quickly 
	(means it is possible to obtain a decent approximation to any desired gate using gates from the generating set). We say that a subset $\mathcal{S}$ of a topological space $X$ is dense if every point  $x \in X$ either belongs to $\mathcal{S}$  or is a limit point of $\mathcal{S}$. The special unitary group $SU(N)$ is defined as the set of all $N \times N$ matrices with unit determinant. For example, the Pauli matrices are the three generators of $SU(2)$. There are two universal (approximately) gate sets:
	$\{ H, \rm{CNOT}, \pi/8$\} as shown in Ref.~\cite{Boykin_2000} and $\{ H, \rm{CCNOT}, \pi/4$\} as shown in Ref.~\cite{Kitaev_1997} and CNOT plus all single-qubit gates \cite{Barenco95}. Suppose we consider the gate set $\{ H, \rm{CNOT}, T (\pi/8)$\} and ask how efficiently a given unitary transformation $U$
	can be implemented. This situation might arise due to the 
	limitation of applying only single-qubit gate most prominently when one required fault-tolerant quantum computation where the 
	features of fault-tolerance are available for selected gates in Clifford group and $\pi/8$ gate. 
	We would like to have a polynomial (in precision and number of qubits)
	number of gates selected from this set. We ask this question because in general
	most unitary transformations cannot be approximated precisely. 
	The SK theorem can be useful in this case. A direct consequence of this theorem is that a quantum circuit of $N$ constant-qubit gates can be approximated to 
	$\epsilon$ error (in operator norm) by a quantum circuit of $\mathcal{O}(N \log^{c} (N/\epsilon))$ gates from a desired finite universal gate set. Different proofs of the theorem give different values for $c$, in Ref.~\cite{dawson2005}, 
	it was shown to be $ \sim 3.97$.  
	This theorem computes how many gates are needed to approximate a given operation for 
	the specified accuracy $\epsilon$ and is polylogarithmic in $1/\epsilon$. 
	We now write down the formal statement as below:
	
	$\emph{Solovay-Kitaev theorem}$: If $\mathcal{S}$ is a finite set of 1-qubit gates which is universal and if for any gate $g \in \mathcal{S}$, 
	the inverse $g^{-1}$ can be achieved by finite sequence of gates in $\mathcal{S}$, then any 1-qubit 
	gate and hence (any unitary transformation) can be approximated using $\mathcal{O}(\log^{c} (1/\epsilon))$ gates with 
	$c < 4$. The basic algorithm is given below: 
	\begin{mybox2}
	\begin{itemize}
	\item	\texttt{function Solovay-Kitaev/SK (UnitaryGate $U$, depth $n$)} \\
	\# Read gate to be approximated and accuracy desired ($\epsilon \to 0$ as $n \to \infty$) 
	\item \texttt{if ($n = 0$), return basic approximation to $U$} \\ 
	\# The function is recursive, so that to obtain an $n$ approximation to $U$, it will call itself to obtain $n-1$ -approximations to the unitary.
	\item \texttt{else set $U_{n - 1}$ = SK(U,n-1)  set $V,W$ = GC-Decompose($UU^{\dagger}$), \\
	set $V_{n-1}$ = SK(V,n-1), set $W_{n-1}$ = SK(W,n-1), 
	return $U_{n} = V_{n-1} W_{n-1} V^{\dagger} W^{\dagger}_{n-1}$} 
	\end{itemize} 
	\end{mybox2}
	
	A single-gate set of universal quantum gates can also be formulated using the three-qubit Deutsch gate given by:
	\begin{equation} 
		\mathbb{D} = \begin{pmatrix}
			1 & 0 & 0 & 0 & 0 & 0 & 0 & 0 \\ 
			0 & 1 & 0 & 0 & 0 & 0 & 0 & 0 \\ 
			0 & 0 & 1 & 0 & 0 & 0 & 0 & 0 \\ 
			0 & 0 & 0 & 1 & 0 & 0 & 0 & 0 \\ 
			0 & 0 & 0 & 0 & 1 & 0 & 0 & 0 \\ 
			0 & 0 & 0 & 0 & 0 & 1 & 0 & 0 \\ 
			0 & 0 & 0 & 0 & 0 & 0 & i  \cos(\theta) & \sin(\theta)  \\ 
			0 & 0 & 0 & 0 & 0 & 0 &  \sin(\theta)  & i  \cos(\theta)  \\ 
		\end{pmatrix}. 
	\end{equation} 
	
		\begin{mybox}
		\textsc{$\blacktriangleright$ Question 5:} What does the following quantum circuit do? By what name do we know the final state known as?
		
		\[ \Qcircuit @C=1em @R=.7em {
			\lstick{\vert 0 \rangle} & \gate{H} & \ctrl{1} & \qw & \qw \\
			\lstick{\vert 0 \rangle} & \qw & \targ & \ctrl{1} & \qw \\
			\lstick{\vert 0 \rangle} & \qw & \qw & \targ & \qw
		}\]
	\end{mybox}
	
		\begin{mybox}
		\textsc{$\blacktriangleright$ Question 6:} Check that CNOT results in the following transformations given below, i.e., $\text{CNOT}(X \otimes \mathbb{1}) \to (X \otimes X) \text{CNOT}$ or alternatively $\text{CNOT}(X \otimes \mathbb{1})\text{CNOT}^{\dagger} \to (X \otimes X)$ and so on. 
		
		\begin{itemize} 
			\item $X \otimes \mathbb{1} \to X \otimes X$ 
			\item $ \mathbb{1} \otimes X \to \mathbb{1} \otimes X$ 
			\item $ Z \otimes \mathbb{1}  \to Z \otimes \mathbb{1} $ 
			\item $ \mathbb{1}  \otimes Z   \to Z \otimes Z $ 
		\end{itemize} 
	\end{mybox}
	It is also known that this gate can be constructed from a sequence of two-qubit gates \cite{1994cond.mat..9111D}. It can also be constructed from a two-qubit gate known as `Barenco gate' \cite{Barenco:1995dx} given by:
	\begin{equation} 
		\mathbb{B} = \begin{pmatrix}
			1 & 0 & 0 & 0 \\ 
			0 & 1 & 0 & 0 \\ 
			0 & 0 & e^{i \alpha} \cos(\theta)  & -i e^{i (\alpha-\phi)} \sin(\theta)  \\ 
			0 & 0 & -i e^{i (\alpha+\phi)} \sin(\theta)  & e^{i \alpha} \cos(\theta) \\ 
		\end{pmatrix}
	\end{equation}
	In fact, it was shown in Ref.~\cite{2002quant.ph..5115S} that 
	the set $\{T, H\}$ is universal and was shown to be true 
	by a short proof in Ref.~\cite{2003quant.ph..1040A}. 
	The fact that $T$ can perform all classical reversible computation 
	makes it clear that $H$ gate which is in fact a quantum Fourier transform (QFT) over
	$\mathbb{Z}_{2}$ is the real `quantum' difference!

	\begin{mybox}
		\textsc{Question 7:} Show that $H$ can be written for one-qubit as:
		\[ H = \frac{1}{\sqrt{2}} \Big[(\ket{0} + \ket{1})\bra{0} +  (\ket{0} - \ket{1})\bra{1} \Big]. \] 
		Show that this transform for $n$-qubits i.e., $H^{\otimes n}$ can be written as:
		\[ H^{\otimes n} = \frac{1}{2^{n/2}} \sum_{x,y} (-1)^{x \cdot y} \ket{x} \bra{y} \]
	\end{mybox}


	\begin{mybox}
		\textsc{$\blacktriangleright$ Question 8:} Consider a control line in a superstition state and target in $\vert 0 \rangle$. Pass this through the CNOT gate and write the result (a Bell state). Show that the circuit given below performs a measurement in the basis of Bell states. This exercise is reproduced from Exercise 4.33 of \cite{Qinfobook}. Find the measurement operators. \newline 
		\begin{equation}
			\Qcircuit @C=1em @R=.7em {
				& \ctrl{1} & \gate{H} & \meter \\
				& \targ & \qw & \meter 
			}
		\end{equation}
	\end{mybox}
	
	
	

	

	\begin{mybox}
		\textsc{$\blacktriangleright$ Question 9:} Show that the circuit below is another way of implementing Toffoli gate. Find another decomposition of Toffoli gate? \newline 
		
		\begin{equation}
			\Qcircuit @C=1em @!R {
				\lstick{\ket{x}}   &   \qw        &   \qw                  &   \ctrl{1}   &   \qw                   &  \ctrl{1}  & \ctrl{2}             &   \qw        &   \rstick{\ket{x}}   \qw                     \\
				\lstick{\ket{y}}   &   \qw        &   \ctrl{1}             &   \targ      &   \ctrl{1}              &  \targ     & \qw                  &   \qw        &   \rstick{\ket{y}}   \qw                     \\
				\lstick{\ket{z}}   &   \gate{H}   &   \gate{\mathbb{C}(\pi/2)}   &   \qw        &   \gate{\mathbb{C}(3\pi/2)}   &  \qw       & \gate{\mathbb{C}(\pi/2)}   &   \gate{H}   &   \rstick{\ket{(x \cdot y) \oplus z}} \qw
			}
		\end{equation}
		
	\end{mybox}
	\begin{figure}
		\centering 
		\includegraphics[width=0.66\textwidth]{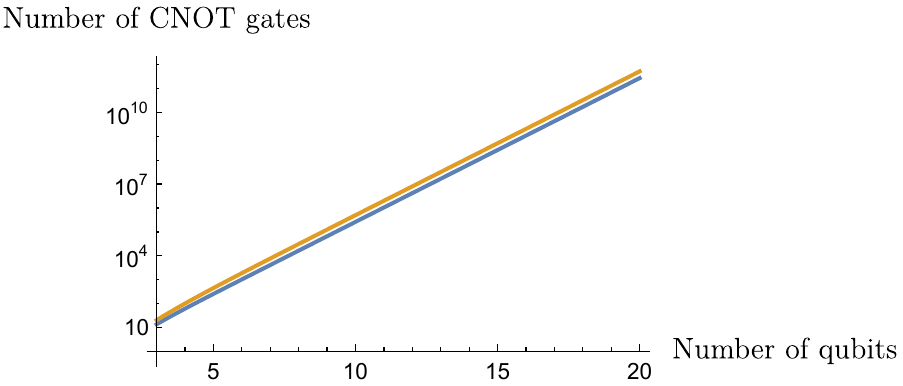}
		\caption{\label{fig:qsd}The theoretical lower bound on the number of CNOTs needed to approximate arbitrary unitary matrix. The upper bound is the
		cost attained using QSD in Ref.~\cite{Shende2006}.}
	\end{figure}
	We now discuss an important ingredient of the quantum circuit and it is related to the 
	decomposition in terms of CNOT and single-qubit gates. It is a well-known result that we can decompose 
	any $2^n \times 2^n$ unitary matrix in terms of single-qubit unitaries and CNOT gates. Every single-qubit 
	unitary gate has 3 real parameters. So, for $n$ qubits, we have $3n$ parameters. Once we add a CNOT gate, 
	it appears that we can add twice one-qubit unitaries resulting in 6 real parameters. However, since the $R_{z}$ gate commutes
	with the control line qubit and $R_{x}$ commutes with the target qubit, we can pass them across 
	and this reduces the number of effective parameters added by CNOT to 4. Suppose we need
	$N$ CNOTs to approximate a $2^n \times 2^n$ matrix, then we have $3n + 4N \ge 4^n - 1$ 
	which gives:
	\begin{equation}
	N \ge \frac{1}{4}\Big( 4^n - 3n - 1\Big).
	\end{equation}
	Using quantum Shannon decomposition (QSD), it was shown in Ref.~\cite{Shende2006} that one 
	can get the cost down to:
	
	\begin{equation}
	N \ge \frac{23}{48} 4^n -\frac{3}{2} 2^n + \frac{4}{3}.
	\end{equation}
	The difference between the lower bound and the implemented cost (about factor of 2) 
	is shown in Fig.~\ref{fig:qsd}.

	\subsection{General quantum operations}
	Let us consider an orthonormal basis $\ket{\phi_k}$, we can write a state as $\kp = \sum_{k} \alpha_{k} \ket{\phi_k}$. 
	von Neumann (vN) measurement of $\kp$ with respect to the $\phi$ basis is described by orthogonal projectors 
	$\{\ket{\phi_k}, \bra{\phi_k}\}$ and will have the output  $k$ with probability:
	\[ \text{Tr} ( \kp \bp  \ket{\phi_k} \bra{\phi_k} )  = \vert \alpha_{k} \vert^{2}. \] 
	von Neumann (vN) measurements are a special kind of projective measurement (which is also known as 
	L\"{u}ders measurement) which is complete. This is often used in quantum computing and communication. 
	We give the circuit which can implement a vN measurement below. 
	
	\begin{figure}[h]
		\centering
		\leavevmode
		\large{
			\Qcircuit @C=1em @R=1.1em {
				& \multigate{2}{U} & \qw &  \ctrl{3} & \qw &  \qw & \qw & \qw  & \multigate{2}{U^{-1}} & \qw \\
				\lstick{ \kp = \sum_{k} \alpha_{k} \ket{\phi_k}} & \ghost{U}  & \qw  & \qw & \ctrl{3}  & \qw &  \qw & \qw & \ghost{U^{-1}} & \qw & \rstick{\ket{\phi_k}} \\
				& \ghost{U} & \qw  &  \qw & \qw & \qw & \ctrl{3} & \qw & \ghost{U^{-1}} & \qw  \\
				\lstick{\ket{0}} & \qw  & \qw & \targ & \qw  & \qw &  \qw & \qw &  \multigate{2}{M} & \qw \\
				\lstick{\ket{0}} & \qw  & \qw & \qw & \targ  & \qw &  \qw & \qw &  \ghost{M}  & \qw & \rstick{\ket{k}} \\  
				\lstick{\ket{0}} & \qw  & \qw &  \qw & \qw  & \qw & \targ  & \qw &  \ghost{M} & \qw 
			}
		}
		\caption{One way of implementing vN measurement through ancillary register. The measurement $M$ measures $\ket{k}$ with probability, $|\alpha_{k}|^2$ while the map through $U^{-1}$ gives the state $\ket{\phi_{k}}$ in the main register. One can also directly measure instead of introducing ancillary register. Note that $U$ is a unitary transformation which implements basis change to the computational basis. The use of CNOT gate to copy to the ancillary register might seem illegal but note that we are just doing reversible transformation copying the computational basis states, we are not cloning/copying arbitrary superposition state. }
		\label{fig:DJ1}
	\end{figure}
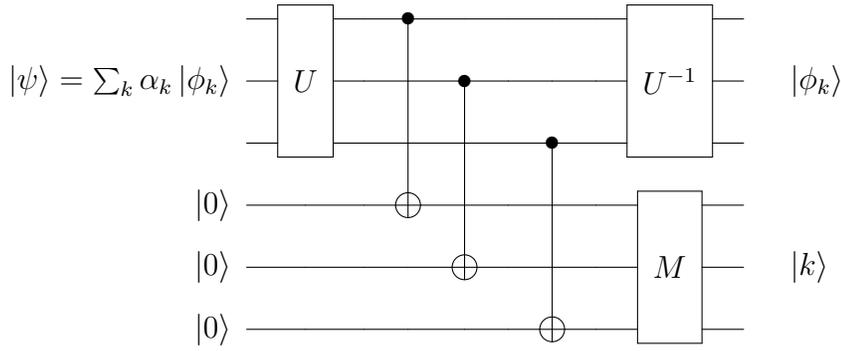
	
	In fact, the idea of projective measurements is related to identifying which 
	state the system is in from the possible set of mutually exclusive states. 
	Suppose, the atom is either in state $\ket{a}$ or $\ket{b}$, through projective measurement
	we can find out the state it is in. These are based on projection operators $P$
	which are Hermitian operators ($P = P^{\dagger}$, $P^2 = P$). We refer to two 
	projection operators $P_1$ and $P_2$ as being orthogonal if $P_1P_2 \kp = 0$. 
	The probability of obtaining output $k$ (as discussed above) is $\texttt{Prob.}(k) = \langle \psi \vert P_{k} \vert \psi \rangle 
	= \rm{Tr}(P_{k} \kp \bp $. These measurements lead to the collapse of the wave function and the state after measurement 
	$ \psi^{\prime}$ is given by: 
	\begin{equation}
	\psi^{\prime} = \frac{P_k \kp }{\sqrt{\langle \psi \vert P_{k} \vert \psi \rangle}}. 
	\end{equation}
	Projective measurements (of which von Neumann is a special case) can be generalized
	as well but we will not discuss it here and refer the reader to textbooks that discuss
	quantum measurements in depth.


	\begin{mybox2}
		\textsc{Example 4:} What is the output of the quantum circuit given below? Assume $U$ is some unitary gate.  \newline 
		
		\begin{equation}
			\Qcircuit @C=1em @R=.7em {
				\lstick{\ket{0}} & \gate{H} & \ctrl{1} & \gate{H} & \qw \\
				\lstick{\ket{\psi_{\text{in}}}}& \qw & \gate{U} & \qw & \qw 
			}
		\end{equation}
		
	\end{mybox2}
	$\square$ The initial state is given by: 
	$ \ket{\psi_0} = \ket{0} \ket{\psi_{\text{in}}}$. Once we 
	apply the $H$ gate it becomes, 
	$ \ket {\psi_1}$ which is given by:
	\[ \ket{\psi_1} = \frac{1}{\sqrt{2}} \Big( \ket{0} \ket{\psi_{\text{in}}}+ \ket{1} \ket{\psi_{\text{in}}}\Big) \]
	Then, if we act on this with the $U$ as given in the Figure, we get:
	\[ \ket{\psi_2} = \frac{1}{\sqrt{2}} \Big( \ket{0} \ket{\psi_{\text{in}}}+ \ket{1} U \ket{\psi_{\text{in}}}\Big) \]
	and now we again act with the $H$ gate,
	\[ \ket{\psi_3} = \frac{1}{2} \Big((\ket{0}  + \ket{1}) \ket{\psi_{\text{in}}} + (\ket{0} - \ket{1})U \ket{\psi_{\text{in}}}\Big) \]
	and this is equal to:
	\[ = \frac{1}{2}  \Big( \ket{0} (\mathbb{1} + U) \ket{\psi_{\text{in}}} + \ket{1} (\mathbb{1} - U) \ket{\psi_{\text{in}}} \Big)\]
	$\square$
	
	
	
	\subsection{Entanglement} 
	
	One of the distinguishing features of quantum mechanics is `entanglement'\footnote{This famously prompted Einstein to complain about spooky action at a distance (in German `spukhafte Fernwirkungen')}.  
	Though there exist various definitions of entanglement of varying complexity, for our purposes, a simple one will suffice. We say that the state is entangled when it cannot be written as a product state. It is often useful to check whether a given state is separable (not entangled) or not. In order to show whether the state is entangled or not, it is easiest to compute the reduced density matrix and compute $\rm{Tr}{\rho_A}^{2}$ and see if it is 1 or not (for normalized states). It is a separable state (not entangled) iff $\rm{Tr}{\rho_A}^{2} = 1$. 
	Suppose we have a Bell state: $|\Psi\rangle=\frac{1}{\sqrt{2}}(|00\rangle+|11\rangle)$, then we have:
	\[ \rho_A=\text{Tr}_B(|\Psi\rangle\langle\Psi|)=\frac12\left(|0\rangle\langle 0|+|1\rangle\langle 1|\right)=\frac12\mathbb{1} \]
	$\text{Tr}(\rho_A^2)=\frac14\text{Tr}(\mathbb{1}\cdot\mathbb{1})=\frac12$. 
	Hence, this is entangled. We have used the shorthand notation: $ \rm{Tr}_{B}(\rho_{AB}) = \rho_A$ above. 
	
	\begin{mybox}
		\textsc{$\blacktriangleright$ Question 10:} Compute $\text{Tr}(\rho_A^2)$ and check if the state given below is separable? 
		\begin{equation}
		\label{eq:Q011}
			\ket{\psi} = \frac{1}{2} \Big( \ket{00} + \ket{10} - \ket{01} - \ket{11} \Big). 
		\end{equation}
		Once you find it is separable, write it in terms of $\ket{\pm} = \frac{1}{\sqrt{2}} (\ket{0} \pm \ket{1})$. Now consider a general state $\ket{\psi} = a \ket{00} + b \ket{10} + c \ket{01} + d \ket{11}$ and find the condition iff the state is to be a separable one. Now think of a measure using $a,b,c,d$ of how we can quantify the extent of entanglement. Clearly for this case it should vanish. In later part of these notes, we will redo this exercise using \QIS, see Question~\ref{mybox:Q999}. In the later part of the notes, we will encounter a name for the function which can be built out of $a,b,c,d$. 
	\end{mybox}

	The above arguments can be systematically understood as follows. Consider a density matrix $\rho$ written as:
	
	\begin{equation}
		\rho = \sum_{i} p_{i} |\psi_{i}\rangle \langle \psi_{i}|.  
	\end{equation}
	If the system is in a pure state then all the $p_{i}$, except one, is zero, and we get $\rho = |\psi \rangle \langle \psi|$. If the $\rho$ is a pure state, then the following conditions are both sufficient and necessary to prove that it is in such a purity: $\rho = \rho^{\dagger}$,  $\mathrm{Tr} \rho = \mathrm{Tr} \rho^{2}$. 
	Usually one takes the $\rho$ to be appropriately normalized such that $\mathrm{Tr} \rho = 1$
	which is equivalent to saying that $ \mathrm{Tr} \rho = \mathrm{Tr} \Big(    \sum_{i} p_{i} |\psi_{i}\rangle \langle \psi_{i}|              \Big)  = 1$
	or $ \mathrm{Tr} \rho = \sum_{n} \langle n | \rho | n \rangle   =   \sum_{n} \langle n | \psi \rangle   \langle \psi | n \rangle = 1$ as per taste. It is straightforward to show that $\mathrm{Tr} \rho = \mathrm{Tr} \rho^{2}$. 
	We will just show that $\rho = \rho^2$, which can be done by writing: 
	$ \rho^2 = |\psi \rangle \langle \psi|\psi \rangle \langle \psi| = \rho$, where the sums have been suppressed. 
	Now consider a mixed state, and we will show that $ \mathrm{Tr} (\rho_{\text{mixed}})^{2} < \mathrm{Tr} \rho_{\text{mixed}} $. 
	We also need to show that if $\rho^2 = \rho$, the state is pure. We first note that since $\rho$ is Hermitian, the eigenvalues are real and the corresponding eigenvectors can be made orthonormal, the proof for this proceeds as follows, 
	\begin{align}
		\rho & = \sum_{i} \lambda_{i} |\lambda_{i}\rangle \langle \lambda_{i}|     ~~~; ~ \text{Spectral decomposition}  \\ 
		& = \sum_{i} \lambda_{i}^{2} |\lambda_{i}\rangle \langle \lambda_{i}|   ~~~; ~  (\because \rho = \rho^{2})
	\end{align}
	This implies that the eigenvalues are either 0 or 1. Hence, $\lambda_{i} = 1$, for some $ i = p$ and 0 for $ i \neq p$. 
	Hence, $\rho = \sum_{i} \lambda_{i} |\lambda_{i}\rangle \langle \lambda_{i}|  = |\lambda_{p}\rangle \langle \lambda_{p}|$
	which implies that $\rho$ is pure. For mixed states, $\mathrm{Tr} (\rho_{\text{mixed}})^{2} < \mathrm{Tr} (\rho_{\text{mixed}})$. 
	The other properties still hold such as, $\rho = \rho^{\dagger}$, $\mathrm{Tr} \rho = 1$ and $\rho \ge 0$ (positivity). 
	The measure of $\mathrm{Tr} \rho^{2}$ from 1 is a good measure of how mixed the state is. For a maximally mixed state, 
	we have $\mathrm{Tr} \rho^{2} = 1/d$, where $d$ is the dimension of the system. This is also called as 
	a maximally mixed density matrix. A density matrix corresponding to a maximally mixed state also has the largest entropy
	as: 
	\begin{align}
		S_{\text{EE}} &= - \mathrm{Tr}_{A} \Big(\rho \ln \rho \Big)  \\
		&= - d ~~ \frac{1}{d} ~~ \mathrm{ln}\Big(\frac{1}{d}\Big)  = \mathrm{ln} ~ d. 
	\end{align}
	Any unitary transformation preserves the pureness of the state intact, i.e., a pure state remains pure. 
	\begin{align}
		\mathrm{Tr} \Big[(U \rho U^{\dagger})^{2}\Big] &= \mathrm{Tr} \Big(U \rho \underbrace{U^{\dagger}U}_{\mathbb{1}} \rho U^{\dagger} \Big) \\
		&=   \mathrm{Tr} \Big(U \rho^{2} U^{\dagger} \Big) \\
		&=   \mathrm{Tr} \rho^{2}. 
	\end{align}
	This definition of entropy (called entanglement entropy or von Neumann entropy) for pure states of a bipartite system is the
	most widely used measure to quantify entanglement. 
	As the astute reader might have noticed before, we can also identify if the state is separable 
	by computing the Schmidt number. Consider a state $\ket{\psi} \in \mathscr{H}_{A} \otimes \mathscr{H}_{B}$ to be a pure state. Then we have an 
	expansion of the form:
	\[ \kp = \sum_{i} \lambda_{i} \ket{a_i} \ket{b_i},\]
	where $\ket{a_i}$ and  $\ket{b_i}$ are orthonormal states belonging to $\mathscr{H}_{A}$
	and $\mathscr{H}_{B}$ respectively. The $\lambda_i \ge 0$ are Schmidt coefficients and 
	satisfy $ \lambda_{i}^{2} = 1$. These are computed by constructing a matrix $ \rm{Tr}_{B} (\kp \bp)$
	whose eigenvalues are $ \lambda_{i}^{2}$. The Schmidt number is defined as the number of non-zero $\lambda_{i}$. The state is separable if this is 1 and entangled if it is greater than 1.

	\subsubsection{Partial trace method}
	
	In this subsection, we will discuss the idea which facilitates the computation of bipartite entropy. The partial trace, $\rm{Tr}_{B}$ is a map from the density matrix $\rho_{AB}$ 
	on some composite system  with Hilbert space $\mathscr{H}_{A} \otimes \mathscr{H}_{B}$
	onto density matrices $\rho_{A}$ on $\mathscr{H}_{A}$. 
	Let us assume that $\{\vert \alpha_{i}\rangle \}$ and $\{\vert \beta_{i}\rangle \}$
	are the basis of $\mathscr{H}_{A}$ and, $\mathscr{H}_{B}$ respectively. Then a 
	density matrix, $\rho_{AB}$ can be decomposed as:
	\begin{equation}
		\rho_{AB} = \sum_{ijkl} c_{ijkl} \vert \alpha_{i}\rangle \langle \alpha_{j} \vert \otimes \vert \beta_{k}\rangle \langle \beta_{l} \vert
	\end{equation}
	and the partial trace is given by\footnote{We note that $\rm{Tr} \vert \beta_{k}\rangle \langle \beta_{j} \vert = \langle \beta_{j} \vert \beta_{k} \rangle $}:
	\begin{equation}
		\label{eq:ptrace1} 
		\rm{Tr}_{B} \rho_{AB} = \sum_{ijkl} c_{ijkl} \vert \alpha_{i}\rangle \langle \alpha_{j} \vert \langle \beta_{l} \vert \beta_{k} \rangle 
	\end{equation}
	\begin{equation}  
		\rm{Tr}_{B} \begin{pmatrix}
			\rho_{11}    ~   &  \rho_{12}  ~ & \rho_{13}  ~ & \rho_{14}    \\
			\rho_{21}   ~    &  \rho_{22} ~  & \rho_{23} ~  & \rho_{24}  \\
			\rho_{31}    ~   &  \rho_{32}  ~ & \rho_{33}  ~ & \rho_{34}  \\
			\rho_{41}    ~   &  \rho_{42}  ~ & \rho_{43} ~  & \rho_{44}  \\
		\end{pmatrix} = \begin{pmatrix}
			\rho_{11}+ \rho_{22}      ~~ &  \rho_{13}  +  \rho_{24}  \\
			\rho_{31} + \rho_{42}     ~~   &  \rho_{33}  +  \rho_{44}\\
		\end{pmatrix}
	\end{equation} 
	while the partial trace over $A$ is given by:
	\begin{equation}  
		\rm{Tr}_{\scaleto{A}{4pt}} \begin{pmatrix}
			\rho_{11}   ~    &  \rho_{12} ~  & \rho_{13} ~  & \rho_{14}    \\
			\rho_{21}  ~     &  \rho_{22}  ~ & \rho_{23} ~  & \rho_{24}  \\
			\rho_{31}   ~    &  \rho_{32}  ~ & \rho_{33}  ~ & \rho_{34}  \\
			\rho_{41}   ~    &  \rho_{42} ~  & \rho_{43}  ~ & \rho_{44}  \\
		\end{pmatrix} = \begin{pmatrix}
			\rho_{11}+ \rho_{33}     ~~  &  \rho_{12}  +  \rho_{34}  \\
			\rho_{21} + \rho_{43}   ~~    &  \rho_{22}  +  \rho_{44}\\
		\end{pmatrix}
	\end{equation} 
	A two-qubit state can be expanded in the orthonormal basis $\{\vert 00 \rangle , \vert 01 \rangle, \vert 10 \rangle, \vert 11 \rangle\}$
	given by:
	\begin{equation}
		\rho_{{\scaleto{AB}{4pt}}} = \rho_{11} \vert 00 \rangle \langle 00 \vert + \rho_{12} \vert 00 \rangle \langle 01 \vert + \cdots + \rho_{44} \vert 11 \rangle \langle 11 \vert
	\end{equation}
	and the partial trace is given by $\ref{eq:ptrace1}$:
	\begin{equation}
		\rho_{{\scaleto{A}{4pt}}} = (\rho_{11} + \rho_{22}) \vert 0 \rangle \langle 0 \vert + (\rho_{13} + \rho_{24}) \vert 0 \rangle \langle 1 \vert + (\rho_{31} + \rho_{42}) \vert 1 \rangle \langle 0 \vert + (\rho_{33} + \rho_{44}) \vert 1 \rangle \langle 1 \vert 
	\end{equation}
One of the other ideas related is - purification. The fact that given mixed density matrix can always be thought of as being obtained from the partial trace of some bigger Hilbert space. This idea is closely related to the phrase -- there exists a `church of bigger Hilbert space' supposedly meaning that there always is a place where one can be purified. 
	Suppose $ |i\rangle \in \mathcal{H}_{A}$ such that: 
	\begin{equation}
		\rho_{A} = \sum_{i} \lambda_{i}  |i\rangle \langle i| 
	\end{equation}
	Extent $\mathcal{H}_{A} \to \mathcal{H}_{A} \otimes \mathcal{H}_{B}$, then we can write 
	$\rho_{A}$ as, 
	\begin{equation}
		\rho_{A} = \mathrm{Tr}_{B} | \psi \rangle_{AB} \langle \psi|_{AB}
	\end{equation}
	which is a pure state where $ | \Psi \rangle_{AB} = \sum_{i} \sqrt{\lambda_{i}} |i\rangle_{A} |i\rangle_{B}$
	Suppose we have a density matrix, $\rho_{123}$, and we purify this state by adding a fourth Hilbert space,
	$\mathcal{H}_{4}$. Then we have, $S_{1234}=0$ and from this we get, $S_{123} = S_{4}$ and 
	$S_{234} = S_{1}$ and $S_{12} = S_{34}$.

	\begin{mybox}
		\textsc{$\blacktriangleright$ Question 11:}  Consider the state given by: 
		\begin{equation}
		\kp = \cos(\alpha) \ket{00} + \sin(\alpha) \ket{11}, ~~~~ 0 < \alpha < \pi/4
		\end{equation}
		Find the eigenvalues of $\rho$ and $\rho_2$ and compute the von Neumann entropy. 
	\end{mybox}
	In addition to von Neumann entropy, there is another notion of entropy often used. This is 
	referred to as `relative entropy'. For two density operators $\rho$ and $\sigma$, it is defined as:
	\begin{equation}
	\label{eq:REN1} 
	S_{b}(\rho\vert\vert\sigma) = \mathrm{Tr} (\rho \log_{b}(\rho) - \rho \log_{b}(\sigma)) 
	\end{equation}
	
	\subsubsection{\label{sec:ent_mstate}Entangled mixed states} 
	 
	 The von Neumann entropy is not a good measure when we deal with entanglement in mixed quantum states. We say a `mixed state' is entangled
	 if it cannot be represented by any mixture of pure states. Unlike the case for pure entangled case, there exists majorly three different definitions of entanglement. These are entanglement of formation (EOF), distillable entanglement (DE), and relative entropy of entanglement (REE)\cite{Hill_1997, Wootters_1998}. 
	 For example, these definitions have distinguishing features. If we consider EOF of a density matrix, it is zero iff (if and only if) $\rho$ can be written as a mixture of product states while DE can still be non-zero for this. We will consider only entanglement of formation in this article. The computation of entanglement of formation for a mixed
	 bipartite state ($\rho_{AB} = \sum_{j} p_j  \ket{\Phi_j} \bra{\Phi_j}$ is given by:
	 \begin{equation}
	 E_{f}(\rho_{AB}) = \text{min.} \sum_{j} p_{j} E (\Phi_j),
	 \end{equation}
	 where the minimum is taken over all pure state decomposition of $\rho_{AB}$. 
	 In addition to EOF, there is another quantitative measure of entanglement known as
	`concurrence'.
	 The entanglement of formation is closely related to another quantity, concurrence (C)
	 through the following relation~\cite{Wooters2001lk}: 
	  \begin{equation}
	  E_{f}(C) = h \Bigg(\frac{1 + \sqrt{1-C^2}}{2} \Bigg),    
	  \end{equation}
	  where $h$ is defined as $ - x \log_{2}(x) - (1-x) \log_{2}(1-x)$. 
	In the appendix, we compute this quantity using \QIS~for the interested reader. 
	These computations can also be done using \MAT~ and we give a brief sample code using
	\href{https://github.com/rogercolbeck/QI}{\texttt{QI Mathematica package}} in the Appendix. 
	Alternatively, one can also use the \href{https://www.wolframcloud.com/obj/wolframquantumframework/DeployedResources/Paclet/Wolfram/QuantumFramework/}{\texttt{QuantumFramework}} package
	from Wolfram. 
	
	\vspace{4mm} 
	
	\begin{mybox}
		\textsc{$\blacktriangleright$ Question 12:}  Using \MAT~and the package mentioned above, answer the following:
		\begin{itemize}
		\item Pick two random density matrices and compute the relative entropy (as defined in \ref{eq:REN1}) between them?
		\item Consider a density matrix given by:
		\begin{equation}
		\rho = \begin{pmatrix}
			5/12   ~~  &  1/6 & 1/6  \\
			1/6   ~~  &  1/6 & 1/6  \\
			1/16   ~~  &  1/6 & 5/12  \\
		\end{pmatrix}.
		\end{equation}Is this a pure or mixed state? Compute the von Neumann entropy. 
		
		\item Create a random density matrix and compute the von Neumann entropy. Do this several times. 
		\end{itemize}
	\end{mybox}

	\subsection{Quantum Fourier Transform (QFT)} 
	
	The effectiveness of QFT\footnote{We understand that this acronym might be unsuited for field theorists}
	was emphasized in a 1994 paper \cite{DJ1994} 
	which eventually was used in Shor's factoring algorithm. 
	\begin{align*}
		\rm{QFT} \ket{x} &= \frac{1}{\sqrt{2}} \Big( \sum_{y=0}^{N-1} e^{2\pi ixy/N} |y\rangle \Big) \\[1em]
		& =  \frac{1}{\sqrt{2}} \Big(e^{2\pi ix\cdot 0/2}|0\rangle + e^{2\pi ix\cdot 1{/}2}|1\rangle\ \Big) \\[1em]
		& =  \frac{1}{\sqrt{2}} \Big( \ket{0} + e^{i \pi x} \ket{1} \Big). \\[1em]
	\end{align*}
	It is easy to check that $\rm{QFT} \ket{0} = \ket{+}$  and $\rm{QFT} \ket{1} = \ket{-}$. Hence, QFT on a single qubit is just like Hadamard gate. Now let us consider that $\{0, 1, \cdots, N-1\}$ forms an orthonormal basis and let $\ket{\alpha} = \sum_{i=0}^{N-1} \ket{j}$ denote a state with $N = 2^n$ for $n$ qubits. 
	Then QFT transforms the state as:
	\begin{equation} 
		\ket{\alpha}	= \sum_{i=0}^{N-1} \ket{j} \to \sum_{i=0}^{N-1} \frac{1}{\sqrt{N}} 
		\sum_{k=0}^{N-1} \gamma^{-jk} \ket{k}
	\end{equation}
	where $\gamma = e^{-i 2 \pi /N}$. Therefore, we have a representation for the QFT as:
	\begin{equation}
		\label{eq:QFT00} 
		Q_N = \frac{1}{\sqrt{N}} \sum_{j,k}^{N-1} \gamma^{-jk} \ket{k} \bra{j}
	\end{equation}
	
	\begin{mybox}
		\textsc{$\blacktriangleright$ Question 13:}  Show that quantum Fourier transform defined by (\ref{eq:QFT00}) is a unitary operation.
	\end{mybox}
	To see a more general working of QFT, let us assume that $\ket{x} = \ket{x_1} \otimes \ket{x_2} \otimes  \cdots  \otimes \ket{x_N}$, then if we perform quantum Fourier transformation on this, we get: 
	\begin{align}
		\rm{QFT} \ket{x} = & \frac{1}{\sqrt{N}}\sum_{y=0}^{N-1} e^{2\pi ixy/N} |y\rangle \nonumber \\ 
		= & \frac{1}{\sqrt{N}}\sum_{y=0}^{N-1} e^{2\pi ix \sum_{k=1}^n y_k/2^k} \nonumber \\ 
		= & \frac{1}{\sqrt{N}}\sum_{y=0}^{N-1} \prod_{k=1}^n e^{\frac{2\pi ixy_k}{2^k}} |y_1 y_2 \dots y_n\rangle \nonumber \\ 
		= & \frac{1}{\sqrt{N}}\sum_{y_1=0}^{1} \sum_{y_2=0}^{1} \dots \sum_{y_n=0}^{1}  \prod_{k=1}^n e^{\frac{2\pi ixy_k}{2^k}} |y_1 y_2 \dots y_n\rangle \nonumber \\ 
		= & \frac{1}{\sqrt{N}} \left(|0\rangle + e^{\frac{2\pi ix}{2^1}}|1\rangle\right) \otimes \left(|0\rangle + e^{\frac{2\pi ix}{2^2}}|1\rangle\right) \otimes \dots \otimes \left(|0\rangle + e^{\frac{2\pi ix}{2^n}}|1\rangle\right) 
	\end{align}
	We can see that at the qubit level, it maps each of them as: $\ket{x_k} \to \frac{1}{\sqrt{2}}\Big(e^{\frac{2\pi ix\cdot 0}{2^k}}|0\rangle + e^{\frac{2\pi ix\cdot 1}{2^k}}|1\rangle\Big) =  \frac{1}{\sqrt{2}}\Big(|0\rangle + e^{\frac{2\pi ix}{2^k}}|1\rangle\Big) $. The complexity of QFT is $\mathcal{O}(n^2)$ where $n$ is the number of qubits.

	
	\subsection{Quantum rules and limitations}  
	
	Since quantum gates implement the principles of quantum mechanics, they are reversible
	and unitary. The fact that gates are reversible is clear by noting that they have the same number of inputs
	and outputs. This is not so in classical logic gates like AND, NOR, where there is a single output
	and the input cannot be created by examining the output. 
	We will now mention several rules and `no-go' theorems which are exclusively quantum in nature. These are important ingredients of the remainder of this article. 
	There are several features of quantum computation that makes it much more interesting than
	classical counterpart. One of these is the property that arbitrary (unknown) quantum states
	cannot be cloned. This is known as `no-cloning' theorem \cite{Wootters_1982,Dieks_1982} 
	It is a simple theorem that has far-reaching consequences. In fact, there is also a reverse theorem known as `no-deleting theorem' \cite{Kumar_Pati_2000}. 
	This is also a no-go theorem, which states that given two copies of some arbitrary quantum state, it is impossible to delete one of the copies. It is a time-reversed dual of the no-cloning theorem. These theorems follow from the linearity of quantum mechanics. The no-cloning theorem makes the theory of quantum error correction (discussed later) highly non-trivial since this means that to correct quantum errors due to computation, we cannot simply make backup copies of the quantum state we would like to preserve. Instead, we must 	protect the original from any likely error for as long as we can. 
	
	\begin{mybox}
		\textsc{$\blacktriangleright$ Question 14:}  The process of cloning means that there is a unitary transformation $U$ such that $ U \ket{\psi} \ket{0} \to \ket{\psi} \ket{\psi}$. Show that this is not possible for an arbitrary state. 
	\end{mybox}
	The proof proceeds as follows. Lets take two cases where $\ket{\psi}$ = $\ket{0}$ and $\ket{\psi}$ = $\ket{1}$, 
	then we have $ U \ket{10} = \ket{11}$ and $ U \ket{00} = \ket{00}$ respectively. By linearity we can write:
	\[ U \Bigg(  \frac{\ket{00} + \ket{10}}{\sqrt{2}}  \Bigg) = \frac{1}{\sqrt{2}} \Bigg(\ket{00} + \ket{11}  \Bigg),\]  
	which is equivalent to: 
	\begin{equation}
		\label{eq:NC2} 
		U \Bigg(  \frac{\ket{0} + \ket{1}}{\sqrt{2}} \Bigg) \ket{0}   = \frac{1}{\sqrt{2}} \Bigg(\ket{00} + \ket{11}  \Bigg),   
	\end{equation}
	but the left hand side of (\ref{eq:NC2}) is (by the definition of the $U$ operator): 
	\[  \Bigg(  \frac{\ket{0} + \ket{1}}{\sqrt{2}} \Bigg) \Bigg(  \frac{\ket{0} + \ket{1}}{\sqrt{2}} \Bigg) \]
	and hence we have a contradiction, i.e., 
	\begin{equation}
		\Bigg(  \frac{\ket{0} + \ket{1}}{\sqrt{2}} \Bigg) \Bigg(  \frac{\ket{0} + \ket{1}}{\sqrt{2}} \Bigg) \neq 
		\frac{1}{\sqrt{2}} \Bigg(\ket{00} + \ket{11}  \Bigg).  
	\end{equation}
Hence, there cannot be a general $U$ operator which can clone an arbitrary state. Note that there are some cases where the theorem doesn't apply, 
which is often a source of confusion. The `no-cloning' theorem does not apply in two cases. The first is if we have the knowledge of how the state was prepared and the second is with reference to classical information. Suppose we have states spanned by orthonormal basis (for example, using CNOT gate, 	we can do $ \ket{a} \otimes \ket{b} \to \ket{a} \otimes \ket{a \oplus b}$ and setting $b = 0$ gives $ \ket{a} \otimes \ket{0} \to \ket{a} \otimes \ket{b}$, so we can clone a bit alone.


	Now we talk about another theorem which states that 
	it is impossible to distinguish non-orthogonal quantum states. 
	The `non-distinguishable' theorem can be proved by contradiction 
	as well. Let us suppose that there is a measurement operator $M$
	which has eigenvalues $m_i$ and has projection operators $P_i$
	of some observable and that it allows us to distinguish between
	two non-orthogonal states $\ket{\psi_1}$ and $\ket{\psi_2}$
	with $\langle \psi_1 \vert \psi_2 \rangle \neq 0$.
	Then we can write for some state $\ket{\chi}$ orthogonal to 
	$\ket{\psi_1}$:
	\begin{equation}
	\ket{\psi_2} = \alpha \ket{\psi_1} + \beta \ket{\lambda} 
	\end{equation}
	It is easy to show that $P_{2} \ket{\psi_1} = 0$ and, 
	\begin{equation}
	1 = \langle \psi_2 \vert P_2 \vert \psi_2 \rangle =  \langle \psi_2 \vert P_2 P_2 \vert \psi_2 \rangle = |\beta|^2  \langle \lambda \vert P_2 \vert \lambda \rangle \le  |\beta|^2.
	\end{equation}
	The last inequality follows from the completeness property. This implies $|\beta|^2=1$ which means that  $\langle \psi_1 \vert \psi_2 \rangle = 0$ which is contradiction.
	So, it is impossible to unambiguously differentiate between non-orthogonal quantum states.	

	
	\section{\label{sec:TQA}Towards Quantum algorithms}
	
	One major area of present research in quantum computing is building algorithms (quantum) that have substantial speedup\footnote{In simple terms, the speedup depends on how much quantum mechanics is available.}compared to its fastest known classical counterpart. This area of research is probably the most interesting and difficult. After close to three to four decades of effort, we only have a handful of algorithms that demonstrate a clear advantage over classical algorithms. By quantum algorithm, we mean a sequence of unitary steps that manipulate the initial state $\vert i \rangle$ such that a measurement of the final state $	\vert f \rangle$ provides the desired solution. In this section, we will focus on some well-known algorithms and later show how they can be implemented in \QIS. We will start with probably the simplest of algorithms and one we are well familiar with from an undergraduate course in conventional electronics, i.e., full-adder.

	\subsection{\label{full_add}Full-adder}
	
	The addition of classical bits using adder circuits is one of the first things one learns in the  undergraduate digital electronics course. 
	We will not provide a recap here, but interested readers can refer to the classic textbook by Malvino and Leach \cite{ML_DE}. The quantum full-adder is similar to the classical full-adder except that now we have an identical number of inputs and outputs because we need quantum circuits to be reversible. Therefore, we define a 4-qubit input, where the input qubits are $A$, $B$, $C_i$ (Carry in) and null ($Z$). The output qubits are $A$, $B$, $S$ (Sum) and $C_o$ (Carry out). The table will be: 
	
	\begin{center}
		\begin{tabular}{ |c|c|c|c||c|c|c|c| } 
			\hline
			$A$ & $B$ & $C_i$ & $Z$ & $C_o$ & $S$ & $A$ & $B$ \\
			\hline
			0 & 0 & 0 & 0 & 0 & 0 & 0 & 0\\  \hline 
			0 & 0 & 1 & 0 & 0 & 1 & 0 & 0\\  \hline 
			0 & 1 & 0 & 0 & 0 & 1 & 0 & 1 \\  \hline 
			0 & 1 & 1 & 0 & 1 & 0 & 0 & 1 \\   \hline
			1 & 0 & 0 & 0 & 0 & 1 & 1 & 0\\  \hline 
			1 & 0 & 1 & 0 & 1 & 0 & 1 & 0\\  \hline 
			1 & 1 & 0 & 0 & 1 & 0 & 1 & 1 \\  \hline 
			1 & 1 & 1 & 0 & 1 & 1 & 1 & 1 \\ 
			\hline
		\end{tabular}
	\end{center}
	Once we have prepared the inputs $A$ and $B$ it can be fed to the quantum circuit which will implement the full-adder. The circuit is given below:
	\begin{equation}
		\Qcircuit @C=1.4em @R=1.3em {
			\lstick{A} & \ctrl{3} & \ctrl{1} & \qw & \qw & \ctrl{1} & \qw & \rstick{A} \\
			\lstick{B} & \ctrl{2}  & \targ & \ctrl{2} & \ctrl{1} & \targ & \qw & \rstick{B} \\
			\lstick{C_i} & \qw & \qw & \ctrl{1} & \targ & \meter & \qw & \rstick{S} \\
			\lstick{Z} & \targ & \qw & \targ & \qw & \meter & \qw & \rstick{C_o}
		}
	\end{equation}
	The implementation of this circuit using \QIS~is given in the Appendix, and we test this for some input 
	from the truth table mentioned above.

	\subsection{Phase kickback}
	
	One of the main ingredients for several quantum algorithms is the idea of - `phase kickback'.  Consider the transformation: $ U \ket{\psi} = e^{i 2 \pi \phi} \ket{\psi}$. For the case where $U$ acts on more than one qubit, we might need more ancillas. We consider two qubits to convey this example. The three parts of this procedure can be summed as follows, 1) An extra qubit which is known as ancilla (control) qubit, 2) Performing Hadamard transform on control qubit(s), and 3) Perform controlled-unitary operation (c-U). It is straightforward to see that the final state will be:
	\begin{equation}
		\label{eq:phase_kb1} 
		\frac{1}{\sqrt{2}} \Big( \ket{0} + e^{i \phi} \ket{1} \Big) \ket{1},   
	\end{equation}
	where we note that controlled-U acts like $U \ket{u} = e^{i 2 \pi \phi} \ket{u}$. 
	The phase has been kicked back. This is what we set out to do, i.e., getting a phase rotation to our control qubit whereas the phase rotation gate was applied to the bottom qubit. This is different to the common belief that the control bit mostly remains the same and the controlled (target) bit transforms. This is the reason this procedure is called `phase kickback'.
	We can do higher powers of $U$ to proceed in the same manner. So, if we instead do, $U^{2^{k}}$ then we will obtain a phase of $e^{i 2 \pi 2^{k} \phi}$ kicked back. The \QIS~implementation of this algorithm is straightforward, and we do this in the Appendix for the interested reader. Note that sometimes we cannot obtain the exact answer. For example, if we implement the code in the Appendix for 
	$\theta = 2\pi/3$, we expect to obtain $\sim 0.3333$
	but indeed get output either to be $0.25$ or $0.375$. 
	This is part of the algorithm, and it is known that any exact answer will only be obtained with some probability $p$. However, we can always increase the number of qubits to get a more precise answer. 
	

	
	\subsection{Deutsch-Jozsa Algorithm}
	
	The first quantum algorithm which showed a speedup over classical algorithm was proposed by
	Deutsch and later generalized to $n$-qubits by Deutsch and Jozsa (DJ) in 1992 \cite{1992DJ} and further improved by Cleve et al. \cite{Cleve_1998}. We note that this doesn't really solve a real problem but is just an example of explicitly showing the quantum advantage. This algorithm makes use of Fourier transform like several other quantum algorithms such as Simon's algorithm (which we do not discuss in these notes) and several others. All these developments eventually led to Shor's algorithm - the most famous quantum algorithm today. We will first explain the Deutsch's algorithm and leave the generalization for the reader to explore from other excellent textbooks and notes available. In simple terms, the statement of the problem is: Suppose we have two boxes and each contains either a red or blue ball inside. We would like to know whether two boxes have the same colored ball. Mathematically, consider the Boolean function $f : \{0,1\} \to \{0,1\}$, then the problem is to find whether $ f(0) = f(1)$ or $f(0) \neq f(1)$. We consider the oracle (or black box): $U_{f} \ket{x,y} = \ket{x,y \oplus f(x)}$ with $ x,y \in \{0,1\}$
	
	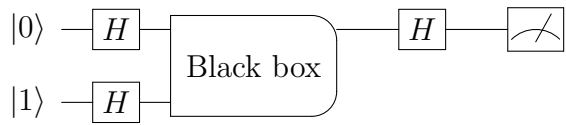
\begin{figure}[h]
		\centering
		\leavevmode
		\large{
			\Qcircuit @C=1em @R=1em {
				\lstick{\ket{0}}& \gate{H} & \multimeasureD{1}{\text{Black box}} & \qw & \gate{H} & \qw &  \meter \\
				\lstick{\ket{1}} & \gate{H}  & \ghost{\text{Black box}} 
		}}
		\caption{Circuit implementing the Deutsch algorithm.}
		\label{fig:DJ1}
	\end{figure}
	
	In this case, the same color means that the function is constant while a different result implies it is balanced. 
	So, $f(0) \oplus f(1) = 0 $ if $f(0) = f(1)$ and 1 otherwise. 
	The sequence of steps as given in the circuit diagram are summarized below:

	\begin{align}
		\ket{0,1} & \xrightarrow{H \otimes H} \ket{+,-} = \frac{1}{\sqrt{2}} \Big( \ket{0} + \ket{1} \Big) \otimes  \ket{-}, \\
		& \xrightarrow{U_{f}} \Bigg[\frac{1}{\sqrt{2}} (-1)^{f(0)} \ket{0} +  \frac{1}{\sqrt{2}} (-1)^{f(1)} \ket{1}\Bigg] \otimes  \ket{-}, \\
		& \xrightarrow{H \otimes \mathbb{1}}  \frac{1}{2}\Bigg[(-1)^{f(0)} + (-1)^{f(1)}\Bigg] \ket{0, -} + 
		\frac{1}{2}\Bigg[(-1)^{f(0)} - (-1)^{f(1)}\Bigg]\ket{1, -}. 
	\end{align}
	Hence, if $f(0) = f(1)$, we will measure $\pm \ket{0,-}$ while if $f(0) \neq f(1)$ then we 
	will measure $\pm \ket{1,-}$. Thus, one call to the oracle (i.e., $U_{f}$ has been used only once above) solves the problem. 
	If we ask how many queries to the oracle must be classically made to determine $f(0) \oplus f(1)$, it is simple to see that we must make two queries. This corresponds to opening both boxes and checking the color of the ball. 
	This algorithm is capable of achieving a speedup to any classical algorithm. 
	This becomes drastic with more qubits. This algorithm 
	can be readily extended to $n$-qubit and to encourage the reader to generalize this algorithm, 
	we give the circuit diagram as an appetizer below for the DJ algorithm in Fig.~\ref{fig:DJ1} below. 
	
	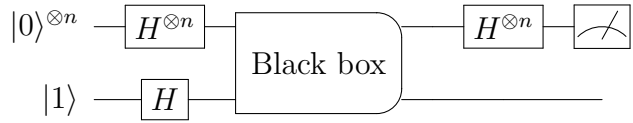
\begin{figure}[h]
		\centering
		\leavevmode
		\large{
			\Qcircuit @C=1em @R=1em {
				\lstick{\vert 0 \rangle ^{\otimes n}}& \gate{H^{\otimes n}} & \multimeasureD{1}{\text{Black box}} & \qw & \gate{H^{\otimes n}} & \meter \\
				\lstick{\vert 1 \rangle} & \gate{H}    & \ghost{\text{Black box}} & \qw  & \qw & \qw & 
		}}
		\caption{Circuit implementing the DJ algorithm on $n$-qubits.}
		\label{fig:DJ1}
	\end{figure}
	We might think what makes the single query to oracle possible? The reason is the parallelism achieved due to rules of quantum mechanics, which enables to perform multiple evaluations simultaneously.	
	\begin{mybox}
		\textsc{$\blacktriangleright$ Question 15:}  Apply the DJ algorithm to input $\ket{0} \ket{0} \ket{1}$ with 
		$f(00)=f(01)=0, f(10)=f(11)=1$ and write down the output. 
	\end{mybox}
	
	
	
	\subsection{Bernstein-Vazirani algorithm}
	
	The Bernstein-Vazirani algorithm is sort of an extension of the Deutsch-Jozsa algorithm discussed above. 
	It main goal of the algorithm is to show that there can be advantages in using a quantum computer as a computational tool for far 
	more complex problems than those covered by the Deutsch-Jozsa problem. The action of the oracle $U_{f}$ is 
	$ \ket{x} \to (-1)^{f(x)} \ket{x}$ with $f(x) = x \cdot s$. 
	
	The main steps of the algorithm can be summarized below:
	
	\begin{itemize}
	\item Suppose the $n$ qubit state is given by $\ket{a}$ usually initialized as $\ket{00000 \cdots 000}$. We first start by applying $H^{\otimes n}$ i.e., 
	\begin{equation}
	\ket{a}  \xrightarrow{H^{\otimes n}} \frac{1}{2^{n/2}} \sum_{x \in \{0,1\}^{n}}  \ket{x}
	\end{equation}
	\item One now applies the oracle to get: 
	\begin{equation}
	\frac{1}{2^{n/2}} \sum_{x \in \{0,1\}^{n}} \ket{x} \mapsto \frac{1}{2^{n/2}} \sum_{x \in \{0,1\}^{n}} (-1)^{f(x)} \ket{x}
	\end{equation} 
	\item The second application of $H^{\otimes n}$ then returns:
	\begin{equation}
	\frac{1}{2^{n/2}} \sum_{x \in \{0,1\}^{n}} (-1)^{f(x) + x \cdot y} \ket{x} = \ket{s} 
	\end{equation}
	\end{itemize} 

	We leave the deduction of final step to get $\ket{s}$ for the interested reader. The similarity between DJ and BV algorithms are: 
	1) Takes in $n$ qubits and outputs 0 or 1, 2) Use of $H$ gate to examine all superposition states. 
	The difference lies in the evaluation and the initial step. DJ determines whether the function is constant or 
	balanced while BV determines the value of function (and search of string, $s$). 
	
	
	\begin{mybox}
		\textsc{$\blacktriangleright$ Question 16:} Apply the BV algorithm for the three qubit case to find `011' using \QIS.
	\end{mybox}

	\subsection{Grover's algorithm} 
	
\begin{figure}
		\centering 
		\includegraphics[width=0.46\textwidth]{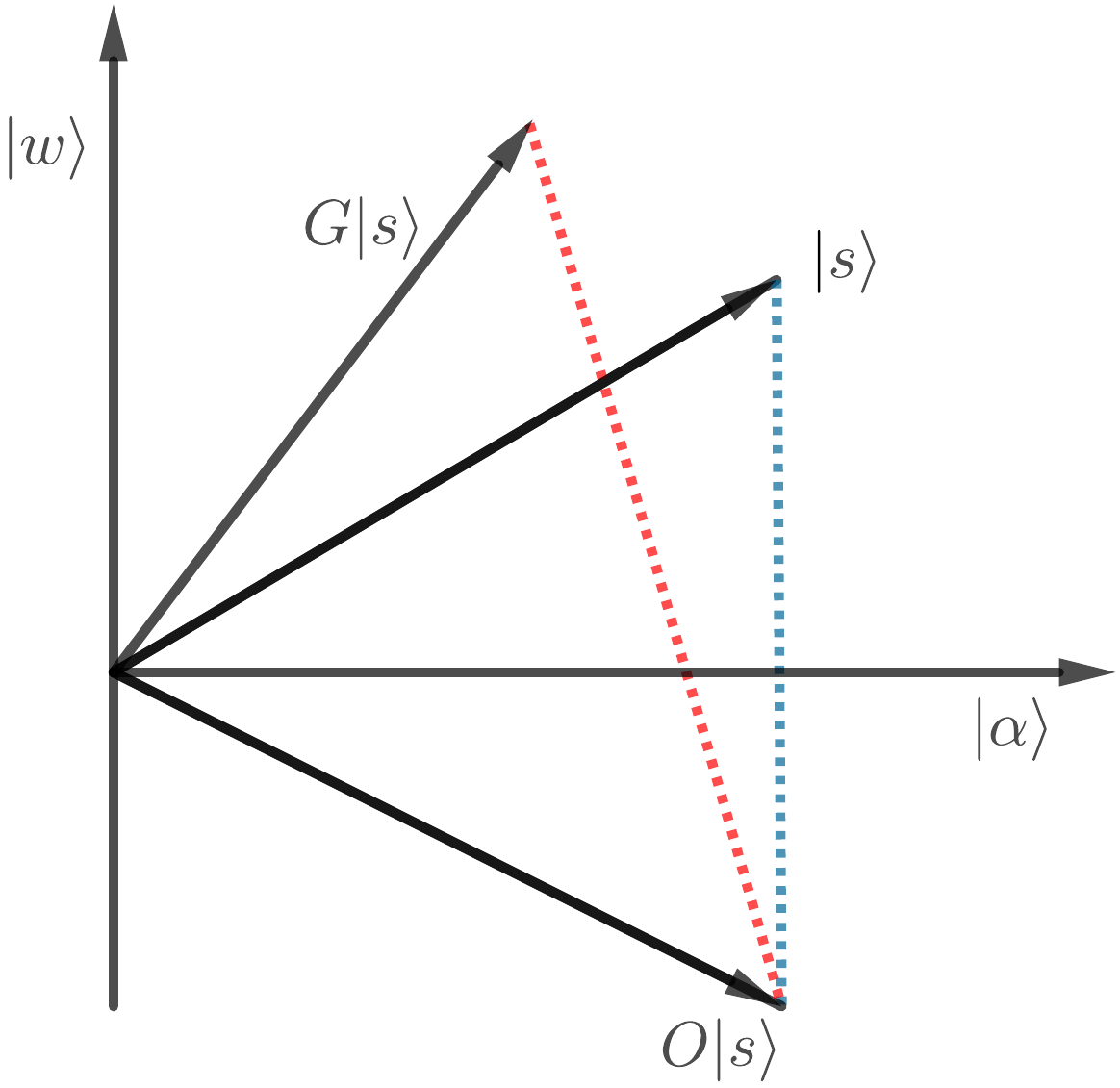}
		\caption{\label{fig:Grover_geo1}Diagrammatic representation of the Grover's algorithm.}
	\end{figure}

	This quantum algorithm was proposed by Grover for structured searching \cite{Grover96, Grover_1997}. The goal of this algorithm is to find the solution to the search problem by 
	making calls to the oracle. Suppose we have a list of $N = 10$ numbers arranged in increasing order from 0 to 9 and
	the goal is to find 5. On average, it will take $\mathcal{O}(N)$ calls to find this. 
	It was shown by that quantum mechanics can help us solve this problem in 
	$\mathcal{O}(\sqrt{N})$ attempts. This is a polynomial speedup over the classical case
	and works by a method called `amplitude amplification'. We will see later that Shor's algorithm offers
	exponential speedup compared to the classical case. 
	This quantum search algorithm consists of repeated application of 
	Grover's operator $G$. There are four steps involved in one iteration (applying $G$ once). 
	We list them below:
	
	\begin{itemize}
		\item Apply the oracle $O$
		\item Apply the Hadamard transformation on $n$ qubits denoted by $H^{\otimes n}$
		\item Perform phase shift unitary operator where every state in computational basis receives a shift. This operator is given by $2 \ket{0} \bra{0} - \mathbb{1}$
		\item Apply the Hadamard again, i.e., $H^{\otimes n}$
	\end{itemize}
	The Hadamard transform (where $N = 2^n$) puts the input state in an equal superposition state written as:
	\begin{equation}
		\kp = \frac{1}{\sqrt{N}} \sum_{x=0}^{N-1} \ket{x}. 
	\end{equation}
	Using this, we can write the combined effect of all four steps as:
	\begin{equation}
		G = H^{\otimes n} (2 \ket{0} \bra{0} - \mathbb{1})  H^{\otimes n} O = (2 \kp \bp - \mathbb{1})O  = D \cdot O.
	\end{equation}
	In short, the oracle operation $O$ reflects the state about $\ket{\psi}$\footnote{There is also an alternate way of achieving amplitude amplification when the state $\kp$ is not known. 
	This is called oblivious amplitude ampliication (OAA)}
	and then $2 \kp \bp - \mathbb{1}$ (also called `diffuser' $D$) reflects it about $\ket{\psi}$. These are the product of two reflections, and the net result is a rotation. One step of Grover's iteration can be diagrammatically represented by Fig.~\ref{fig:Grover_geo1}. We also note that $D =  H^{\otimes n} (2 \ket{0} \bra{0} - \mathbb{1}) H^{\otimes n} = H^{\otimes n} X^{\otimes n} (\text{MCZ}) X^{\otimes n} H^{\otimes n}$. Hence, this can be built from $H$ gates, $X$ gates, and single  multi-controlled $Z$ gate. 
We give the code to implement this algorithm using \QIS~in Sec.~\ref{sec:last}.

	\subsection{Quantum phase estimation (QPE) algorithm} 
	
	The QPE algorithm introduced by Kitaev is an important quantum computation algorithm. 
	It plays a central role in several other algorithms. The goal of this algorithm is to estimate 
	$\theta$ in $U\kp = e^{i2\pi\theta} \kp$ for a unitary operator $U$. Since the operator 
	is unitary, all eigenvalues have a norm of unity. There are several steps involved in QPE 
	which we describe below:
	
	\begin{itemize}
		\item \kp~is one set of qubit registers (the one below in the figure) and the other set is $n$ qubits counting (first) register. We have $\ket{\psi_0} = \ket{0}^{\otimes n} \kp $
		\item Apply $n$-bit Hadamard gate on the first register to get $\ket{\psi_1} = \frac{1}{2^{n/2}}(\ket{0} + \ket{1})^{\otimes n} \kp$
		\item Apply controlled-U (CU) gate (or rather controlled phase with some $\theta$)\footnote{Note that this is what we discussed before in `phase kickback'} to get
		\begin{align}
			\ket{\psi_2} & =\frac {1}{2^{\frac {n}{2}}} \left(|0\rangle+{e^{2\pi i \theta 2^{n-1}}}|1\rangle \right) \otimes \cdots \otimes \left(|0\rangle+{e^{2\pi i \theta 2^{1}}}\vert1\rangle \right) \otimes \left(|0\rangle+{e^{2\pi i \theta 2^{0}}}\vert1\rangle \right) \otimes \kp,  
		\end{align} 
		where we note that the rightmost qubit in the first register is $\ket{q_0}$ (see figure) 
		\item Apply inverse quantum Fourier transform to get
		\begin{align}
			\vert\psi_3\rangle = \frac {1}{2^{\frac {n}{2}}}\sum _{\alpha=0}^{2^{n}-1}e^{2\pi i \alpha \theta}|\alpha\rangle \otimes | \psi \rangle \xrightarrow{\rm{QFT}_n^{-1}} \frac {1}{2^n}\sum _{x=0}^{2^{n}-1}\sum _{\alpha=0}^{2^{n}-1} e^{-\frac{2\pi i \alpha}{2^n}(x - 2^n \theta)} |x\rangle \otimes |\psi\rangle
		\end{align}
	\end{itemize}

	\begin{align*}
		\Qcircuit @C=.7em @R=.7em {
			\lstick{\ket{q_0}}    & \gate{H} & \qw & \ctrl{4} & \qw        & \qw & \cdots & & \qw                & \multigate{3}{\rm{QFT}_n^{\dagger}} & \qw    & \meter & \cw \\
			\lstick{\ket{q_1}}    & \gate{H} & \qw & \qw      & \ctrl{3}   & \qw & \cdots & & \qw                & \ghost{\rm{QFT}_n^{-1}}	& \qw  & \meter & \cw \\
			\lstick{\vdots\ \ } & \vdots   &     &          &            &     &        & &                    & \pureghost{\rm{QFT}_n^{\dagger}}    &        & \vdots &     \\
			\lstick{\ket{q_{n-1}}}    & \gate{H} & \qw & \qw      & \qw        & \qw & \cdots & & \ctrl{1}           & \ghost{\rm{QFT}_n^{\dagger}}	& \qw  & \meter & \cw \\
			\lstick{\ket{\psi}} & /^m \qw  & \qw & \gate{U} & \gate{U^2} & \qw & \cdots & & \gate{U^{2^{n-1}}} & \qw
		}
	\end{align*}

	\begin{mybox}
		\textsc{$\blacktriangleright$ Question 17:}  Consider the following circuit (this is Problem 5.3 in Ref.~\cite{Qinfobook}):
		\[
		\Qcircuit @C=1.2em @R=0.9em @! {
			\lstick{\ket{0}} & \gate{H} & \ctrl{1} & \gate{H} &  \meter & \\
			\lstick{\ket{u}^{\otimes n}} & \qw & \gate{U} &  \qw & \qw & 
		}
		\]
		where $\ket{u}$ is an eigenstate of $U$ with eigenvalue equal to $e^{i \theta}$. Show that the top qubit is measured to be 0 with probability equal to $ \cos^{2}(\theta/2)$
	\end{mybox}

	\subsection{Shor's factoring algorithm} 
	
	One of the central goals of quantum computation is to identify a problem
	where the quantum offers a significant advantage over classical methods. 
	In this regard, the pioneering work of Ref.~\cite{Shor:1994jg}
	showed that it is possible to factor a number using a quantum computer in a polynomial
	time, unlike the requirement of exponential time with classical computers. 
	This problem is much more interesting than it appears to be at first glance. It is a known fact that 
	multiplying two numbers (say 13 and 73) is much easier than finding prime factors of 949
	as it would take more time. This problem is very hard (and sometimes intractable) 
	with large numbers. This was one of the first examples which explicitly showed that quantum algorithms can do
	exponentially better than classical ones. The factoring algorithm is the flagship result in quantum algorithms of the 
	last three decades. The algorithm efficiently factors a given integer into two integers. 
	For example, we can factor 1,624,637,792,837 into the primes 15,485,867
	and 104,911 respectively faster than any classical algorithm. This is a very practical application because in today's world many modern cryptography systems, such as RSA protocol heavily rely upon the fact that factoring such large integers is computationally difficult for any given computer in a reasonable time. 	However, we are far from practical usage of this algorithm since one would ideally need several thousand qubits to perform such a task and is \emph{well}-beyond the NISQ era. In 2001, it was shown how to successfully factor 15 via Shor’s algorithm on a 7-qubit nuclear magnetic resonance (NMR) quantum computer 
	\cite{Vandersypen_2001}. In fact, a few years ago, it was argued in Ref.~\cite{2019arXiv190509749G} 
	that it would take about 8 hours to factor 2048-bit RSA integers using 20 million noisy qubits!
	We can summarize the algorithm in the following steps:
	\begin{enumerate}
		\item If $N$ is even, return the factor 2  (of course this most likely won't happen!) 
		\item Check whether $N = a^b$ for integers $a, b$. Return $a$ as a factor if it is true. 
		\item Pick a random integer $p$ between 2 and $N$.  If $\texttt{gcd(N,p)} \neq 1$, we are done and can return the factor. If not, we continue to Step 4 (likely will always happen!).
		\item Use an algorithm to find period $r$ such that $a^r \rm{mod}~N = 1$
		\item If $r$ is even and $a^{r/2} \rm{mod}~N \neq -1$, then evaluate $\texttt{gcd}(a^{r/2} \pm 1, N)$. If we find non-trivial factor, return that value as a factor otherwise we return to Step 3. 
	\end{enumerate}
	Note that Euclid's algorithm can be used to find $\texttt{gcd(a,b)}$ efficiently.\footnote{In his book, `The Art of Computer Programming, Vol. 2: Seminumerical Algorithms', Knuth comments that this algorithm is the granddaddy of all algorithms because it is the oldest nontrivial algorithm that has survived to the present day. It was written down by Euclid circa 300 B.C.} Let us clarify the step of period finding for the reader. Consider: 
	$a^{r} = 1 (\text{mod N})$ where $r$ is the smallest positive integer. 
	Now suppose we have, $N=15, a=7$ then $r= 4$. This is well-defined 
	only if $N$ and $a$ are co-prime or relatively prime (i.e., they have no common factor except 1). 
	In \MAT~, this can be evaluated by doing:
	$\texttt{MultiplicativeOrder[7, 15]}$. The fact that finding period can actually help us find prime factorization 
	can be obtained from lemmas that follow from Fermat's little theorem (1640) i.e., for every $a$ and prime $p$, 
	we have $a^{p} = a (\text{mod}~p)$. An alternate statement of this theorem is also often mentioned: 
	$a^{p-1} = 1 (\text{mod}~p)$ where $p$ is prime and $a$ is any integer not divisible by $p$.

	
	\section{\label{sec:VQE}Variational Quantum Eigensolver (VQE)}
	
	One of the main motivations why quantum computing is required is the necessity to simulate complicated quantum many-body systems or solve large scale linear algebra problems. These are very challenging (and sometimes not possible) for classical computers due to the high computational cost involved. Even though quantum computers offer a promise to the solution, the error correction and feasibility are still not well understood. One of the major areas of research with respect to how algorithms can be developed is Variational Quantum Algorithms (VQAs). We refer the interested reader to Ref.~\cite{cerezo2020variational} to start the reference trail. These algorithms make use of classical optimizer and along with quantum circuits are argued to have a role in wide-ranging areas where quantum computation might be useful in the future. But, in particular, the earliest and well-known of these VQAs is an algorithm that has drawn considerable interest in the last decade. This is known as VQE and makes use of the well-known variational principle to compute the ground state energy of a given Hamiltonian which is important for many problems in quantum chemistry, condensed matter physics, and quantum many-body systems. One of the main advertised advantages of this method is that it can be used to deal with complex wave functions of a wide class of systems in polynomial time when classical methods grow exponentially. Though this method has not yet truly outclassed the classical algorithms, it has incredible potential once we enter beyond NISQ era. This algorithm has well-known drawbacks since this is a non-linear optimization problem. We refer the interested reader to Ref.~\cite{Tilly:2021jem} for references. The Variational Quantum Eigensolver (VQE) was originally developed in Ref.~\cite{Peruzzo_2014}
and is among the most promising examples of NISQ algorithms. It aims to compute an upper bound for the ground state energy of some given Hamiltonian. VQE is a hybrid (mixture of classical and quantum) algorithm which can be implemented on NISQ devices
and used to obtain the ground state of the quantum many-body of given Hamiltonian 
$\mathcal{H}$ using some ansatz. The basic steps of the algorithm are as follows:
	
	\begin{enumerate} 
		\item Prepare an initial quantum state $\ket{0}$ on a quantum computer. 
		\item Generate a good ansatz quantum state $ \ket{\Psi (\Theta)}$
		by applying a suitable  unitary transformation (quantum circuit composed of quantum gates)
		as $\ket{\Psi(\Theta} = \hat{U}(\Theta) \ket{0}$
		\item The energy expectation value is measured on a quantum computer as $E_{\Theta} = \langle \Psi(\Theta)|\mathcal{H}|\Psi(\Theta) \rangle$ by 
		measuring each Pauli string separately and then summing the individual terms. 
		\item Update the parameters $\Theta$ to get smaller energy $E(\Theta)$ by using a classical optimizing algorithm.
		\item Repeat steps 2-5 until desired convergence is attained. 
	\end{enumerate} 
	Within VQE, the cost function is defined as the expectation value of the Hamiltonian computed in the trial state. The ground state of the target Hamiltonian is obtained by 
	performing an iterative minimization of the cost function. The optimization is carried out by a classical optimizer, which leverages a quantum computer to 
	evaluate the cost function and calculate its gradient at each optimization step. The local optimization 
	within the vector space generated by ansatz 
	can suffer due to the problem of vanishing gradients. This can often result in a phenomenon known as 
	`barren plateaus' \cite{McClean2018} which are a well-known issue in VQE computations but 
	there have been some proposed methods to overcome
	this issue \cite{PRXQuantum.1.020319}. 
	A detailed discussion is beyond the scope of this article. 
	
	The choice of ansatz is usually determined by hardware availability (known as hardware-efficient ansatz)
	and efficiency of considering qubits as logical qubits.\footnote{Usually when we have 
	qubits, they are prone to errors and can make computation incorrect.
	We want qubits that are ideal or at least not prone to many errors, they are known as `logical qubits'. 
	The problem is that usually for every logical qubit we need multiple physical qubits which can sometimes be hundreds of them.}
	We refer the interested reader to Ref.~\cite{Kandala} for an elaborate discussion on hardware-efficient trial states. The VQE that 
	accompanies \QIS~comes with some given choices such as RX, RYRZ, $SU(2)$ etc. with some user-defined number of repetitions. 
	To measure the energy expectation value in the third step above, we must have a qubit 
	Hamiltonian of size $2^n \times 2^n$. This is usually not an easy task, however, this can be simplified 
	considerably by writing the original Hamiltonian in terms of the sum of Pauli operators (X,Y,Z, $\mathbb{1}$) and their tensor products.  
	However, not all Hamiltonians (even though they are sparse) admit an efficient Pauli representation. For these kind of systems, 
	alternate methods have also been proposed \cite{Kirby2021}. 
	This can also be understood as the first step even before doing any computation and has to be done only once.  
	The choice of the right optimization method is crucial for the problem. We can choose from a range of methods such as 
	BFGS (Broyden–Fletcher–Goldfarb–Shanno) optimizer method, COBYLA\footnote{Note that this is non-gradient based optimization, 
	we do not need to know
	the derivative}(Constrained Optimization By Linear Approximation optimizer), 
	SLSQP (Sequential Least SQuares Programming optimizer), Nelder-Mead or any other depending on the requirement and efficiency required. 
	For example, another example of a local optimizer in QISKIT is Powell method, which performs unconstrained optimization; 
	it ignores bounds or constraints and is a conjugate direction method. More details can be found in Ref.\cite{QISKIT_TB}. Another interesting method to 
	construct the ground state of an interacting Hamiltonian is the adiabatic state preparation (ASP) method. It works 
	reasonably well when there is no level crossing (i.e., the excited state of the Hamiltonian do not cross each other and ground state). 
	Recently, there have been some interesting work combining the ideas of VQE approach and ASP approach \cite{Schiffer:2021xiv}.  

	
	When we deal with physical systems, we can often express a given Hamiltonian as the linear combination of the tensor product of Pauli matrices
	$\sigma_{x}$, $\sigma_{y}$, and $\sigma_{z}$ which we call `Pauli strings' i.e., 
	$\hat{P} = \otimes_{i=1}^{n} \hat{\Sigma}$
	where $\hat{\Sigma} = \{\sigma_{x},\sigma_{y},\sigma_{z}, \mathbb{1}\}$. Note that as the Pauli strings form a complete basis, any Hamiltonian of size
	$2^N \times 2^N$ can be decomposed into a
	linear combination of Pauli strings. For example, let us consider the Hamiltonian of some system to be given by:
	\[ H = \left(
	\begin{array}{cccccccc}
		\frac{1}{2} & 0 & -\frac{i}{2} & 0 & 0 & 0 & 0 & -\frac{1}{2}-\frac{i}{2} \\
		0 & \frac{1}{2} & 0 & \frac{i}{2} & 0 & 0 & -\frac{1}{2}+\frac{i}{2} & 0 \\
		\frac{i}{2} & 0 & \frac{1}{2} & 0 & 0 & -\frac{1}{2}+\frac{i}{2} & 0 & 0 \\
		0 & -\frac{i}{2} & 0 & \frac{1}{2} & -\frac{1}{2}-\frac{i}{2} & 0 & 0 & 0 \\
		0 & 0 & 0 & -\frac{1}{2}+\frac{i}{2} & \frac{1}{2} & 0 & \frac{i}{2} & 0 \\
		0 & 0 & -\frac{1}{2}-\frac{i}{2} & 0 & 0 & \frac{1}{2} & 0 & -\frac{i}{2} \\
		0 & -\frac{1}{2}-\frac{i}{2} & 0 & 0 & -\frac{i}{2} & 0 & \frac{1}{2} & 0 \\
		-\frac{1}{2}+\frac{i}{2} & 0 & 0 & 0 & 0 & \frac{i}{2} & 0 & \frac{1}{2} \\
	\end{array}
	\right) \]
	which can be written in terms of Pauli operators as: $H = \frac{1}{2}(\sigma_z \otimes  \sigma_y \otimes  \sigma_z  - \sigma_x \otimes  \sigma_x \otimes  \sigma_x - \sigma_y \otimes  \sigma_y \otimes  \sigma_y  + \mathbb{1}  \otimes  \mathbb{1}
	\otimes  \mathbb{1})$. The program to do this transformation can be found in the subroutine of the VQE program in Sec.~\ref{sec:last}. 
	One of the drawbacks for VQE are handling models for whose Hamiltonian there is no known efficient decomposition in 
	terms of Pauli strings. In such a case, the advantage of VQE is reduced considerably. This is an important question
	and some work along this direction has already been done. We refer the reader to Ref.~\cite{2021PhRvL.127k0503K} 
	for extended discussion. 
	
	One important issue with computations in the NISQ-era is the noise associated with the calculation. 
	Strictly speaking, the theory of fault-tolerant quantum computing is still several decades away and hence there
	is need to understand and mitigate errors in calculations from current era. The quantum error mitigation (QEM) 
	is a technique to understand the errors and reduce them such that we can still draw meaningful conclusions from the computations
	which otherwise seems just like noise and unphysical. There are several methods developed for QEM. We will mention 
	few methods below and refer the reader to Ref.~\cite{Cai2022} for extended list. 
	
	\begin{itemize}
	\item Zero noise extrapolation (ZNE): This technique mitigates errors in noisy quantum computations without using additional quantum resources. This is based on understanding how 
	the noise scales and how we can extrapolate to zero noise. In ZNE, a quantum program is deliberately altered to run at different levels of noise from processors. 
	\item Randomized compilation (RC): This is a protocol which is constructed 
	to overcome this issue by converting coherent errors into stochastic noise. 
	Randomized compiling makes use of twirling groups 
	while preserving the overall unitarity of the circuit.  
	Pauli twirling is a method which was introduced for the mapping of states into 
	canonical form in entanglement purification studies in the 
	quantum information literature. It was also used in reducing the exponential cost
	in quantum tomography. In QEM, it is used as a technique to improve the 
	performance on NISQ devices. The nature of noise in quantum systems can be broadly classified as either coherent (unitary) or incoherent. 
	The coherent noise are more dangerous since they accumulate quickly and can lead to worst-case error. 
	One of the most effective methods of dealing with coherent noise/error is to convert them to Pauli errors. 
	For this one often uses `Pauli twirling'. Pauli twirling convert error channels to Pauli channels for estimation of error threshold. 
	Precisely, it converts the coherent noise into incoherent Pauli channels by supressing the
	off-diagonal terms. In addition, if the twirling is performed with Clifford group, it is called Clifford twirling. 
	This converts the error channels into depolarising channels. In general, any arbitrary channel
	can be converted to Pauli channel. 
	\item Learning based error mitigation: This method was introduced in Ref.~\cite{2021PRXQ....2d0330S}. The basic idea of this method is 
	to start with a noisy expectation value $E(\textbf{P})$ for some primary circuit $\textbf{P}$ and achieve the noiseless value 
	$E_{0}(\textbf{P})$ passing through series of optimizations of the expectation value which depends on some set of parameters $\vec{\theta}$. 
	The optimal choice of $\vec{\theta}$ when obtained through learning-based methods based on temporary circuit 
	$\textbf{T}$ which is close enough to the primary circuit. 
	
	\end{itemize}

	\subsection{\label{subsec:AHO}Quantum Anharmonic Oscillator}
	
	We will now discuss how to apply VQE to solve (i.e., find the ground state energy) quantum anharmonic oscillator. This example provides sufficient exposure to the basis working of VQE. This problem was also studied in \cite{miceli2018quantum} and we will 
	follow their discussion closely. For the ansatz, we choose $\texttt{EfficientSU2}$ circuit from \QIS~circuit library, which is a hardware-efficient, heuristic ansatz with alternating rotation and entanglement layers. Note that it is possible to calculate excited states of a Hamiltonian based on a technique known as variational quantum deflation (VQD) proposed recently in Ref.~\cite{Higgott_2019}. Let us write the Hamiltonian as:
	\begin{equation}
		\hat{H}^{\prime} = \hat{H} - g \hat{X}^{3} - h \hat{X}^{4}. 
	\end{equation}
	For the cubic oscillator, if we consider three qubits, $\hat{H}^{\prime}$ is a $8 \times 8$ matrix which for $g = 0.02$ is given by:
	\begin{small} 
		\begin{align}
			\hat{H}^{\prime} = & 4\mathbb{1}_8 - 0.152956\mathbb{1}_4X -0.5 \mathbb{1}_4Z - 0.12289 * \mathbb{1}XX - 0.0629948 \mathbb{1}YY -1\mathbb{1}Z\mathbb{1} +  0.0237627\mathbb{1}ZX -0.0280252X\mathbb{1}X \nonumber \\  & -0.0561195XXX + 0.0287333XYY + 0.0107047XZX -0.0280252Y\mathbb{1}Y -0.0287333YXY -0.0561195YYX \nonumber  \\  &
			+ 0.0107047YZY -2Z\mathbb{I}_4 +  0.0872346Z\mathbb{1}X + 0.0842295ZXX + 0.041655ZYY +0.0207442ZZX
		\end{align}
	\end{small}
	The energy levels for cubic anharmonic oscillator ($h =0$) which we take as example is known to be:
	\begin{equation}
		E_{n} = -\frac{g^2}{8}\left(30n^2 + 30n + 11 \right) + \mathcal{O}(g^4), 
	\end{equation}
	and the ground state energy level upto $\mathcal{O}(g^6)$ is given by (see Table (3.1) of Ref.~\cite{Kleinert:2004ev}) 
	\begin{equation}
		E_{0} = \frac{1}{2} - \frac{11}{8} g^2 - \frac{465}{32} g^4 - \frac{39709}{128} g^6 
	\end{equation}
	We provide the code in \QIS~in the Appendix for the interested reader though admittedly it can definitely be improved and optimized of which we have not taken any care except checking that it works. We need to create $H$ matrix for this system which can be done in \MAT~as below \footnote{Note that there are other ways to do this, but this is probably the cleanest} which will be then converted to Pauli string and input to VQE to find $E_{0}$: 
	
	\begin{mdframed}[backgroundcolor=celadon!6] 
		\begin{lstlisting}[language=Mathematica]
SetDirectory[NotebookDirectory[]];
nbits = 3; (* number of qubits *)
n = 2^nbits; 
lattice = Table[(2 a - n - 1)/2, {a, n}]; (* Lattice for position operator *)
X = DiagonalMatrix[Sqrt[(2 \[Pi]/n)] lattice];(* Position operator *) 
F = (1/Sqrt[n]) Table[N@Exp[(-I 2 Pi lattice[[j]] lattice[[k]])/n], {j, n}, {k, n}];(* Discrete Fourier transform *)
P = Chop[F\[ConjugateTranspose] . X . F]; (* Momentum operator) *)
A = Sqrt[0.5] (X + I P); (* Annihilation operator *)
A1 = Table[If[(j - i) == 1, i^.5, 0], {i, n}, {j, n}] ;  (* Annihilation operator in energy (E) basis *)
X1 = Sqrt[0.5] (A1\[ConjugateTranspose] + A1); (* Position operator in E basis *)
P1 = I Sqrt[0.5] (A1\[ConjugateTranspose] - A1); (* Momentum operator in E basis *)
g = 0.02; 
h = 0.04; 
H0[X_, P_] := .5 P . P + .5 X . X;  (* Harmonic oscillator Hamiltonian *)
H1[X_, P_] := .5 P . P + .5 X . X - g MatrixPower[X, 3]; (* cubic anharmonic Hamiltonian *)
H2[X_, P_] := .5 P . P + .5 X . X + h MatrixPower[X, 4];  (* quartic anharmonic Hamiltonian *)
H0A[A_] := A\[ConjugateTranspose] . A + .5 IdentityMatrix[n]; (* Harmonic oscillator Hamiltonian in E basis *)
H1A[A_, X_] := H0A[A] - g MatrixPower[X, 3]; (* Cubic anharmonic Hamiltonian in E basis *)
H2A[A_, X_] := H0A[A] + h MatrixPower[X, 4]; (* Quartic anharmonic Hamiltonian in E basis *)
H = H1A[A1, X1]; (* Make Hamiltonian to export *)
hamName = "HO"; (* Set Hamiltonian name for file *)
Export["ham_" <> hamName <> ".txt", H, "Table"]; (* Export Hamiltonian to file which would be read by our QISKIT program! *)
		\end{lstlisting}
	\end{mdframed}
In fact, an astute reader might ask: What is the point of doing this VQE computation when I can easily find the ground state energy of an $8 \times 8$ Hamiltonian in three lines by doing:	
	\begin{mdframed}[backgroundcolor=celadon!6] 
		\begin{lstlisting}[language=Mathematica]
from numpy import linalg as LA
w, v = LA.eig(H)
print ("Ground state energy", w[0])
		\end{lstlisting}            
	\end{mdframed}
To address this, we note that the interest in VQE algorithms is because for a Hamiltonian of size $2^n \times 2^n$ where $n$ is a large number and $H$ is complicated enough that it is not very sparse like it is in this case, then it will take classical computing/exact diagonalization an exponentially longer time and a large amount of memory. This is often given the name `solution does not scale with system size'. However, if it is assisted through VQE (creating quantum state ansatz and optimizing classically) then the growth with $n$ and sparseness of $H$ will be much more efficient (at most polynomials!). It must be noted that VQE has already been able to substantially get close to some gold standard in quantum chemistry, however, the discussion of this lies beyond the scope of this article. We just point out one reference of the computation done in quantum chemistry that was for quantum simulation of the deuteron binding energy on quantum processors such as IBM QX5 (with 16 SC qubits) and Rigetti 19Q (19 SC qubits) \cite{dumitrescu2018cloud} which can be used to start a reference trail.   

\subsection{$O(3)$ non-linear sigma model}
We mention another toy model (familiar to most physicists), known as $O(3)$ 
non-linear sigma model in 1+1-dimensions. This was studied using VQE recently in the Ref.~\cite{Araz:2022tbd}
on about 10 lattice sites in the weak and strong coupling limit with a topological $\theta = \pi$ term. 
The results obtained with two different choices of ansatz showed that VQE does not 
correctly reproduce the ground state energy for large $\beta$ and works best for $\beta \ll 1$.
The increase in number of parameters to be optimized did not allow the authors to explore VQE for 
$l_{\rm{max.}} >1/2$. Note that the allowed values of $l$ with $q=1/2$ which corresponds to $O(3)$ model with $\theta = \pi$ are half-integer values. 
The eigenbasis is denoted by $\ket{qlm}$ rather than the usual $\ket{lm}$.
If $q=\theta=0$, the $l$ takes integer values starting at 0. 

Based on results from classical tensor network computations based on MPS, 
it is clear that to approach the asymptotic scaling regime where the continuum physics can be extracted, 
one needs $ 1.0 < \beta < 1.4$ and it is not possible without including higher charge representations.
It would be  interesting to study this system with improved hybrid methods like VQE for larger $\beta$ and find
agreement with tensor network results. In this subsection, we will not pursue any `quantum' or hybrid calculations and 
just show that one can use exact diagonalization (ED)\footnote{We thank Kostas Orginos and Felix Ringer for some discussions related to this}
to compute the ground state energy for small number of sites in the strong and weak coupling limits. 
The Hamiltonian of the model is given by:
\begin{equation}
\label{eq:Ham}
    \hat{H} = \frac{1}{2\beta} \sum_{i} \textbf{L}^{2}_{i} - \beta \sum_{\langle i,j \rangle} \textbf{n}_{i} \cdot \textbf{n}_{j},
\end{equation}
where $i$ and $j$ are the nearest neighbour sites on a spatial line, $\textbf{n}$ is a unit 3-vector at site 
$i$, $\textbf{L}$ is the angular momentum and $\beta$ is the coupling. 

The dimension of the local on-site Hilbert space is given by $ \sum_{0}^{l_{max.}} (2l+1)$ which is $d=2$ with $l_{\rm{max.}} = 1/2$. 
Therefore, $H$ will be a $16 \times 16$ matrix for $N=4$. We can exactly diagonalize (ED) the Hamiltonian 
by expressing (\ref{eq:Ham}) in terms of Pauli matrices and compare to the results 
from Ref.~\cite{Araz:2022tbd}. We find that for $\beta=1/10$ and $N=4$, we get $E_{0}/N = 3.72778$
while for $\beta=10$, we get $E_{0}/N = -2.1847$ at fixed truncation of 
$l_{\rm{max.}} = 1/2$ at each site. The result at large $\beta$ is inaccurate (not close to convergence) 
since the truncation effects are important. One obtains $E_{0}/N = -5.62$ if  $l_{\rm{max.}} = 3/2$
is used with $N=4$ sites with $H$  given by a $6^4 \times 6^4$ matrix. For VQE, we would need
to work with 11 qubits for which optimisation might not be straightforward. 
The code for implementing this with $l_{\rm{max.}} = 1/2$ is given below. 
\begin{mdframed}[backgroundcolor=celadon!6] 
		\begin{lstlisting}[language=Mathematica]
X = {{0, 1}, {1, 0}}; 
Y = {{0, -I}, {I, 0}}; 
Z = {{1, 0}, {0, -1}};  
Nplus =  (X + I Y) ; 
Nminus = (X - I Y ); 
SITES = 4; 
HH[\[Beta]_, N_] := N (3/(8 \[Beta] )) IdentityMatrix[16] + \[Beta] (KroneckerProduct[Nplus , Nminus, IdentityMatrix[2], IdentityMatrix[2]] + KroneckerProduct[Nminus, Nplus, IdentityMatrix[2], IdentityMatrix[2]]  + KroneckerProduct[Z , Z, IdentityMatrix[2], IdentityMatrix[2]]/9) +  \[Beta]  (KroneckerProduct[IdentityMatrix[2], Nplus , Nminus,  IdentityMatrix[2]] + KroneckerProduct[IdentityMatrix[2], Nminus, Nplus, IdentityMatrix[2]]  + KroneckerProduct[IdentityMatrix[2] , Z, Z, IdentityMatrix[2]]/9) + \[Beta]  (KroneckerProduct[IdentityMatrix[2], IdentityMatrix[2], Nplus , Nminus ] + KroneckerProduct[IdentityMatrix[2], IdentityMatrix[2], Nminus, Nplus]  + KroneckerProduct[IdentityMatrix[2] , IdentityMatrix[2], Z, Z ]/9)  + \[Beta]  (KroneckerProduct[Nminus, IdentityMatrix[2], IdentityMatrix[2], Nplus ] + KroneckerProduct[Nplus, IdentityMatrix[2], IdentityMatrix[2], Nminus]  + KroneckerProduct[Z, IdentityMatrix[2] , IdentityMatrix[2], Z ]/9); 
HH[0.1, SITES]/SITES // Eigenvalues // Min // N
(* Exercise: Write a general $N$ site program. Try ED for 6 sites and compare.*) 
		\end{lstlisting}
	\end{mdframed}
Note that going beyond the $l_{\rm{max.}} = 1/2$ is non-trivial. 
The size of $H$ with $l_{\rm{max.}} = 5/2$ will be $12^4 \times 12^4$
and we would need 14-15 qubits to represent it. If we take $\theta = 0$, then the allowed
values of $l$ will be $0, \cdots l_{\rm{max.}}$ with size of Hamiltonian 
given by $(l_{\rm{max.}}+1)^{2N} \times (l_{\rm{max.}}+1)^{2N}$ 
for $N$-sites.  

There is an alternative view to think about quantum computation of this model. 
Rather than thinking about qubits, we can represent (\ref{eq:Ham}) in terms of
two distinct harmonic oscillators (qumodes) at each site using the Schwinger representation. 
This representation is straightforward for the kinetic term 
(since it is just the number operator) but one can also express the nearest-neighbor interaction term in oscillator basis. 
The schematic form of the Hamiltonian is:  
\begin{equation}
H = \sum_{\langle ij \rangle} f(a^{\dagger}_{i},b^{\dagger}_{i}, a_{i}, b_{i}) f(a^{\dagger}_{j},b^{\dagger}_{j}, a_{j}, b_{j}) + \sum_{i} f(n_{i}), 
\end{equation} 
where the two oscillators at each site are denoted by $a,b$ and $n_{i} = a^{\dagger}_{i}a_{i}  + b^{\dagger}_{i}b_{i}$.
The Hamiltonian of this model has some similarity to the Bose-Hubbard
for which mapping from bosonic infinite-dimensional Hilbert space to qubits
have been developed \cite{Somma2003, Sawaya2019}. 
In addition to the 1+1-dimensional $O(3)$ model, the 
(hybrid) quantum algorithms
have also been applied to other models (one spatial dimension) 
such as two-flavor Gross-Neveu model \cite{Asaduzzaman:2022bpi}
and to Schwinger model \cite{Shaw:2020udc}. 
It will be interesting to understand these 
models from the perspective of a continuous variable (bosonic) quantum 
computing methods and some work along these lines have already been done
using Xanadu's simulator \cite{Yeter-Aydeniz:2021mol} where the ground state energy
for scalar field theory was computed using the quantum imaginary time evolution (QITE) algorithm
proposed with and without use of ansatze in Refs.~\cite{2019npjQI...5...75M,2020NatPh..16..205M}.

\section{\label{sec:QEC}Quantum Error Correction (QEC)} 

Quantum computers often offer drastic speedup to solutions of certain types of problems by using principles of quantum mechanics such as: superposition and entanglement. However, the use of qubits in place of classical bits also makes the quantum computer susceptible to errors unlike anything we know in classical computers. The origin of these errors is primarily due to decoherence - a process in which the environment interacts with the qubits and changes the quantum states, which eventually leads to corrupted results. For example, the failure rate is about one error in $10^{17}$ operations for classical computers, which for quantum computers is much severe. This partly makes the theory of quantum error correction more subtle 
than classical counterpart, in addition to several reasons below: 
	\begin{itemize}
		\item The act of measurement or observation in quantum mechanics destroys
		the quantum state and makes its recovery impossible. For example, if we have a superposition 
		state, and we perform a measurement, we lose the superposition by choosing one outcome. 
		\item No-cloning theorem prohibits the copying of any arbitrary quantum state, and this means 
		that repetition code obtained by just duplicating the bit will not work. 
		\item Unlike classical bit, qubit is continuous valued and hence we need infinite precision to locate the error exactly in order to correct it. 
	\end{itemize}
	The goal of the quantum error-correcting (QEC) code is to protect some subspace of the entire Hilbert space $\mathcal{H}$ 
	from some errors. More precisely, suppose we want to 
	preserve a $k$-dimensional subspace (coding space) against some known errors. 
	This is achieved by mapping the states into a larger, $n$-dimensional Hilbert space. 
	In that case, we can refer to it as the $(n,k)$-quantum code. Sometimes, an alternate definition is also used, where by $[n,k,d]$ we denote a quantum error-correcting code that uses $n$ qubits to encode $k$ qubits with distance $d$. There is a well-known Knill-Laflamme theorem \cite{Knill:1996ny} for the conditions of quantum error correction. 
	It is stated as follows: Let $\mathcal{C}$ be a quantum-error correcting code defined as a subspace of the $n$-dimensional Hilbert space, 
	$\mathcal{H}_2^{\otimes n}$, and let $\mathcal{E} \subset \mathfrak{C}^{2^n\times2^n}$ be a set of errors. Then $\mathcal{C}$ can correct 
	$\mathcal{E}$ if and only if 
	
	\begin{equation}
	\langle\psi_i|E_a^\dagger E_b|\psi_j\rangle = c_{ab}\delta_{ij} ,
	\end{equation} 
	where {$|\psi_i\rangle$} is a base of the subspace that defines the code 
	$\mathcal{C} \subset \mathcal{H}_2^{\otimes n}$ and $E_a$, $E_b$ $\in \mathcal{E}$. Equivalently, we can also express the above statement as: Let $C$ be a quantum code and let $P$ be a projector onto
	$C$. Suppose a list of operation elements is represented by $\{E_{i}\}$. Then, a necessary and sufficient condition
	for the error-correcting scheme correcting error on $C$ is:
	\begin{equation}
		P E_{i}^{\dagger}E_{j}P = \alpha_{ij}P,
	\end{equation}
	for some Hermitian matrix $\alpha_{ij}$. 
	A quantum error-correcting code is a pair (C, R) consisting of a quantum code and a recovery operator. 
	We will see one example (the Shor's nine-qubit code) for error correction toward the end of this section. 
	In order to implement QEC, one would like to know which gate set are sufficient. The answer to this is given by the Rains-Solovay theorem.\footnote{This theorem is not often referred to by this name, we call it this following the discussion at: \href{https://cstheory.stackexchange.com/questions/11308/universal-sets-of-gates-for-su3}{https://cstheory.stackexchange.com/questions/11308/universal-sets-of-gates-for-su3}} It states that: The Clifford group (denoted below by $C$) together with any gate not in this group is universal for quantum computation. However, there are some combinations that are most frequently used and proven to be universal. Such examples are $\{C + \text{CCNOT}\}$, $\{C + \pi/8\}$, and $\{C + \text{controlled}-\pi/4\}$.

	In 1995, Shor \cite{PhysRevA.52.R2493} discussed a way to reduce the rate of decoherence in quantum memory
	and hence paved the way for the practical usage of quantum computers,
	This work led to a lot of interesting developments in the field of `quantum error correction'. 
	This was a remarkable result since it implied that a quantum state 
	even though it cannot be cloned due to the `no-cloning' theorem can be 
	protected in a noisy environment. One would at this point think - 
	why do we need to correct errors at all? The error correction in quantum
	computers is more delicate than classical errors. This is partly because error correction
	is needed to protect the quantum superposition and entanglement, both of which are
	crucial ingredients of any quantum computation. In the original work, 
	Shor proposed to store the quantum information not in a single qubit, but
	it was later shown that it can be reduced to five, which is the minimum required
	\cite{Laflamme:1996iw}. 

	\subsection{Bit-flip error} 
	
	Bit flip is defined when we have $\ket{0} \to \ket{1}$ or vice versa. Suppose we are interested in sending out a single qubit $\ket{\psi} = \alpha \ket{0} + \beta \ket{1}$
	through during which a flip i.e., $X \ket{\psi} = \beta \ket{0} + \alpha \ket{1}$ can occur with 
	some probability $p$. Such a process is referred to as `bit-flip'. One way to correct this 
	is to do repetition as follows of the classical bit $\ket{0}_{L} \to \ket{000}$, where we mean logical qubit by subscript $L$. 
	Assuming that the probability of 
	bit-flip as $p$, we can show that the total probability of error persisting (i.e., remaining uncorrected or
	wrongly corrected) is $3p^{2}(1-p) + p^3 = 3p^{2} - 2p^{3}$. 
	\begin{mybox}
		\textsc{$\blacktriangleright$ Question 18:}  Compute the minimum fidelity attained by the three-qubit bit flip? \newline 	
	\end{mybox}
	The circuit to correct bit-flip is given by:
	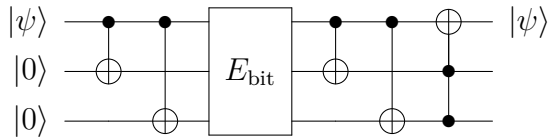
\begin{figure}[h]
		\centering
		\leavevmode
		\large{
			\Qcircuit @C=1em @R=.7em {
				\lstick{\ket{\psi}} & \ctrl{1} & \ctrl{2} & \multigate{2}{E_{\text{bit}}} & \ctrl{1} & \ctrl{2} & \targ     & \rstick{\ket{\psi}} \qw\\
				\lstick{\ket{0}}    & \targ    & \qw      & \ghost{E_{\text{bit}}}        & \targ    & \qw      & \ctrl{-1} & \qw \\
				\lstick{\ket{0}}    & \qw      & \targ    & \ghost{E_{\text{bit}}}        & \qw      & \targ    & \ctrl{-2} & \qw
		}}
		\caption{Bit flip circuit}
		\label{fig:BP1}
	\end{figure}
	For the example, we will take $ E_{\text{bit}}$ to be a $X$ gate for now. 
	We have $Z = \ket{0} \bra{0} - \ket{1} \bra{1}$
	\begin{align}
		Z_1 Z_2 \mathbb{1} = \Big( \ket{00} \bra{00} + \ket{11} \bra{11} \Big) \otimes \mathbb{1} - 
		\Big( \ket{01} \bra{01} + \ket{10} \bra{10} \Big) \otimes \mathbb{1}
	\end{align}
	If both first and second qubit are same, we get +1 or -1 if they are different.  Similarly, we can 
	compute $Z_2 Z_3$. Then assuming that there is a single bit flip, we can deduce the following table. 
	\begin{center}
		\begin{tabular}{ |c|c|c|c| } 
			\hline
			$Z_1 Z_2$ & $Z_2 Z_3$ & Which bit flipped? \\
			\hline
			+1 & +1 & None ($\mathbb{1}$)  \\  \hline 
			+1 & -1 & Third  ($X_3$) \\  \hline 
			-1 & +1 & First  ($X_1$)  \\  \hline 
			-1 & -1 & Second ($X_2$)  \\
			\hline
		\end{tabular}
	\end{center}
	
		\begin{figure}[h]
		\[
		\Qcircuit @C=.3em @R=.3em @! {
			\lstick{\ket{1}} & \ctrl{3} & \qw & \qw & \qw & \qw & \qw \\
			\lstick{\ket{2}} & \qw & \ctrl{2} & \ctrl{3} & \qw & \qw & \qw \\
			\lstick{\ket{3}} & \qw & \qw & \qw & \ctrl{2} & \qw  & \qw   \\ 
			\lstick{\ket{\rm{AC}}} & \targ & \targ & \qw & \qw  &  \meter \\ 
			\lstick{\ket{\rm{AC}}}& \qw & \qw & \targ & \targ  &  \meter
		}
		\]
		\caption{A circuit to identify the bit flip with two ancillary bits. This is the error-detection part of the
			bit flip code. This is followed by a recovery procedure.}
		\label{fig:gates1}
	\end{figure}
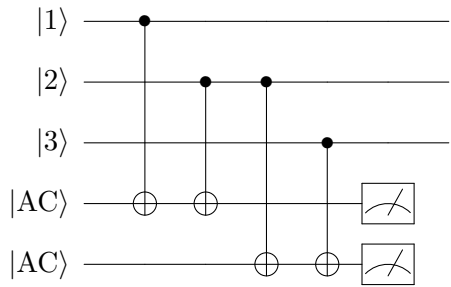
	This step detected the error. Now, in the recovery step, one simply applies $X$ gate on the flipped bit 
	to complete the error correction process. One important thing is that these measurements ($ZZ$) do not spoil the superposition of quantum states that we wish to preserve since they never give any information about $a, b$ (amplitudes). The bit flip code is similar in spirit to the classical error-correcting codes. One more advanced code is the `phase-flip' code which has no classical counterpart and hence is 
	more interesting as a quantum code. But note that there are ways to describe `phase-flip' into `bit-flip'. 
	We can always go from `quantum' to `classical' but not the other way around!

	\subsection{Phase-flip error}
	
	We now discuss another error that can occur. Instead of the operator $X$ which was applied to the bit flip case, 
	consider now the action of $Z$ operator. Then the state $a \ket{0} + b \ket{1}$ is transformed to $a \ket{0} - b \ket{1}$. Now suppose we rather work in the basis $\ket{\pm}$, then $ Z \ket{+} \to 	\ket{-}$. Hence, phase flip is like bit flip in a different basis! This presents itself that all the basic parts of the error correction i.e., encoding, error-detection, and recovery can be applied as bit flip code. The encoding circuit for bit flip will then be:
	\begin{figure}[h]
		\[
		\Qcircuit @C=.3em @R=.3em @! {
			\lstick{\kp} & \ctrl{1} & \ctrl{2} & \gate{H} &  \qw \\
			\lstick{\ket{0}} & \targ & \qw &  \gate{H} &  \qw \\
			\lstick{\ket{0}} & \qw & \targ &  \gate{H} & \qw  
		} 
		\]
		\caption{Circuit to encode the phase flip code. }
		\label{fig:FP1}
	\end{figure}
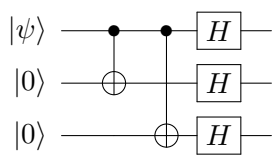
	
	\subsection{Shor's nine qubit code} 
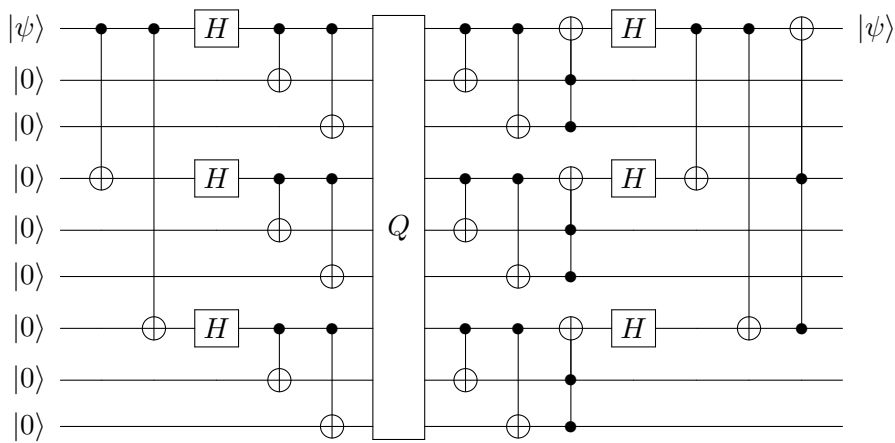
\begin{figure}[h]
		\[ 
		\Qcircuit @C=1em @R=.7em {
			\lstick{|\psi\rangle} & \ctrl{3} & \ctrl{6} & \gate{H} & \ctrl{1} & \ctrl{2} & \multigate{8}{Q} & \ctrl{1} & \ctrl{2} & \targ     & \gate{H} & \ctrl{3} & \ctrl{6} & \targ     & \rstick{\ket{\psi}} \qw\\
			\lstick{\ket{0}}      & \qw      & \qw      & \qw      & \targ    & \qw      & \ghost{Q}        & \targ    & \qw      & \ctrl{-1} & \qw      & \qw      & \qw      & \qw       & \qw \\
			\lstick{\ket{0}}      & \qw      & \qw      & \qw      & \qw      & \targ    & \ghost{Q}        & \qw      & \targ    & \ctrl{-2} & \qw      & \qw      & \qw      & \qw       & \qw \\
			\lstick{\ket{0}}      & \targ    & \qw      & \gate{H} & \ctrl{1} & \ctrl{2} & \ghost{Q}        & \ctrl{1} & \ctrl{2} & \targ     & \gate{H} & \targ    & \qw      & \ctrl{-3} & \qw \\
			\lstick{\ket{0}}      & \qw      & \qw      & \qw      & \targ    & \qw      & \ghost{Q}        & \targ    & \qw      & \ctrl{-1} & \qw      & \qw      & \qw      & \qw       & \qw \\
			\lstick{\ket{0}}      & \qw      & \qw      & \qw      & \qw      & \targ    & \ghost{Q}        & \qw      & \targ    & \ctrl{-2} & \qw      & \qw      & \qw      & \qw       & \qw \\
			\lstick{\ket{0}}      & \qw      & \targ    & \gate{H} & \ctrl{1} & \ctrl{2} & \ghost{Q}        & \ctrl{1} & \ctrl{2} & \targ     & \gate{H} & \qw      & \targ    & \ctrl{-6} & \qw \\
			\lstick{\ket{0}}      & \qw      & \qw      & \qw      & \targ    & \qw      & \ghost{Q}        & \targ    & \qw      & \ctrl{-1} & \qw      & \qw      & \qw      & \qw       & \qw \\
			\lstick{\ket{0}}      & \qw      & \qw      & \qw      & \qw      & \targ    & \ghost{Q}        & \qw      & \targ    & \ctrl{-2} & \qw      & \qw      & \qw      & \qw       & \qw
		}
		\] 
		\caption{The circuit to implement Shor's code where $Q$ is a channel which can corrupt single qubit.} 
		\label{fig:shor_9_1}
	\end{figure}

	Though it is nice to correct the bit and phase flip separately, one desires a code to correct any arbitrary error using a single code. The ability to correct arbitrary errors on a \emph{single} qubit can be achieved by Shor's nine-qubit code and is shown in Fig.~\ref{fig:shor_9_1}. 
	The basic step of Shor's construction is to note that:
	\begin{align}
	\ket{0} & = \ket{0}_{L} = \ket{+++}, \\ 
	\ket{1} & = \ket{1}_{L} = \ket{---}, 
	\end{align}
	will control the phase errors, while, 
	\begin{align}
	\ket{+} &= \frac{1}{\sqrt{2}} \Big(\ket{000} + \ket{111} \Big), \\
	\ket{-} &= \frac{1}{\sqrt{2}} \Big(\ket{000} - \ket{111} \Big),
	\end{align}
	will correct against the bit flip error. Combining these two, we can write:
	
	\begin{equation}
	\ket{0} \to \frac{1}{2\sqrt{2}} \Big( \ket{000} + \ket{111} \Big) \otimes \Big( \ket{000} + \ket{111} \Big) \otimes \Big( \ket{000} + \ket{111} \Big). 
	\end{equation} 
	Suppose the error is modelled by a unitary transformation $U$ such that:
	\begin{equation}
	U  = c_0 \mathbb{I} + c_1 X + c_2 Y + c_3 Z
	\end{equation}
	Now, if U is equal to $\mathbb{I}$, then no error has occurred. If $U = X$, then a bit flip occurs, if $U = Z$ then a phase flip occurs. Lastly, 
	if $U = iY$ then both a bit flip error and a sign flip error occur. 
	\begin{mybox}
		\textsc{$\blacktriangleright$ Question 19:}  Show that if the probability of a single qubit being affected (either phase or bit flip or mixture of the two) is $p$, then the
		probability of having more than one qubit having an error in Shor's nine qubit code is $36p^2$.  \newline 	
	\end{mybox}
	The quantum error-correcting conditions are a set of equations which have to be satisfied by a quantum error-correcting code to protect against a particular
	type of noise. 
	In the next paragraph, we will see what the stabilizer set is for Shor's code. But before this, we need to define the definition of a stabilizer of a state. 
	We say that a unitary, $U$, stabilizes a pure state $\kp$ if $ U \kp = \kp$. Note that the phase is important and if we have 
	$U \kp = -\kp$, then it does not stabilizes.

	\begin{mybox}
		\textsc{$\blacktriangleright$ Question 20:}  Show that $\{II, XX, -YY, ZZ\}$ forms a stabilizer group for the Bell state given by $(\ket{00} + \ket{11})/\sqrt{2}$.
			\end{mybox}

	Now, we do a short example that will clarify this further. Suppose we are asked whether the three-qubit phase flip code $\ket{0_L} = \ket{+++}$, $	\ket{1_L} = \ket{---}$.
	satisfied the conditions for the set of error operators
	$\{I, Z_{1}, Z_{2}, Z_3\}$. 
	The stabilizer set of Shor's code is\footnote{Note that $Z_{1}Z_{2} = ZZIIIIIII$ ~or~ $Z \otimes Z \otimes I_{128}$}: \\
	\begin{equation} 
	\{Z_{1}Z_{2}, Z_{2}Z_{3}, Z_{4}Z_{5}, Z_{5}Z_{6}, Z_{7}Z_{8}, Z_{8}Z_{9}, X_{1}X_{2}X_{3}X_{4}X_{5}X_{6}, 
	X_{4}X_{5}X_{6}X_{7}X_{8}X_{9}\}
	\end{equation} 
	The Shor's code provides the window to the more complicated quantum error-correcting (QEC)
	codes. However, this code is not the most optimal and Steane 7-qubit code \cite{Steane:1995vv} 
	is better. This code has six stabilizer generators compared to eight in Shor's code. 
	These are given by:
	\begin{equation}
	\{IIIXXXX, IXXIIXX, XIXIXIX, IIIZZZZ, IZZIIZZ, ZIZIZIZ\}
	\end{equation}
	All these QEC codes are best explained by the idea of stabilizer codes. For example, one famous example is the toric code, which is a topological quantum error correcting code and is an example of a stabilizer code/circuit \footnote{A stabilizer circuit is a quantum circuit in which every gate is 
	selected from a stabilizer set i.e.,$\{H, \text{CNOT}, P\}$ or a 1-qubit measurement gate}. It is defined on a two-dimensional spin-lattice and has $\mathbb{Z}_2$ topological order. It reduces to a $\mathbb{Z}_2$ lattice gauge theory well known for decades in an appropriate limit. In this code, the stabilizer operators are defined on the spins around each vertex and on the plaquette and are known as $A$ and $B$ operators defined as: $A_v = \prod_{i \in v} \sigma_{i}^{x}$ and 
	$B_p = \prod_{i \in p} \sigma_{i}^{z}$ respectively. We often refer to a QEC code as $[[n, k, 2t+1]]$ code \footnote{The two braces denote that is a quantum code
	unlike classical code such as Hamming code which uses single parentheses} and Shor's code is an example of
	[[9,1,3]] stabilizer code. Sometimes, $2t+1$ is also called `distance', and code is referred to as $[[n, k, d]]$ code. 
	Soon after nine-qubit code, Steane found a seven-qubit code that can correct the same error as Shor's code. 
	In fact, both these codes are `degenerate' which loosely means that it can be further improved. One of the ways we can optimize the code is to introduce the quantum Hamming bound
	\cite{2009arXiv0904.2557G}. If a given code uses $n$ qubits, corrects error on up to $t$ qubits and encodes $k$ qubits, then
	the following inequality holds:
	\begin{equation}
	\Bigg( \sum_{j=0}^{t}3^{j} {n \choose j}  \Bigg) 2^{k} \le 2^n.  
	\end{equation}
	It is easy to check that with $k=1, t=1$, the lower bound is saturated with $n=5$ hence 
	there exists a 5-qubit code which is [[5,1,3]] which is nondegenerate. In general, it is 
	clear that the stabilizer circuits/codes are very important for performing quantum error-correction
	and play an important role in fault-tolerant circuits. 	The idea of stabilizer circuits 
	is also important for a theorem known as `Gottesman-Knill' (GK) theorem~\cite{Gottesman:1998hu}. The theorem implies that any stabilizer circuits can be perfectly simulated in polynomial time 
	on a (probabilistic) classical computer. Hence, they offer no quantum advantage by themselves. 
	This theorem originated in Ref.~\cite{Gottesman:1998hu} where author refers to private communication with Knill. This result does not explain the reason why quantum computing can provide exponential speedup over classical computing, but it does tell us that quantum algorithms which make use of entanglement built using Hadamard and CNOT gates do not 
	offer any advantage. This result is contrary to the natural expectation that no quantum gates can be simulated efficiently by classical computers since $n$-qubit quantum circuit operates in a $2^n$-dimensional Hilbert space. 
	But admittedly the situation if very different for non-Clifford gates which cannot be simulated efficiently. 
	If it was shown someday that even non-Clifford gates can be simulated by classical computers, then it would imply
	a further (and significant) lack of usefulness of quantum computers.
	
	However, even after different codes to perform quantum error correction, this has 
	been not possible to do in practice.  One of the issues is that encoding of a single logical 
	qubit in $N$ physical qubits also raises the possibility of error occurring by that factor. 
	The apparatus which fixes the single-qubit error must therefore be reasonably fast 
	to not incur other errors in one of those $N$ qubits to reach a reasonable trade-off. 
	This is not an easy experimental problem and hence there have been various other 
	proposals to perform error corrections. One of the main alternatives is the use 
	of `continuous variables' degrees of freedom (quantum harmonic oscillator). 
	In such a setting, the logical qubits will be photon (Fock) states that span a
	two-dimensional subspace from the potentially infinite harmonic oscillator states. 
	Though, we have discussed the qubit-based quantum computing in this article, the
	quantum computation and information based on continuous variables (CV) are also 
	extensively used in quantum teleportation, error correction, and cryptography. 
	For the CV approach to quantum computation, the main role is played by 
	Bose operators $b_i^{\dagger}, b_i$ known as creation and annihilation operators
	satisfying following commutation relations (also known as `Heisenberg-Weyl algebra): 
	\begin{align}
	\label{eq:BCR} 
	& [b_i^\dagger, b_j^\dagger] = 0, ~~~ [b_i, b_j] = 0, ~~~ [b_i, b^\dagger_j] = \delta_{ij} \mathbb{1} . 
	\end{align}
	The inner product space must be of infinite dimension for (\ref{eq:BCR}) to hold. One defines
	the vacuum state of $N$ qumodes (similar to $N$ qubits) as $\textbf{\ket{0}} = \ket{000\cdots0}$. 
	The normalized number state (or Fock state) can be written as:
	
	\begin{equation}
	\frac{1}{\sqrt{n_{1}! n_{2}! \cdots n_{k}!}} (b_{j_1}^{\dagger})^{n_1}  (b_{j_2}^{\dagger})^{n_2} \cdots (b_{j_k}^{\dagger})^{n_k} \textbf{\ket{0}} = \ket{n_1, n_2, \cdots, n_k}.  
	\end{equation}
	We refer the reader to Appendix \ref{sec:QCCV} and 
	excellent review \cite{Braunstein:2005zz} for more technical details and to start
	the reference trail. 
	
	\section{\label{sec:last}Looking at NISQ and beyond}
	
	One of the motivations for building the quantum computer, at least for physicists, is to solve the time evolution of some complicated quantum mechanical system. This was also the motivation given by Feynman around 1982. However, it was 
	only in 1996 that an explicit quantum-inspired algorithm for simulating the time
	evolution of a quantum system with \emph{local} Hamiltonian was given for the first time \cite{Seth1996}. The algorithm made use of the Lie-Trotter formula to express the unitary time evolution as a sequence of exponentials and had a complexity of $\mathcal{O}(t \vert \vert H \vert \vert)^{2}$
	where $t$ is the time and $\vert \vert H \vert \vert$ is the spectral norm (also called two-norm)
	of the Hamiltonian. 
		
	However, it is clear that most of the physically interesting quantum models cannot be simulated on NISQ devices with just a few hundred/thousand noisy qubits. However, there is still a lot to understand and this is an active area of research. It is already evident that even for non-relativistic quantum systems the size of Hilbert 
	space is very large and adding the relativistic behavior when studying quantum field theories makes it worse. One has to understand the reduction of the degrees of freedom properly while keeping the 
	symmetries (such as `gauge symmetry') intact. There have been various proposals such as $D$-theory/quantum link models \cite{Chandrasekharan:1996ih}, dual variable using character expansions with sum over irreducible representations of the gauge group \cite{Bazavov:2019qih} which are then truncated, quantum groups such as $SU(2)_{q}$ approach, loop-string-hadron formalism \cite{Raychowdhury:2019iki}, light front dynamics method, and truncating the continuous symmetry groups by 
	discrete point symmetry such as in Ref.~\cite{Alexandru:2021jpm}. All these methods are based on finding smaller subspace of
	the vector space to be represented by qubits. There is also a method based on continuous variables (CVs) which plays an important role in analog quantum computing and has several advantages over the discrete methods. These schemes have their merits and drawbacks and the effectiveness in obtaining the continuum limit is still debatable. Some qubit regularization schemes	
	also exits for $O(3)$ non-linear sigma model by exploiting equivalence to $SU(2)$ and associated truncation 
	over irreducible representations but it is by no means a general method and works only for this specific group. 
	We do not know of any work which attempts $O(N)$ models with $N \ge 4$. Even for scalar field theories such
	as $\phi^4$ theory in 1+1-dimensions, the quantum simulation and extracting of real-time properties and correlators 
	are still not fully understood. There are approaches based on field variables following JLP	\cite{Jordan:2011ci}, harmonic oscillator basis \cite{Klco:2018zqz}, continuous variable 
	approach \cite{Marshall:2015mna}, and single-particle basis \cite{Barata:2020jtq} to name a few. Here, we will simply focus on some simple examples where one desires time evolution using some Hamiltonian with circuit depths 
	\footnote{The depth of a circuit is the longest path between the data input and the output, where each 
		intermediate gate counts as one. The circuit depth can be easily found in \QIS~using 
		$\textcolor{BlueNCS}{\texttt{qc.depth()}}$
		where $\texttt{qc}$ is the quantum circuit.} 
	that may be performed on the NISQ devices. 
	
	We can perceive any quantum circuit as a simulation of some Hamiltonian $H$. This follows from the standard observation that any (noiseless) 
	the quantum circuit can be built up of pieces of unitary transformations whose product is also unitary. 
	Once we have a given unitary transformation, we can find 
	$H$ such that $U = e^{-iHt}$ where $t$ is some real parameter. 
	The time evolution of any given state can be defined by this $U$. 
		
	In addition to the time evolution of a given quantum state with some $H$, this unitary evolution also
	plays an important role in the HHL algorithm \cite{PhysRevLett.103.150502, 2021arXiv210809004M} which is a quantum algorithm to solve a linear system of equations i.e., $A \vec{x} = \vec{b}$ where $A$ is a Hermitian $n \times n$ matrix which is $d$-sparse. 
	HHL algorithm started a mini-revolution in various areas of research including quantum machine learning (QML). The reason was 
	in the fact that it addresses one of the most important problems of linear algebra - solving linear systems. HHL is expected to 
	have time complexity $\mathcal{O}(\log(n)d^{2}\kappa^{2}/\epsilon)$ and exponential speedup over the best known classical algorithm, conjugate gradient \cite{Hestenes1952}, with complexity $\mathcal{O}(nd \sqrt{\kappa} \log(1/\epsilon))$. 
	
	However, there are some conditions that must be satisfied for this to hold. These conditions related to loading $\vec{b}$ quickly, applying 
	$e^{-iAt}$ with $A$ being $d$-sparse \footnote{A matrix is $d$-sparse if it contains at most $d$ nonzero entries per row/column for some $d \ll n$
	}, well-behaved condition number $\kappa$ of matrix $A$, and the lack of obtaining all elements of $\vec{x}$ \cite{Aaronson2015}. 
	It is clear that these might be difficult to satisfy all at once in practice and hence the speedup might be lost. And even if we 
	satisfy all these conditions, we do not have full access to the $\vec{b}$ like the classical case but only to $\ket{A^{-1} b}$. 
	With the state in hand, we can make measurements to know the features of the output state. It might have deep implications 
	in the future, but, in practice, HHL does not appear to be of much use in the NISQ era. Also, the HHL algorithm 
	complexity severely depends (more than the best classical) on condition number 
	and sparsity. 
	
	\subsection{Hamiltonian Simulation} 

	We now return to the main question of interest -- how to implement $U = e^{-iHt}$ given some hermitian matrix $H$ of size $N \times N$ which is $d$-sparse on a quantum initial state $\kp$ i.e., how to do: $U \kp \to \ket{\psi(t)}$. We say that we can 
	efficiently simulate any $N$-qubit Hamiltonian $H$ if we can approximate $U$ for any $t$ in $\text{poly}(N)$ 
	using at most polynomial number of one- and two-qubit gates. It is widely believed that wide range of 
	Hamiltonians cannot be simulated efficiently. However, if we are given that we can 
	simulate some smaller Hamiltonians $H_{1}, \cdots, H_{k}$, then their sum $H$ can also be simulated
	using Lie-product formula which we discuss later in the section. In addition, 
	if we can simulate $H_{1}$ and $H_2$, then we can also simulate 
	$[H_1, H_2]$ using the identity:
	
	\begin{equation}
	e^{-i[H_1, H_2]t} = \lim_{{N\to\infty}} \left( e^{-iH_{1}t/\sqrt{N}} e^{-iH_{2}t/\sqrt{N}}
	e^{iH_{1}t/\sqrt{N}} e^{iH_{2}t/\sqrt{N}} \right)^N.
	\end{equation}
	Another important result is the unitary conjugation i.e., if we can 
	simulate $H$ and can perform (efficiently) the unitary operation $U$, 
	then we can simulate $UHU^{\dagger}$. This is true because of the identity
	in (\ref{eq:UC1}).


	Let us consider a simple two-site model with $H = \sigma_z \otimes \sigma_z = Z_1 \otimes Z_2$. We could express this term: $\exp(i \sigma_z 			\otimes \sigma_z t)$ 
	as $e^{i \sigma_z \otimes \sigma_z t} = \cos(t) \mathbb{1} + i \sin(t) \sigma_z \otimes \sigma_z$\footnote{Note that we have
		$\exp(\pm i \theta U) = \cos(\theta) \mathbb{1} \pm i \sin(\theta)U$ if $U^{2} = \mathbb{1}$.} which can be written by 
		some manipulation (show this) as $\mathrm{CNOT} \left(\mathbb{1} \otimes e^{i \sigma_z  t}\right) \mathrm{CNOT}$ since $\mathrm{CNOT} = 
	\ket{0} \bra{0} \otimes \mathbb{1} + \ket{1} \bra{1} \otimes \sigma_{x}$. The circuit to implement \cite{Whitfield_2011} this term is given below: 
	
	\begin{equation}
	\label{eq:ZZcir} 
		\Qcircuit @C=1.3em @R=1.1em {
			& \ctrl{1} & \qw & \ctrl{1} & \qw \\
			& \targ & \gate{R_z} & \targ & \qw 
		}
	\end{equation}
	where the argument of $R_{z}$ gate is $-2t$. This is sometimes also written 
	as: $\exp(it Z_j \otimes Z_k) = \rm{CNOT}_{jk} (\mathbb{I}_{j} R_z(-2t)_{k}) \otimes \rm{CNOT}_{jk}$
	with $j,k$ denoting the qubits on which the gates act. 
	Similarly, we can construct a circuit for the three-qubit term, 
	$\sigma_z \otimes \sigma_z \otimes \sigma_z$ as:

	\begin{equation}
		\Qcircuit @C=1.1em @R=1.1em {
			& \ctrl{1} & \qw      & \qw & \qw & \ctrl{1} & \qw\\
			& \targ    & \ctrl{1}  & \qw & \ctrl{1} & \targ  & \qw\\ 
			& \qw      & \targ    & \gate{R_z} & \targ & \qw & \qw
		}
	\end{equation}
	Sometimes, for brevity, we refer to the Pauli matrices just by $X,Y, Z$ respectively. Now, suppose we have different Pauli matrices
	term in the Hamiltonian as: $ \Big(X \otimes X  + Y \otimes Y\Big) $. The corresponding circuit can be used to perform time evolution 
	i.e., it is equivalent to $\exp[i (\alpha/2)(X \otimes X  + Y \otimes Y)]$
	
	\begin{equation}
	\label{eq:XXYYshort} 
		\Qcircuit @C=1.1em @R=1.1em {
			& \ctrl{1} & \gate{H}    & \ctrl{1} & \gate{R_z(-\alpha)} & \ctrl{1} & \gate{H} & \ctrl{1} & \qw \\
			& \targ    & \qw & \targ  & \gate{R_z(\alpha)} & \targ  & \qw & \targ & \qw 
		}
	\end{equation}
	It is easy to check this using \MAT~code below which has been broken down for simplicity. 
	
	\begin{mdframed}[backgroundcolor=celadon!6] 
	\begin{lstlisting}[language=Mathematica]
X = PauliMatrix[1];
Y = PauliMatrix[2];
Z = PauliMatrix[3];
XXplusYY = MatrixExp[I (KroneckerProduct[X, X] + KroneckerProduct[Y, Y]) \[Alpha]/2];
XtimesX = MatrixExp[I (KroneckerProduct[X, X]) \[Alpha]/2];
ZtimesZ = MatrixExp[I (KroneckerProduct[Z, Z]) \[Alpha]/2];
YtimesY = MatrixExp[I (KroneckerProduct[Y, Y]) \[Alpha]/2];
XtimesY = MatrixExp[I (KroneckerProduct[X, Y]) \[Alpha]/2];
CNOT = {{1, 0, 0, 0}, {0, 1, 0, 0}, {0, 0, 0, 1}, {0, 0, 1, 0}};
HAD = {{1/\[Sqrt]2, 1/\[Sqrt]2}, {1/\[Sqrt]2, -(1/\[Sqrt]2)}};
S = {{1, 0}, {0, I}};
S1 = KroneckerProduct[S, S];
HAD1 = KroneckerProduct[HAD, IdentityMatrix[2]];
HAD2 = KroneckerProduct[HAD, HAD];
RZ1 = KroneckerProduct[IdentityMatrix[2], MatrixExp[I (\[Alpha]/2) Z]];
RZ2 = KroneckerProduct[MatrixExp[I (\[Alpha]/2) Z], MatrixExp[I (-\[Alpha]/2) Z]];

(* Check! *) 

XXYY = CNOT . HAD1 . CNOT . RZ2 . CNOT . HAD1 . CNOT // FullSimplify; 
ZZ = CNOT . RZ1 . CNOT // FullSimplify;
XX = HAD2 . ZZ . HAD2 // FullSimplify; 
YY = ConjugateTranspose[S1] . XX . S1 // FullSimplify;
XXplusYY == XXYY 
ZtimesZ == ZZ
XtimesX == XX
YtimesY == YY
HY = {{1/\[Sqrt]2, -I/\[Sqrt]2}, {I/\[Sqrt]2, -(1/\[Sqrt]2)}};
HY . PauliMatrix[3] . HY  == PauliMatrix[2] ;  (* True *) 
HAD . PauliMatrix[3] . HAD  == PauliMatrix[1]; (* True *) 
YYcir = KroneckerProduct[HY, HY] . ZZ . KroneckerProduct[HY, HY] // FullSimplify ; 
YtimesY == YYcir; 
XYcir = KroneckerProduct[HAD, HY] . ZZ . KroneckerProduct[HAD, HY] // FullSimplify ; 
XtimesY = XYcir; 

\end{lstlisting}
\end{mdframed}


It is also useful to note that $e^{it I \otimes \mathcal{P} \otimes I} = I \otimes e^{it \mathcal{P}} \otimes I
$, where $\mathcal{P}$ is single Pauli matrix or a string because $
e^{it I \otimes \mathcal{P} \otimes I} = \cos(t) I \otimes I \otimes I + i\sin(t) I \otimes \mathcal{P} \otimes I = 
 I \otimes \big( \cos(t) I + i \sin(t) \mathcal{P}  \big) \otimes I = I \otimes e^{i \mathcal{P} t} \otimes I
$. Sometimes, it is possible that $H$ has commuting terms. For example, consider 
$H = X \otimes Y + Z \otimes Z$. It is straightforward to see that the terms commute
and therefore $\exp(-i Ht) = \exp(-it X \otimes Y) \exp(-it Z \otimes Z)$. The circuit for 
the first term is:
\begin{equation}
		\Qcircuit @C=1.1em @R=1.1em {
			& \gate{H_{y}} & \ctrl{1}    & \qw & \ctrl{1} & \gate{H_{y}} \\
			& \gate{H}    & \targ & \gate{R_z(-2t)} & \targ & \gate{H}
		}
\end{equation}
where $H_{y} = \frac{1}{\sqrt{2}} 
\begin{pmatrix}
1 & -i  \nonumber \\
i & - 1
\end{pmatrix}$ 
while the second term is given by (\ref{eq:ZZcir}). Combining, we get for one step, 
\begin{equation}
		\Qcircuit @C=1.1em @R=1.1em {
			& \ctrl{1}  & \qw & \ctrl{1}  & \gate{H_{y}} & \ctrl{1}    & \qw & \ctrl{1} & \gate{H_{y}} & \qw \\
			& \targ  & \gate{R_z(-2t)} & \targ & \gate{H}   & \targ & \gate{R_z(-2t)} & \targ & \gate{H} & \qw 
		}
\end{equation}
consisting of four two-qubit gates and six one-qubit gates. But, this might not always be efficient. 
Consider the circuit in (\ref{eq:XXYYshort}) for $H = \Big(X \otimes X  + Y \otimes Y\Big)$. 
Since the terms commute, we can also use alternate circuit as given below in the $\MAT$ code, 
however it requires many more gates. 

\begin{mdframed}[backgroundcolor=celadon!6] 
	\begin{lstlisting}[language=Mathematica]
CT = KroneckerProduct;
X = PauliMatrix[1];
Y = PauliMatrix[2];
Z = PauliMatrix[3];
CNOT = {{1, 0, 0, 0}, {0, 1, 0, 0}, {0, 0, 0, 1}, {0, 0, 1, 0}};
HAD = {{1/\[Sqrt]2, 1/\[Sqrt]2}, {1/\[Sqrt]2, -(1/\[Sqrt]2)}};
HY = {{1/\[Sqrt]2, -I/\[Sqrt]2}, {I/\[Sqrt]2, -(1/\[Sqrt]2)}};  
XXYYv1 = CNOT . CT[HAD, IdentityMatrix[2]] . CNOT . CT[MatrixExp[I (\[Alpha]/2) Z], MatrixExp[I (-\[Alpha]/2) Z]] . CNOT . CT[HAD, IdentityMatrix[2]] . CNOT // FullSimplify; 
XXYYv2 =  (CT[HAD, HAD] . CNOT . CT[IdentityMatrix[2], MatrixExp[I (\[Alpha]/2) Z]] . CNOT . CT[HAD, HAD]) . (CT[HY, HY] . CNOT . CT[IdentityMatrix[2], MatrixExp[I (\[Alpha]/2) Z]] . CNOT . CT[HY, HY] ) // FullSimplify; 
XXYYv1 == XXYYv2 == MatrixExp[I (KroneckerProduct[X, X] + KroneckerProduct[Y, Y]) \[Alpha]/2]
\end{lstlisting}
\end{mdframed}

	\begin{mybox}
		\textsc{$\blacktriangleright$ Question 21:} Show that any Pauli string is 1-sparse. 
	\end{mybox}
	

	\begin{mybox}
		\textsc{$\blacktriangleright$ Question 22:}  Consider the three-qubit Hamiltonian given by: 
		\[H = \sigma_y \otimes  \sigma_y \otimes  \sigma_x  - \sigma_z \otimes  \sigma_z \otimes  \sigma_z  - \sigma_z  \otimes  \mathbb{1}
	\otimes  \mathbb{1}.\]
	Construct the quantum circuit for the unitary operator ($\exp(i H)$) corresponding to it. It might be useful to remember that
	$ X = H \cdot Z \cdot H$ and $Y = H_{y} \cdot Z \cdot H_{y} $ where  $H_{y} = \frac{1}{\sqrt{2}} \begin{pmatrix}
1 & -i  \nonumber \\
           i & - 1
\end{pmatrix}$.
	\end{mybox}
	
	However, the method we just described in terms of Pauli strings is highly inefficient
	and in fact for most cases, as worse as classical methods. We will now consider some 
	state-of-the-art Hamiltonian simulation techniques which are better suited for 
	quantum computers. These are based on sparse representation rather than local 
	representation in terms of Pauli operators. 

	We recall some useful definitions about norms 
	of vectors and operators since they will appear later in the section. 
	To measure the distance between two elements in an arbitrary vector space, we can use -
	Manhattan norm (also called `l1 norm'): $\vert \vert x \vert \vert_{1} = \sum_{i=1}^{n} \vert x_i \vert$, or 
	the Euclidean norm (also called `l2 norm'): $\vert \vert x \vert \vert_{2} = \Bigg(\sum_{i=1}^{n} x_i^{2}\Bigg)^{1/2}$ or
	the Supremum norm: $\vert \vert x \vert \vert = \text{max.}  \vert x_i \vert$ where $x = (x_1, \cdots, x_n) \in \mathbb{R}^{n}$. 
	For matrices (operators), let us denote by $\sigma(A)$, the vector whose elements are singular values of positive operator
	$\vert A \vert$ which is defined as: $(A^{\dagger} \cdot A)^{1/2}$. Then the Schatten $p$-norm (where $1 \le p < \infty$) is defined as:
	\begin{equation}
	\vert \vert A \vert \vert_{p}  = \Bigg(\sum_{i=1}^{N} \sigma_{i}^{p}\Bigg)^{1/p}. 
	\end{equation}
	This norm is unitary invariant. When $p=2$, it is called the Frobenius norm or the Hilbert–Schmidt norm. 
	We also note that Frobenius norm is $\vert \vert A \vert \vert_{F} = \sqrt{\mathrm{Tr} A^{\dagger}A} = \sum_{i,j} A_{ij}^{2}$. 
	And that $\vert \vert UA \vert \vert_{F}^{2}  = \vert \vert A \vert \vert_{F}^{2}$. We can use this and 
	the equivalent result $\vert \vert AV \vert \vert_{F}^{2}  = \vert \vert A \vert \vert_{F}^{2}$ for unitary matrices $U,V$
	to reduce it to the form of singular values. First we write SVD of $A$ as $USV^{T}$. Then we have 	
	 $\vert \vert A \vert \vert_{F} = \vert \vert USV^{T} \vert \vert_{F} = \vert \vert SV^{T} \vert \vert_{F} = 
	 \vert \vert S \vert \vert_{F} = \sqrt{\sum_{i=1}^{N} \sigma_{i}^{2}}$. The $p=\infty$ limit is called the spectral norm. In some cases, people also refer to the 
	spectral or operator norm (Schatten $\infty$-norm) by induced 2-norm which is square root of maximum eigenvalue of $(A^{\dagger} \cdot A)^{1/2}$.

	In general, there are two most-used ways of decomposing the unitary operator 
	(exponential of Hermitian matrix $H$): 
	1) Lie-product method, and 2) Taylor Series. We discuss them below. 

	\begin{itemize} 
	\item Lie-product decomposition approach: This is sometimes called divide and conquer method since the Hamiltonian is broken into small pieces
	as $H = \sum_{j=1}^{m} \alpha_{j} H_{j}$ which are then evolved. The unitary operator is given by: 
	\begin{equation}
	\exp(-iHt) =  \Bigg(\prod_{j=1}^{m} \exp(-iH_{j}t/N)\Bigg)^{N} + \mathcal{O}\Bigg( \sum_{j,k} \bigg\vert\bigg\vert [H_j, H_k] \bigg\vert\bigg\vert  t^{2}/N\Bigg)
	\end{equation} 
	So, if we have all the terms in decomposition of $H$ nearly-commuting, then the Trotter error is reduced greatly
	and vanishes if they commute. Therefore, one can improve on errors by making use of 
	various commutative relations. Based on the structure of $H$, some Hamiltonian simulations might be 
	very simple and some can be notoriously hard.  
	
	\item Taylor Series: This approach proceeds following the standard decomposition:
	\begin{equation}
	U(t) = e^{-iHt} = \sum_{k=0}^{\infty} \frac{(-itH)^{k}}{k!}. 
	\end{equation}
	However for all practical purposes, we have to truncate this infinite sum at some order
	$K$ as:
	
	\begin{align}
	\widehat{U}(t) &= \sum_{k=0}^{K} \frac{(-itH)^{k}}{k!} \nonumber \\
	&= \sum_{k=0}^{K} \sum_{l = 1}^{L} \frac{t^k}{k!} \prod \alpha_{l} (-i)^{k} \prod H_{l}  \label{eq:SSum} \\
	&= \sum_{i=0}^{N} \widetilde{\alpha_i} \widetilde{H_i}~. 
	\end{align}
	where $\tilde{H} = (-i)^k \prod H$ and coefficients $\tilde{\alpha} = t^k/k! \prod \alpha$. 
	The second sum in (\ref{eq:SSum}) comes from the 
	decomposition\footnote{Recall that any complex matrix can be written as sum of unitary matrices. In this case, 
	sum of $L$ unitaries}
	of $H$ in terms of linear combination of unitaries (LCU). 
	We have absorbed
	the negative imaginary number in the operator such that all 
	$\alpha{l}$ are positive. Similar to the product formula, we divide
	in $r$ segments, such that:
	\begin{equation}
	\label{eq:TrotterErrorLCU} 
	\bigg\vert\bigg\vert \widehat{U}(t/r) - U(t/r) \bigg\vert\bigg\vert \le e^{\alpha t/r}  \frac{(\alpha t/r)^{K+1}}{(K+1)!}, 
	\end{equation} 
	where $\alpha = \sum \alpha_{i}$. If we want this upper bounded error to be $\epsilon$, then
	we need to choose:
	\begin{equation}
	K = \mathcal{O} \Bigg(\frac{\log (\alpha t/\epsilon)}{\log \log (\alpha t/\epsilon)}\Bigg). 
	\end{equation}
	The proof of (\ref{eq:TrotterErrorLCU}) proceeds as follows: 
	\begin{proof} 
	\begin{align}
	\bigg\vert\bigg\vert \widehat{U}(t) - \sum_{k=0}^{\infty} \frac{(-itH)^{k}}{k!}  \bigg\vert\bigg\vert &= \bigg\vert\bigg\vert \sum_{k=K+1}^{\infty} \frac{(-itH)^{k}}{k!}  \bigg\vert\bigg\vert \nonumber \\
	& \le \sum_{K+1}^{\infty} \frac{(t \sum_{l} \alpha_{l})^{k}}{k!} \nonumber \\
	& \le \frac{(t \sum_{l} \alpha_{l})^{K+1}}{(K+1)!} \Bigg[\sum_{j=0}^{\infty} (t \sum_{l} \alpha_{l})^{j} /j! \Bigg] \nonumber \\
	& \le e^{\alpha t/r}  \frac{(\alpha t/r)^{K+1}}{(K+1)!}
	\end{align}
	\end{proof}
 	Note that in the literature, the LCU followed by oblivious amplitude 
 	amplification \cite{2013Berry} is referred to as 
	`truncated Taylor series' (TTS) algorithm \cite{2015BerryTTS}. 
	\end{itemize}

	
	\vspace{10mm}

	The efficiency of the quantum simulation depends strongly on how the terms in Hamiltonian behave and how many qubits they act 
	at a given time. In this regard, one defines the notion of $k$-local Hamiltonian. Suppose we have a Hamiltonian acting on $n$ qubits, i.e., the operator is a $2^n \times 2^n$ matrix, $H = \sum_{a} H_{a}$, it is $k$-local in the sense that each $H_{a}$ 
	acts non-trivially on at most $k$ qubits at a time. A $k$-local Hamiltonian can be read
	in polynomial time and is simpler to handle. 
	If we want to determine the ground state energy of such a Hamiltonian, then it is called the 
	k-local Hamiltonian problem. There is a very close relationship between complexity
	classes and $k$-local Hamiltonian problem. It was shown in several works that it is \textbf{QMA} complete for 
	all $k \ge 2$ \cite{Kempe2004}. A detailed discussion of this lies beyond the scope of this article. In Table, 
	we list all major Hamiltionian simulation algorithms and their complexity. However, note that no one algorithms
	fit all conditions. The choice is based on the nature of Hamiltonian (time-dependent or independent) and symmetries
	which can be exploited and how local the $H$ is in general. In general, it is known that if the size of $H$ is $2^n \times 2^n$, then 
	there exists a decomposition of $H$ $\sum_{j=1}^{m} H_{j}$ where each $H_{j}$ is 1-sparse, such that $m = \mathcal{O}(d^2)$ 
	and each query to any $H_{j}$ can be simulated\footnote{Note that $\log^* n$ implies the iterated algorithm. It is defined as:
\[ 
\log^* n =
\begin{cases}
0 & \text{if } n \le 1 \\
1 + \log^* (\log n) & \text{if } n > 1
\end{cases} \] 
queries to $H$ (which is very optimum!). } by making $\mathcal{O}(\log^* n)$.

\begin{mdframed}[backgroundcolor=blue!2]
\textsc{We denote the size of Hamiltonian by $N$, the evolution time by $t$, the sparsity by $d$ and the maximum allowed error by $\epsilon$ and
$1/2 > \delta > 0$.}  \\ 
\vspace{10mm} 
\begin{tabular}{@{}p{10cm}@{} @{}p{8cm}@{}}
\noindent\rule{15cm}{0.4pt} \\ 
\textbf{Algorithm} & \textbf{Complexity}  \\ \\ 
Lie-trotter (only $k$-local $H$) [1996] \cite{Seth1996} & $\mathcal{O}(\text{poly}(\log N) (\vert\vert H \vert \vert t)^{2}/\epsilon$  \\[.3cm]
Lie-trotter (simulatable $H$) [2007] \cite{Berry2006} & $\mathcal{O}(d^4 (\log^* N) \vert\vert H \vert \vert t)^{1+\delta}/\epsilon^{\delta}$ \\[.3cm]
Lie-trotter (simulatable $H$) [2011] \cite{Childs2011} & $\mathcal{O}(d^3 (\log^* N) \vert\vert H \vert \vert t)^{1+\delta}/\epsilon^{\delta}$ \\[.3cm]
Truncated Taylor Series with Trotter [2014] \cite{2015BerryTTS} & $\mathcal{O}(d^2 \vert\vert H \vert \vert t) \frac{\log(d^2 \vert\vert H \vert \vert t/\epsilon)}{\log\log(d^2 \vert\vert H \vert \vert t/\epsilon)}$  \\[.3cm]
LCU+OAA with quantum walks [2015] \cite{2015arXiv150101715B} & $\mathcal{O}(d \vert\vert H \vert \vert t) \frac{\log(d \vert\vert H \vert \vert t/\epsilon)}{\log\log(d \vert\vert H \vert \vert t/\epsilon)}$  \\[.3cm]
Qubitization/Quantum Walks \cite{2016Low}  & $\mathcal{O}(d \vert\vert H \vert \vert t + \log(1/\epsilon)) $ \\[.3cm]
\end{tabular}
 \end{mdframed}
 
One thing to note is that the Hamiltonian simulation algorithms based on Lie-Trotter 
product formulas do not admit error dependence that is $\text{polylog}(1/\epsilon)$ whereas 
starting with Ref.~\cite{2013arXiv1312.1414B, 2015BerryTTS}. However, for most applications in 
Physics, it seems like the use of Trotter method is preferred since for models with 
local interactions, the complexity of using non-Trotter is not worth the gain in errors. 
We are not aware of a systematic study of a lattice gauge theory model comparing different 
time evolution methods.
 
	
\section{Future}

\begin{figure}
\centering 
\includegraphics[width=0.65\textwidth]{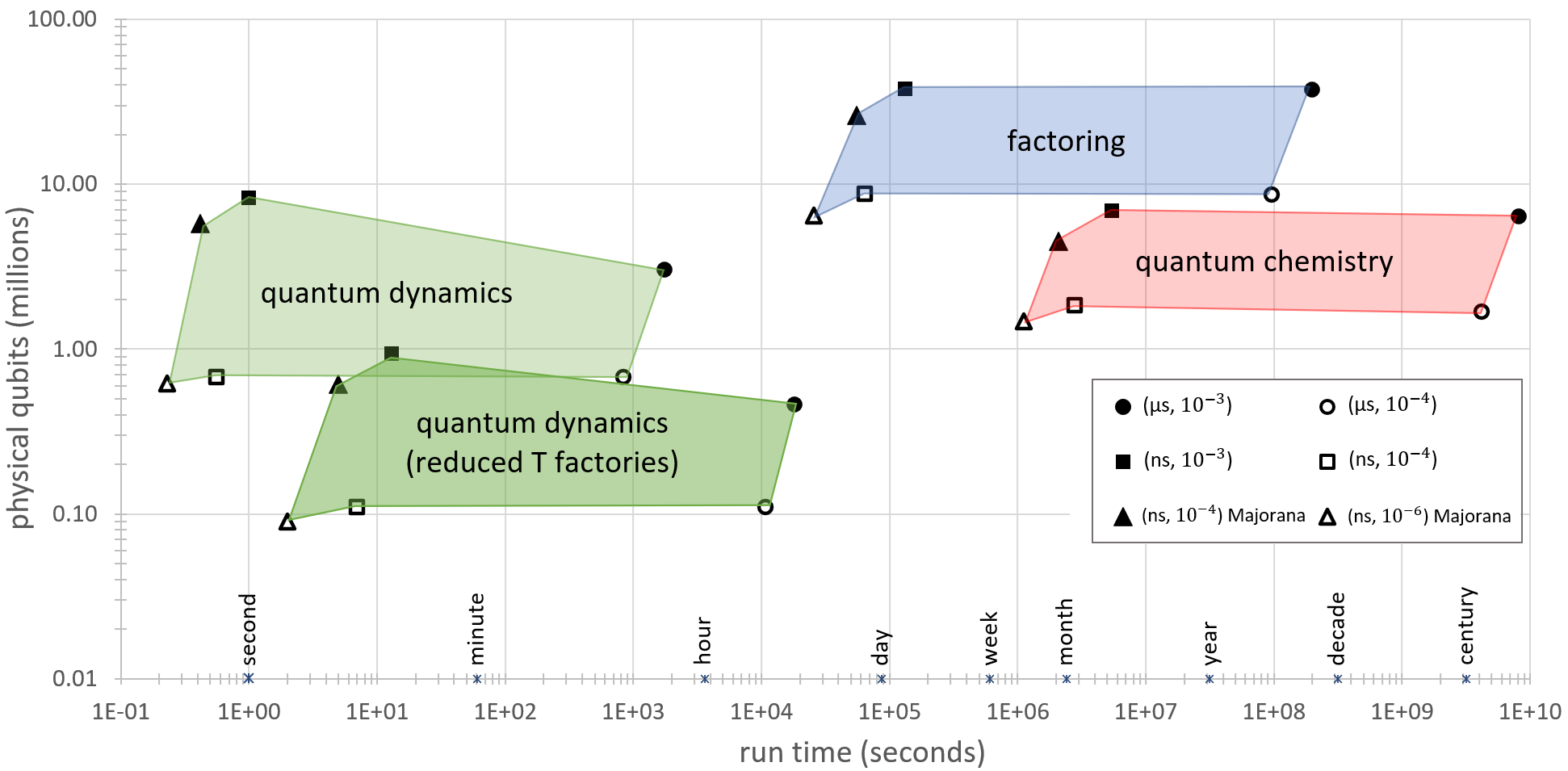}
\caption{\label{fig:EXP}Estimate of the number of qubits 
needed to perform different types of computations. The inset mentions the effect of gate operation time on the estimate.
The figure is taken from Ref.~\cite{2022arXiv221107629B}.}
\end{figure}

The challenges in making scalable quantum computers and performing interesting computations which could change the 
way we perceive the world is still decades away. We have come a long way since 
the first papers in the field were written about 45 years ago, however, lot remains 
to be done both on theoretical and experimental front. For the physically interesting problems physicists 
would really like to solve, it is expected that an order of million noisy 
qubits might be needed as shown in Fig.~\ref{fig:EXP}. The pursuit of this goal will definitely teach us new 
things about computation and more importantly about deep foundations of 
quantum mechanics. One must be hopeful to solve several if not all 
problems.

\section{Acknowledgements}

The research was supported by the U.S. Department of Energy, Office of Science, 
National Quantum Information Science Research Centers, Co-design Center for 
Quantum Advantage under contract number DE-SC0012704.
This material is based upon work supported by the U.S. Department of Energy, Office of Science, Office of 
Nuclear Physics under contract DE-AC05-06OR23177. 
Research at Perimeter Institute was supported
by the Government of Canada through the Department of
Innovation, Science and Economic Development Canada
and by the Province of Ontario through the Ministry of
Research, Innovation and Science. All the quantum circuits 
were made using \texttt{Qcircuit} introduced in Ref.~\cite{eastin2004qcircuit}.

\appendix

\section{\label{sec:QCCV}Continuous variable quantum computing} 

In this article, we have focused only on the qubit-based quantum computing which utilize
the two-state quantum system (qubits) to store and manipulate the information. 
However, not all quantum systems in nature are not restricted to this setup. For example, 
one can have qudit systems where the local Hilbert space is $d > 2$. In fact, one can
go all the way to consider the $d \to \infty$ limit which are formally referred to as 
`qumodes' and constitute the continuous variable (CV) formulation of quantum computing (CVQC). 
One can think of `qumodes' as storing the state of a boson which lives in an infinite dimensional vector space. 
Therefore, a single qumode can be described by a 
quantum-mechanical harmonic oscillator. 
It is these states which one uses to encode the information similar to qubits where only 
$\vert 0 \rangle$ and $\vert 1 \rangle$ are used. In general, 
the state of the qumode is usually represented by two bases:
1) Phase-space basis where the CV gates or operators are written in terms of $\hat{x}, \hat{p}$, 
and 2)  Fock space basis where operators are written in terms of $\hat{a}, \hat{a}^{\dagger}$. 
In the phase-space representation, the computational basis is in terms of strength of 
electric and magnetic fields which are related by Fourier transform. 
In the Fock basis, it is given in terms of 
measurement of the number of bosons (or photons) 
measured in the computational basis represented by $\ket{n}$. 
In practice, one truncates the qumodes at some cutoff $\Lambda$
to obtain a qudit system. We refer the interested reader to Ref.~\cite{2019Sera} for 
more detailed explanation. 

This approach to quantum computing is formally related to the storing and manipulation in terms of quantum 
harmonic oscillators (QHO) and makes use of physical observables
whose numerical values belong to continuous intervals 
(say the $x$ and $p$ quadrature operators). The CV formulation has a very natural 
experimental realisation in terms of photon and the protocols of 
quantum information can be implemented using 
the techniques of quantum optics. Every mode of the electromagnetic
field is a single QHO with its own set of creation and annihilation operators
which are related to the `quadratures'. It has been well-known now for decades that the theory of quantum information based on
this alternate CV formalism provides an alternative ground for quantum computing
\cite{2005Brau,2012Weed}. In particular, they have been shown to be useful towards long-term 
goal of fault-tolerant quantum computing \cite{2001Gott, 2014Mirr}. 
The CV quantum systems is also significant for quantum machine learning since it can provide an arena for 
realization of quantum neural networks \cite{2019Kill}. 
There also have been extension of algorithms 
to continuous variables such as Grover's algorithm \cite{Pati2000}, 
Deutsch-Jozsa \cite{Pati2003} and analog of Shor's code to correct an 
arbitrary single wavepacket errors using a 5-wavepacket code 
\cite{Braunstein:1997db}. Before going into further 
details of CV states, gates, and how measurements are done, 
we present a quick summary of the QHO. A one-dimensional quantum
harmonic oscillator is described by the following Hamiltonian:
\begin{equation}
\hat{H} = \frac{\hat{p}^2}{2} + \frac{\hat{x}^2}{2}.
\end{equation}
One can then define a complex operator $\hat{a}$ as:
\begin{equation} 
\hat{a} = \frac{1}{\sqrt{2}} (\hat{x} + i \hat{p}) 
\end{equation} 
and the complex conjugate of this as $\hat{a}^{\dagger}$. Using this, we can write $\hat{x}$ and $\hat{p}$ as:
\begin{align}
\hat{x} = \frac{1}{\sqrt{2}} (\hat{a} + \hat{a}^{\dagger}), \nonumber \\
\hat{p} = i \frac{1}{\sqrt{2}} (\hat{a} - \hat{a}^{\dagger}).
\end{align} 
The operator $\hat{a}$ is not Hermitian, since it is not equal to $\hat{a}^{\dagger}$. 
The action of these operators on the energy eigenstates $\ket{n}$ (also `Fock' states) are given as:
\begin{align}
\hat{a}^{\dagger} \ket{n} =  \sqrt{n+1} \ket{n+1}, \nonumber \\
\hat{a} \ket{n} =  \sqrt{n} \ket{n-1}.
\end{align} 
because of how it adds and removes quanta, these are also known as `creation' and 
`annihilation' operators. Using these relations, it is sometimes useful to define the number
operator $N = \hat{a}^{\dagger} \hat{a}$ which acts as: $ N \ket{n} = n \ket{n}$. Using this, we 
can write the QHO Hamiltonian as: 
\begin{align}
\hat{H} = \hbar \omega \Big(N + \frac{1}{2}\Big).
\end{align} 
We will first discuss some important gates in CVQC and then
move to discuss the CV states and measurements. 

\subsection{CV gates}
Just like the qubit based computation, we have CV gates in terms of
creation and annihilation operators. We describe some Gaussian CV gates 
below: 

\begin{itemize}
\item \underline{Displacement} operator is defined as: 
\begin{equation}
\label{eq:displ1} 
    D(\alpha) = \exp\Big(\alpha \hat{a}^{\dagger} - \bar{\alpha}\hat{a}\Big) 
\end{equation}
where $\alpha \in \mathbb{C}$ and $\hat{a}^{\dagger}$ and $\hat{a}$ are the
creation and annihilation operators respectively. The set of all displacement operators form a 
Heisenberg-Weyl group which is the `qumodes' generalization of Pauli group for qubits. 
This operator affects the position and momentum operators as $D^{\dagger}(\alpha) \hat{x} D(\alpha) = \hat{x} + \sqrt{2 \hbar}~\mathfrak{R} (\alpha) \mathbb{1}$
and $D^\dagger(\alpha) \hat{p} D(\alpha) = \hat{p} +\sqrt{2 \hbar }~\mathfrak{I} (\alpha) \mathbb{1}$
\item \underline{Squeezing} gate (operator) is given as:
\begin{equation}
    S = \exp\Big(\frac{1}{2} \Big(z^* \hat{a}^{2} - z \hat{a}^{\dagger 2}\Big) \Big),  
\end{equation}
where $z = re^{i\phi}$ known as squeezing parameter. 
The operator transforms the position and momentum operators as: $ S^{\dagger}(z) \hat{x} S(z) = e^{-r} \hat{x}$
and $ S^{\dagger}(z) \hat{p} S(z) = e^{r} \hat{p}$
\item \underline{Two-mode squeezing}: The two-mode version of squeezing operator is given by:
\begin{equation}
S_2(z) = \exp\left(z \hat{a}^{\dagger}_1 \hat{a}^{\dagger}_2 - z^* \hat{a}_1 \hat{a}_2 \right),
\end{equation}
where $z$ is as defined above.  
\item \underline{Rotation}: 
\begin{equation} 
R(\phi) = \exp\left(i \phi \hat{a}^{\dagger} \hat{a} \right)
\end{equation} 
and action on $x$ and $p$ as: 
\begin{align}
R^\dagger(\phi)\hat{x} R(\phi) = \hat{x} \cos \phi - \hat{p} \sin \phi, \nonumber \\ 
R^\dagger(\phi)\hat{p} R(\phi) = \hat{p} \cos \phi + \hat{x} \sin \phi.
\end{align}
\item \underline{Beam-Splitter} gate is defined as:
\begin{equation}
\rm{BS}(\theta, \phi) = \exp(\theta (e^{i\phi} \hat{a_i} \hat{a}^{\dagger}_{j} - e^{-i\phi} \hat{a}^{\dagger}_{i} \hat{a}_{j}))
\end{equation}
A 50-50 BS is the one when $\phi=0, \pi$ and $\theta=\pi/4$. 
\item \underline{Controlled-X}: The controlled-X gate (also known as sum/addition gate is a controlled 
position displacement and is given by (in position basis): 
\begin{equation}
\text{CX}(r) = \exp \left({-i r \hat{x}_1 \hat{p}_2}\right).
\end{equation}
It is called addition gate because $\text{CX}(r) \ket{x_1, x_2} = \ket{x_1, x_2+r x_1}$. 
This gate can also be decomposed in terms of single-mode squeezing 
and beamsplitter gate. One can also define double-controlled version \cite{Wang2001} 
of this gate i.e, $\text{CCX}(r) =  \exp(-i r \hat{p}_3(\hat{x}_1 + \hat{x}_2))$. 
These gates have some interesting usage. For example, in the 
qubit formalism, to add, we use full-adder as discussed in Sec.~\ref{full_add}
which makes use of single and double-controlled CNOT gates. 
An equivalent CV version would then make use of the gates discussed above. 
\end{itemize}
There is another useful state known as `displaced squeezed' state which is obtained
$\ket{\alpha,z} = D(\alpha)\ket{0,z} = D(\alpha)S(z)\ket{0}$, with $z$ being the squeezing parameter 
as above. In addition to these gates, there are also non-Gaussian gates such as `cubic' and `Kerr' gates
given by $V(\gamma) = \exp(i \frac{\gamma}{3!} \hat{x}^3)$ 
and $K(\kappa) = \exp(i \kappa \hat{n}^2)$ respectively
where $\kappa$ is some real parameter and $\hat{n}$ is the
number operator. 

\begin{mybox2}
		\textsc{Example 5:} Assume that we set $\phi=0$ such that BS$(\theta) = \exp(\theta (\hat{a} \hat{b}^{\dagger} - \hat{a}^{\dagger} \hat{b}))$. Show that \newline 
		\begin{align}
		B \hat{a} B^{\dagger} &= \hat{a} \cos \theta + \hat{b} \sin \theta, \\ \nonumber  
		B \hat{b} B^{\dagger} &= -\hat{a} \sin \theta + \hat{b} \cos \theta. 
		\end{align}
	\end{mybox2}
	\begin{proof}
	To show this, we have to use the BCH formula given by: 
	\begin{equation}
	\label{eq:BCH1} 
	e^{\alpha X} A e^{-\alpha X} = \sum_{n=0}^{\infty} \frac{\alpha^{n}}{n!} C_{n}, 
	\end{equation}
	where $\alpha$ is a complex number and $A,C_{n},X$ are operators. 
	$C_{n}$ is recursively defined as: $C_0 = A, C_1 = [A,C_0], C_2 = [A,C_1]$ and so on. 
	For proving the first result, we note that $B a B^{\dagger} = e^{\theta X} a e^{-\theta X}$ where $X = ab^{\dagger} - a^{\dagger}b$. 
	We have suppressed hat for convenience from the operators $a, b$. We find that $C_0 = a, C_1 = b, C_2 = a, C_3 = -b$ and so on.
	The general sum in (\ref{eq:BCH1}) can be split in even and odd parts since $C_n = i^{n} a$ for even $n$ and 
	 $C_n = -i^{n+1} b$ for odd $n$ as:
	 \begin{align}
	 B a B^{\dagger} &= \sum_{n, \rm{even}} \frac{(i\theta)^{n}}{n!}a -i \sum_{n, \rm{odd}}\frac{(i\theta)^{n}}{n!} b, \\ \nonumber 
	 &= a \cos \theta + b \sin \theta. 
	 \end{align} 
	\end{proof}
	We leave it for the reader to show the following for general $\phi$, 
	\begin{align}
	B^\dagger(\theta,\phi) a  B(\theta,\phi) &= a \cos \theta -b e^{-i \phi} \sin \theta,\nonumber \\ 
	B^\dagger(\theta,\phi) b B(\theta,\phi) &= b \cos \theta + a  e^{i\phi} \sin \theta.
	\end{align}

\subsection{CV states}

The most basic CV state is the vacuum state (Gaussian) defined as:
\begin{equation}
\ket{0} = \frac{1}{\sqrt[4]{\pi}}\int dx~e^{-x^2/(2)}\ket{x}, 
\end{equation}
Following this, we can define other states by acting with some operator on $\ket{0}$ as: 
\begin{itemize}
\item \underline{Coherent states}: These are minimum uncertainty state and eigenstate of the annihilation operator
($\hat{a} \ket{\alpha} = \alpha \ket{\alpha}$) given by:
\begin{equation}
    \ket{\alpha} = D(\alpha) \ket{0} = e^{-\frac{1}{2}\vert \alpha \vert^{2}} \sum_{n=0}^{\infty} \frac{\alpha^n}{\sqrt{n!}} \ket{n} 
\end{equation}
in the Fock basis. 
\item \underline{Squeezed states}: A squeezed state is a minimum uncertainty state with unequal quadrature 
variances unlike coherent states. They are obtained by the action of  squeezing operator on vacuum
i.e., $\ket{z} = S(z) \ket{0} $
\item \underline{Displaced squeezed (or squeezed coherent) states}: A displaced squeezed state 
is obtained by the action of squeezing operator on vacuum followed by a 
displacement operator
i.e., $\ket{\alpha, z} = D(\alpha) S(z) \ket{0} $
\item \underline{Cat states}: These are superpositions of two coherent states (with opposite phases) 
$\ket{\alpha}$ and $\ket{-\alpha}$ and are given by:
\begin{equation}
    \ket{\rm{cat}} = \mathcal{N} e^{-\vert \alpha \vert^{2}/2} \Big( \ket{\alpha} + e^{i\phi} \ket{-\alpha} \Big),
\end{equation}
when $\phi = 0$, this is known as even cat state 
denoted by $\ket{C^{+}_{\alpha}}$ while an odd cat state which is 
$\ket{\rm{cat}} \propto \ket{\alpha} - \ket{-\alpha}$ is obtained with $\phi=\pi$. 
The normalization $\mathcal{N}$ is equal to $(2 + 2e^{-2\vert \alpha \vert^{2}} \cos(\phi))^{-1/2} 
= (2 \pm 2e^{-2\vert \alpha \vert^{2}})^{-1/2}$ for even (odd) cat states i.e., $\ket{C^{\pm}_{\alpha}}$. 
These states are non-Gaussian in nature. The odd and even cat states 
are quasi-orthogonal: $\langle -\alpha \vert \alpha \rangle = e^{-2\alpha^2}$
and become orthogonal only as $\alpha \to \infty$. 
\end{itemize}

\subsection{Measurements} 

There are two classes of measurements in CVQC namely 
Gaussian and non-Gaussian. The Gaussian class consists 
of two types: homodyne and heterodyne measurements. 
The non-Gaussian measurement is photon counting. 

Homodyne detection (ideal case) is a kind of projective measurement 
onto the eigenkets of the quadrature operator $\hat{x}$. 
While on the other hand, the heterodyne is a kind of projective 
measurement onto the eigenkets of both 
the quadrature operator $\hat{x}$ and $\hat{p}$.
Since they do not commute, they can only be measured 
with some uncertainity. They can also be thought of as
projection onto the coherent states $\ket{\alpha}$. 
The lack of certainity stems from the fact that
coherent states are not orthogonal. 

The non-Gaussian photon counting 
measurement, also known as photon number resolving (PNR)
detector, as the name suggests, counts the number 
of photos. This uses the particle like character of the 
quanta rather than wavelike from the above mentioned
`dyne'-type measurements. This measurements 
projects onto the eigenkets of the number operator $\ket{n}$. 
The process of PNR detection on \emph{single} mode 
of the multi-mode (large number of qumodes) Gaussian state
changes the other modes to non-Gaussian. This is not the 
case for the `dyne'-type measurements which preserves
the Gaussian character of the multimode state. In view of this, 
photon counting measurement is useful for implementing 
non-Gaussian gates like cubic gate discussed before. 

Now, with some background behind us, let us implement
these using some SDK. Similar to the use of \QIS~for qubit based QC, 
there is a photonic simulator using CVs developed
by Xanadu known as \textsc{Strawberry Fields} \cite{Killoran2019}.
We refer to the reader to this reference for additional simulation
details. We give a small code snippet below with this simulator. 

\begin{mdframed}[backgroundcolor=celadon!6] 
	\begin{lstlisting}[language=Python]
import strawberryfields as sf
from strawberryfields.ops import *

with prog.context as q:

    # Prepare the initial state
    Fock(3) | q[0] # 3 photons in qumode 1
    Fock(1) | q[1] # 1 photon in qumode 2
    
    BSgate(theta, np.pi/2) | (q[0], q[1]) # Two-mode BS gate
    Kgate(r)  | q[0] # Kerr 
    Rgate(-r) | q[0] 
    Kgate(r)  | q[1]
    Rgate(-r) | q[1]
    MeasureFock() | q[0] 
    # Photon counting measurement (also known as photon number resolving [PNR] detector)
    # Measures a set of modes in the Fock basis.
    
    # To export circuit in LaTeX format, use prog.draw_circuit()
    # To print circuit, use prog.print()
    
\end{lstlisting}
\end{mdframed}

\section{\label{sec:fwc}Fun with commutator algebra}

The fundamental commutation relation in quantum mechanics is between the position and
momentum operator (we often suppress the hat over $x$ and $p$ and other operators):
\begin{equation}
\label{eq:comm1} 
[x,p] = i \hbar.  
\end{equation}
This commutation relation is strictly obeyed only in infinite-dimensional
Hilbert space. One way to see this is to obtain a `paradox' by assuming that they
belong to a finite vector space. Considering taking trace on both sides of 
(\ref{eq:comm1}). We find that left side is zero using cyclic nature of trace
while the right side is non-zero. This is paradox arising because strictly speaking
$x,p$ are not trace-class operators. We can also consider commutation relation 
between powers of $x,p$ as: 
\begin{align}
[x,p^{n}] & = [x, pp^{n-1}] \nonumber \\
& = p[x,p^{n-1}] + i\hbar p^{n-1}  \nonumber \\
& \cdots \nonumber \\ 
& = p^{n-1}[x,p] + (n-1) i\hbar p^{n-1}  \nonumber \\
& = i\hbar n p^{n-1}, 
\end{align}
using $[A, BC] = [A, B]C + B[A, C]$ in the first step and repeating several intermediate steps $n-1$ times. 
We can also use the relation below to compute the above commutator:
\begin{equation} 
[X^{n},Y] = nX^{n-1}[X,Y] + X^n[Y,X],
\end{equation} 
and use it to show that
\begin{equation} 
[x^{n},p] = i \hbar n x^{n-1},
\end{equation} 
and a more general form i.e., $[f(x),p] = i \hbar f^{\prime}(x)$. 

A widely studied physical system is the quantum harmonic oscillator. 
The reason is that it is quadratic in both $x$ and $p$ operators. 
We define the number operator $N = a^{\dagger}a$
satisfying following relations:
\begin{equation}
[N, a^{\dagger}] = [a^{\dagger}a, a^{\dagger}] = a^{\dagger}[a,a^{\dagger}] = a^{\dagger}
\end{equation} 
using $[AB, C] = A[B, C] + [A, C]B$ and $[a,a^{\dagger}]=1$. 
Similarly, we can prove that $[N, a] = -a$. We also need 
to compute commutator such as $[a^{n}, a]$. For this, it is useful to note 
two relations when $f$ is an analytic function.
\begin{align}
[a^{\dagger}, f(a)] &= -\frac{df(a)}{da}, \nonumber \\
[a, f(a^{\dagger})] &= \frac{df(a^{\dagger})}{da^{\dagger}}. 
\end{align}  
This immediately leads to $[a, (a^{\dagger})^n] = n (a^{\dagger})^{n-1}$, and 
$[a^{\dagger}, a^n] = -n a^{n-1}$. Using these we can then compute: 
\begin{align}
[a^{q}, n] &= (a^q a^{\dagger} - a^{\dagger} a^q)a \nonumber \\
& = -[a^{\dagger}, a^q]a \nonumber \\
& = q a^{q}. 
\end{align}
where $q$ is some integer power. Similarly, we can show that
$[(a^{\dagger})^q, n] = -q (a^{\dagger})^q$. Some other useful 
commutation relations which follows are:
$[\hat{a}, e^{\alpha \hat{a}^{\dagger}}] = \alpha e^{\alpha \hat{a}^{\dagger}}$
and $[\hat{a}^{\dagger}, e^{\alpha \hat{a}}] = -\alpha e^{\alpha \hat{a}}$. 
When dealing with these operators, we should normal order them. 
The process of putting a product into normal order is 
called normal ordering, also known as Wick ordering. Doing this results in all 
creation operators being left of all annihilation operators in the product. 
It is denoted by the symbol $:~:$. For example, $:bb^{\dagger}: = b^{\dagger}b$
and $:b^{\dagger}b: = b^{\dagger}b$. We give a~\MAT~
code to normal order the operators and simplify some 
commutation expressions discussed above. 

\begin{mdframed}[backgroundcolor=celadon!3] 
	\begin{lstlisting}[language=Mathematica]
Unprotect[NonCommutativeMultiply];
A_ ** (B_ + C_) := A ** B + A ** C
(B_ + C_) ** A_ := B ** A + C ** A
A_ ** c_?NumberQ := c A
c_?NumberQ ** A_ := c A
A_ ** (B_ c_?NumberQ) := c A ** B
(A_ c_?NumberQ) ** B_ := c A ** B
A_ ** (B_ c_Rational) := c A ** B
(A_ c_Rational) ** B_ := c A ** B
A_ ** (B_ c_Power) := c A ** B
(A_ c_Power) ** B_ := c A ** B

commutator[A_, B_] := A ** B - B ** A
fundamentalXP[expr_] := ExpandAll[expr //. p[i_] ** q[i_] :> q[i] ** p[i] - I h]
fundamentalHO[expr_] := ExpandAll[expr //. a ** SuperDagger[a] :> SuperDagger[a] ** a + 1 //. b ** SuperDagger[b] :> SuperDagger[b] ** b + 1]
h /: NumberQ[h] = True; 
commutator[p[x], q[x] ** q[x] ** q[x]] // fundamentalXP  (* [p,q^3] = -3 i \hbar q^2 *) 
NCP[x___] := NonCommutativeMultiply[x];
NCP[0.50*SuperDagger[a] ** a + SuperDagger[b] ** b, 0.50 * SuperDagger[a] ** a + SuperDagger[b] ** b] // fundamentalHO
\end{lstlisting}
\end{mdframed}

In dealing with continuous variable approach to quantum computation, 
one often has to manipulate the bosonic operators. We collect some results in this 
section for the convenience of the reader.  The central result is the `BCH formula'. 
It is given by:
\begin{equation}
e^{X}e^{Y} = e^{X+Y+ \frac{1}{2}[X,Y] + \frac{1}{12}([X,[X,Y]] + [Y,[Y,X]]) + \dots}. 
\end{equation}
For example, using this we can write (\ref{eq:displ1}) as:

\begin{align}
D(\alpha) &= \exp\Big(\alpha \hat{a}^{\dagger} - \bar{\alpha}\hat{a}\Big) \nonumber \\ 
& = \exp(-\frac{1}{2} \vert \alpha \vert ^{2})\exp(\alpha \hat{a}^{\dagger}) \exp(-\bar{\alpha}\hat{a})) 
\end{align}

There is another very useful expression often called the dual of 
BCH formula first discussed in Sec.~4 of Ref.~\cite{Magnus:1954zz}. 
This is known as Zassenhaus formula. 
\begin{equation}
e^{t(X+Y)} = e^{tX} \cdot e^{tY} \cdot e^{-\frac{t^{2}}{2}[X,Y]} \cdot e^{\frac{t^{3}}{6}(2[Y,[X,Y]]+[X,[X,Y]])} \cdots 
\end{equation}
In general, it is written as ($t=1$):
\begin{equation}
e^{X+Y} = e^{X}e^{Y} \prod_{k=2}^{\infty} e^{C_{k}(X,Y)} 
\end{equation}
where as written above, the first few terms are $C_{2}(X,Y) =  
-\frac{1}{2}[X,Y]$ and $C_{3}(X,Y) =  \frac{1}{3}[Y, [X,Y]] + \frac{1}{6}[X, [X,Y]] $. 
Though this is not that well-known, the corrolary of this formula is 
widely known as the Lie-product formula given as: 
\begin{equation}
e^{t(A+B)} = \lim_{{N\to\infty}} \left(e^{\frac{At}{N}} e^{\frac{Bt}{N}} \right)^N
\end{equation}, 
or equivalently: 

\begin{equation}
e^{-it(A+B)} = e^{-itA} e^{-itB} + \mathcal{O}(t^2). 
\end{equation}

The error with $p$-order product formula goes as $\mathcal{O}(t^{p+1})$. In fact, rather than
exponential of just sum of two operators, we can consider the general case which is 
often the most useful in quantum simulation problems. In that case, we have the following
result:
\begin{equation}
e^{-i\sum_{j=1}^{m}H_{j}t} = \prod_{j=1}^{m}e^{-iH_{j}t} + O(m^2t^2), 
\end{equation}
with $t \ll 1$. This expansion would have been exact (i.e. no error) if it was just an ordinary exponential
but the error is because $H$ is an operator/matrix. What happens if our simulation time is $t \gg 1$. 
In that case, one just divides the entire evolution into $r$ steps such that $t^{\prime} = t/r \ll 1$,

\begin{equation}
e^{-i\sum_{j=1}^{m}H_{j}t} = \Bigg(\prod_{j=1}^{m}e^{-iH_{j}t^{\prime}}\Bigg)^{r}  + O(m^2t^2/r). 
\end{equation} 
If we desire an error of $\epsilon$, then we must take $ r \sim m^2t^2/\epsilon$. One can also use second-order formula:
\begin{equation}
e^{-i\sum_{j=1}^{m}H_{j}t}  = \Bigg(\prod_{j=1}^{m} e^{-iH_{j}t/2r} \prod_{j=m}^{1}e^{-iH_{j}t/2r} \Bigg)^r + O(m^3t^3/r^2). 
\end{equation}
In this case, we need to scale $r$ as $\sim m^{3/2}t^{3/2}/\sqrt{\epsilon}$. The gate costs, accuracy, and the 
structure of the Hamiltonian determines which choice is better. 

Another frequently used identity is 
\begin{equation}
\label{eq:UC1} 
Ue^{-iHt}U^\dagger = e^{-iUHU^\dagger t}. 
\end{equation} 
To prove the identity, let's start with the left hand side and expand it: 
\begin{equation} 
U\left(\sum_{n=0}^\infty \frac{(-iHt)^n}{n!}\right)U^\dagger, 
\end{equation} 
since the operators U and $U^\dagger$ commute with the scalar coefficients in the series, 
we can rearrange as:
\begin{equation} 
\sum_{n=0}^\infty \frac{(-iUHU^\dagger t)^n}{n!} = e^{-iUHU^\dagger t}
\end{equation} 
which proves the result.  An expression often encountered in Hamiltonian evolution problem is of the form $U^{\dagger}AU$ where $U$ is exponential of
matrix. For this, we have the following result known as `BCH lemma' given by: 
\begin{equation} 
\label{eq:gen1} 
e^{B}Ae^{-B} = A + [B,A] + \frac{1}{2!}[B,[B,A]] + \frac{1}{3!}[B,[B,[B,A]]] + \cdots
\end{equation}
Using (\ref{eq:gen1}), we can show that $e^{-i \theta a^{\dagger} a} a^{\dagger} e^{-i \theta a^{\dagger} a} = a^{\dagger} e^{i\theta}$. 
The proof proceeds as follows:
\begin{align}
e^{-i \theta a^{\dagger} a} a^{\dagger} e^{-i \theta a^{\dagger} a} &= a^{\dagger} + i\theta [a^{\dagger} a, a^{\dagger}] + \frac{(i\theta)^2}{2!} [a^{\dagger} a, [a^{\dagger} a, a]] + \cdots 
&= a^{\dagger}(1 + i\theta + \frac{(i\theta)^2}{2!} + \cdots = a^{\dagger} e^{i\theta}. 
\end{align}
where we have used $N = a^{\dagger}a$ and $[N, a^{\dagger}] = a^{\dagger}$.

We can also show following results which we leave to the reader. 
\begin{align}
\hat{a} e^{-i\theta \hat{n}} = e^{-i\theta} e^{-i \theta \hat{n}}  \hat{a},  \\
\hat{a}^{\dagger} e^{-i\theta \hat{n}} = e^{i\theta} e^{-i \theta \hat{n}} \hat{a}^{\dagger}. 
\end{align}
Suppose now we have two set of oscillators (denoted by $a$ and $b$ respectively) 
with their own creation and annihilation operators. 
Let us define the following:  
\begin{equation}
    L_{+} = a^{\dagger} b, L_{-} = b^{\dagger} a, K_{+} = a^{\dagger} b^{\dagger}, K_{-} = ba
\end{equation}

Then we can compute:
\begin{equation}
    [K_{+}, L_{+}] = a^{\dagger} b^{\dagger}a^{\dagger} b - a^{\dagger} b a^{\dagger} b^{\dagger} = a^{\dagger}a^{\dagger}b^{\dagger}b -
    a^{\dagger}a^{\dagger} b b^{\dagger} = -a^{\dagger}a^{\dagger}([b,b^{\dagger}]) = -a^{\dagger}a^{\dagger}
\end{equation}

\begin{equation}
    [K_{-}, L_{+}] = baa^{\dagger} b - a^{\dagger} bba = 
    aa^{\dagger}bb -a^{\dagger}abb = bb
\end{equation}
We can also compute the commutator of $K_{+}$ and $K_{-}$. 
\begin{align}
[K_{+}, K_{-}] &= a^{\dagger} b^{\dagger}ba - baa^{\dagger} b^{\dagger} \nonumber \\
& = a^{\dagger} b^{\dagger}ba - aa^{\dagger}bb^{\dagger}  \nonumber \\
& = a^{\dagger} b^{\dagger}ba - a^{\dagger}abb^{\dagger} - bb^{\dagger} \nonumber \\
& = a^{\dagger}a(b^{\dagger}b - bb^{\dagger}) - bb^{\dagger} \nonumber \\
& = -a^{\dagger}a - \mathbb{1} - b^{\dagger}b = -2K_{3}.  
\end{align} 
The operators $K_{+}$, $K_{-}$, $K_3$ form a representation of the Lie algebra $\mathfrak{su}(1,1)$
and is the basis of Schwinger's work of rewriting the angular momentum in terms of oscillators.  
\begin{align}
e^ABe^{-A} &= (I + A + \frac{1}{2!}A^2 + \frac{1}{3!}A^3 + \dots)(B - \frac{1}{1!}[A,B] + \frac{1}{2!}[A,[A,B]] - \frac{1}{3!}[A,[A,[A,B]]] + \dots) \\ \nonumber 
&= B + [A,B] + \frac{1}{2!}(A^2B - AB^2 + [A,[A,B]]) + \frac{1}{3!}(A^3B - A^2B^2 + AB^3 - [A,[A,[A,B]]]) + \dots \\ \nonumber 
&= B + \sum_{n=1}^{\infty} \frac{1}{n!}\sum_{k=0}^{n-1}(-1)^k A^kB\underbrace{A^{n-1-k}}_{\text{($n-1-k$) times}} - \sum_{n=1}^{\infty} \frac{1}{n!}\sum_{k=0}^{n-1}(-1)^k[A,[A,[\dots[A,B]\dots]]] \\ \nonumber 
&= B + \sum_{n=1}^{\infty} \frac{1}{n!}\sum_{k=0}^{n-1}(-1)^k A^kB - \sum_{n=1}^{\infty} \frac{1}{n!}\sum_{k=0}^{n-1}(-1)^k[A,B_k] \\ \nonumber 
&= B + \sum_{n=1}^{\infty} \frac{1}{n!}\sum_{k=0}^{n-1}[A,B_k]. 
\end{align}
where we have defined the nested commutator. There is another way to prove this using the idea of \emph{parameter differentiation}. 
For this, consider $f(t)$ defined as $e^{tA} B e^{-tA}$ with $f(0) = B$. 
Then the derivatives are:
\begin{align}
f^{\prime}(t) &= e^{tA}ABe^{-tA} + e^{tA}B(-A)e^{-tA} = e^{tA}[A,B]e^{-tA}, ~~ f^{\prime}(0) = [A,B] \\ \nonumber 
f^{\prime\prime}(t) & =  e^{tA}[A,[A,B]]e^{-tA} ~~ f^{\prime}(0) = [A,[A,B]]. \\ \nonumber
\end{align}
Therefore, we can write:
\begin{equation}
f(t) = f(0) + t f^{\prime}(t) + \frac{f^{\prime\prime}(t)}{2!} t^2 + \cdots 
\end{equation}
and setting $t=1$, we get the desired 
expression. 
\vspace{5mm}
For $[A,B]$ central, i.e., commuting with both X and Y i.e., $[A,[A,B]], [B,[A,B]] = 0$, we can also prove
a general result. Let us denote $f(\epsilon) = e^{\epsilon A }e^{\epsilon B}$, then we have:
\begin{align}
\label{eq:centralCA} 
f^{\prime}(\epsilon) &= A e^{\epsilon A} e^{\epsilon B} + e^{\epsilon A}e^{\epsilon B} B, \nonumber \\
&= (A + e^{\epsilon A}Be^{-\epsilon A})f(\epsilon), \nonumber \\ 
&= (A + B + \epsilon[A,B]) f(\epsilon). 
\end{align}
where we have used $e^{\epsilon A}Be^{-\epsilon A} = B + \epsilon[A,B]$. 
The solution of (\ref{eq:centralCA}) is:
\begin{equation}
f(\epsilon) = \exp(\epsilon(A+B) + \frac{\epsilon^2}{2}[A,B]).
\end{equation}
Setting $\epsilon=1$ above, the special case of BCH formula as:
\begin{equation}
e^{A}e^{B} = e^{A+B + \frac{1}{2}[X,Y]}  .
\end{equation}
Another result that follows from BCH lemma is when the commutator of A and B is a $c$-number i.e., 
$[A,B] = c$ and $[A,[A,B] = 0$. Then we get, 
\begin{equation}
\label{eq:cnumber1} 
e^{\alpha A}Be^{-\alpha  A} = B + \alpha c. 
\end{equation}
If the nested first commutator is non-zero i.e., $[A,[A,B]] = \beta B$ then one can show that:
\begin{equation}
e^{\alpha A}Be^{-\alpha  A} = B \cosh(\alpha \sqrt{\beta}) + [A,B] \frac{\sinh(\alpha \sqrt{\beta})}{\sqrt{\beta}}. 
\end{equation}
Note that this reduces to former result since the $\beta \to 0$ limit of $\sinh$ term is $\alpha$. 
There is another useful theorem (often called similarity theorem):
\begin{equation}
e^{\alpha A} f(B) e^{-\alpha A} = f(e^{\alpha A} B  e^{-\alpha A}) 
\end{equation}
The proof is simple and proceeds as follows:
\begin{proof}
From the identity $[e^{\epsilon A}Be^{-\epsilon A}]^{n} = e^{\epsilon A}B^{n}e^{-\epsilon A}$, 
we find that the theorem follows if $f(B)$ can be expanded in power series. 
\end{proof}
\noindent Using this theorem, we can also obtain the following result:
\begin{align}
\exp(-\alpha a^\dagger)f(a,a^\dagger)\exp(\alpha a^\dagger)  = f(a+\alpha, a^{\dagger})
\end{align}
which also follows from (\ref{eq:cnumber1}) by setting $A = a^{\dagger}$ and $B= a$, and $c=-1$.

\section{\label{sec:last}Sample programs using \QIS}
\input{appendix.tex}

	\newpage
	\addcontentsline{toc}{section}{References}
	\bibliographystyle{utphys}
	\bibliography{ref.bib}
\end{document}

%% file: appendix.tex
As mentioned before, \QIS~(pronounced kiss-kit) is a backend simulator created by IBM where we can build up a code which can then be sent to actual quantum hardware or quantum processing units (QPU) in use by companies like IBM, Amazon, Honeywell etc. This is like how commercial airline pilots train in a simulator when a new Boeing/Airbus aircraft is about to be launched\footnote{In fact, one of the reasons for the
failure of 737 Max aircraft resulting in tragic accidents around 2019 is that 
even though some new systems were introduced, the pilots were never
ordered to take the new simulator training because Boeing cited 
similarity with the ones they had already taken. It was however not true, and the cause of
accidents (MCAS system) was indeed the crucial difference.}. 
Once we have sufficient practice, we can pass the quantum program to the real machines, but until we really need them for some problem, the simulator is good enough to learn and get acquainted with the fundamental working of the computation. We also note that the calculation of computation time and associated charges for quantum computing on real devices is not as straightforward as classical computers. For example, we will see below that to get reasonable output from the simulator, we have to perform a certain step number of times, which is referred to as `shot' i.e., a shot is a single execution of a quantum algorithm on a QPU. Depending on the quality of QPU, the per shot cost is determined. So, it might be that some quantum computing hardware company will charge $1$ for 100 shots and some will charge the same for 1000 shots. The per-shot price is usually not affected by the number or type of gates used in a quantum circuit or the number of variables employed. The idea of shots is closely related to the errors in quantum computers, and it is to be expected that if there is a QPU with a well-defined error correcting procedure, then it will limit the number of shots needed considerably. Another part of the computation cost is based on number of `tasks'. A task is a sequence of successive shots using the same circuit design. \QIS~is made up of four elements, each of which has a specific functionality. These elements are Terra (Earth), Aer (Air), Ignis (Fire), and Aqua (Water). For example, the Aer provider contains a variety of high performance simulator backends for a variety of simulation methods. The available backends on the current system can be viewed using \texttt{Aer.backends()} in a Jupyter notebook. 

This open-source software platform enables one to work with the quantum language  
Open Quantum Assembly Language (Open QASM), used in quantum computers in the IBM. 
We encourage the reader to read more about them if interested in Ref.~\cite{QISKIT_TB}. 
Here, we will briefly summarize the different core modules of \QIS~for convenience. 
The Qiskit open-source development library is composed of four core modules:
\begin{itemize}
\item Qiskit Terra: This contains core elements to create quantum programs at the circuit/pulses level as well as to optimize these core elements by considering the particular physical quantum processor constraints.
\item Qiskit Aer: This provides a C++ simulator framework and tools to develop noise models to conduct realistic simulations in the presence of noise and errors occurring during execution on real quantum devices.
\item Qiskit Ignis: This provides a framework to understand the noise sources in the quantum circuits/devices as well as to develop different noise reduction procedures. 
\item Qiskit Aqua: This contains a library of cross-domain quantum algorithms suitable for application in the NISQ era. 
\end{itemize}
The syntax of \QIS~consists of following high-level steps: 1. Building step in which we design a quantum circuit that represents the problem at hand, 
2. The execution step in which we run experiments on different backends (including both actual quantum hardware and simulators), 
and the last is the analysis step where one summarizes the statistics and visualize the results. 
The SDK we will focus on in these notes deals with systems with finite local Hilbert space of $d = 2$ (qubits) 
or maybe some generalization to qudits too ($d > 2$) but they are often not the best 
way for bosonic states which have infinite sized vector space. For these models, 
it is often more practical to make use of what is called \emph{qumodes} 
which are continuous-variable quantum modes. 
Very recently, an open-source software was introduced in Ref.~\cite{Stavenger:2022wzz} which 
extends \QIS~to admit bosonic degrees of freedom. This enables the simulation 
of hybrid qubit and bosonic systems such as bosonic Hubbard models, Jaynes-Cummings etc.
However, the support for parametrized circuits\footnote{Parameterized quantum circuits consist of quantum 
gates defined through tunable parameters. They are 
fundamental building blocks of majority of 
near-term quantum machine learning algorithms} is still in development phase. 
Another open-source project which is probably better suited for ground state energy
computation of bosonic models is Xanadu's \texttt{Strawberry Fields}, which 
is a full-stack Python library for designing, optimizing, and utilizing photonic quantum computers 
(analog computing). Using continuous variables (CV) instead of the more traditional qubits (digital)
approach has both advantages and disadvantages. Some groups \cite{Yeter-Aydeniz:2021mol}
have explored the use of QITE (quantum imaginary-time evolution) to study the 
ground state energy using CVs. This approach might also be useful in the future for 
other interesting models such as $\phi^4$-theory. Though the scope of this discussion 
lies beyond the scope of the current article, we will briefly mention some differences to qubit approach. 
The CV approach to quantum computing keeps the same computational power of the qubit model. 
This approach utilizes the infinite-dimensional bosonic mode called `qumodes'. 
While the qubits make use of discrete coefficients, the CV model makes use of bosonic harmonic oscillator, 
which is defined by the operators $a$ and $a^{\dagger}$  satisfying $[a, a^{\dagger}] = 1$
and where $\ket{x}$ are the eigenstates of $x$ operator defined in terms of  $a, a^{\dagger}$. 
In this approach, the information about input states are stored in terms of Gaussian states
which are non-classical states with Gauss-Wigner quasi-probabilistic functions.

\subsection{Installing package and quantum hello world} 
In this section, we will provide several codes related to sections in the main text. The easiest way 
is to visit Google's Colab\footnote{\href{https://research.google.com/colaboratory}{https://research.google.com/colaboratory}}and open a new 
Jupyter notebook. We can install \QIS~by doing: $\texttt{!pip install qiskit ipywidgets}$. A simple check that everything is installed properly 
is to write this short piece of code
and run it. 
\begin{mdframed}[backgroundcolor=celadon!6] 
	\begin{lstlisting}[language=Python]
from qiskit import *
qc = QuantumCircuit(2);  # Circuit of two qubits 
# This is equivalent to QuantumCircuit(2,0) i.e., 2 qubits and no classical bits 
qc.h(0);   # Apply Hadamard gate to the first qubit 
qc.draw() # Draw the circuit 
	\end{lstlisting}
\end{mdframed}
We then extend this to write our first `quantum' code given below:
\begin{mdframed}[backgroundcolor=celadon!6] 
	\begin{lstlisting}[language=Python]
# Welcome to "Hello World" for quantum computing
from qiskit import *
from qiskit import QuantumCircuit, assemble, Aer
from qiskit.visualization import plot_bloch_multivector, plot_histogram
sim = Aer.get_backend('aer_simulator')

def x_measurement(qc, qubit, cbit):
	qc.h(qubit) # Apply Hadamard to get superposition 
	qc.measure(qubit, cbit)
	# Measure 'qubit' in the X-basis, and store the result in 'cbit
	return qc

qc = QuantumCircuit(1,1);  # Initialize one quantum and one classical bit. 
x_measurement(qc, 0, 0)
qobj = assemble(qc)  # Assemble circuit to run 
counts = sim.run(qobj).result().get_counts()  # Do the simulation, returning the state vector
print (counts)  # Count the occurences 
plot_histogram(counts)  # Display the output as a histogram 
	\end{lstlisting}
\end{mdframed}

\subsection{Some basic operations using qubits and gates}

Let us initialize two qubits in $\ket{0}$ each. We can apply Hadamard gate to the first qubit and then feed them through CNOT gate. It is easy to check that  it generates the output as: $(1/\sqrt{2})(\ket{00} + \ket{11})$. We provide the code below to compute this out state and check that it agrees with the expectations. We leave this to the reader.

\begin{figure}
	\centering 
	\includegraphics[width=0.45\textwidth]{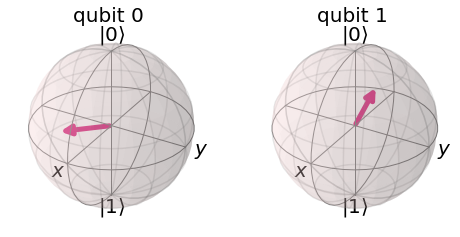}
	\caption{\label{fig:Bloch00}A random two-qubit state depicted on the sphere.}
\end{figure}

\begin{mdframed}[backgroundcolor=celadon!6] 
	\begin{lstlisting}[language=Python]
from qiskit import *
from qiskit.quantum_info import Statevector
from qiskit.visualization import plot_bloch_multivector, plot_histogram, array_to_latex
qc = QuantumCircuit(2);
qc.h(0);
qc.draw()
q2 = qc.cx(0,1);
bell = Statevector.from_instruction(qc)
display(array_to_latex(bell, prefix="\\text{Statevector} = "))
	\end{lstlisting}
\end{mdframed}
It is easy to generate $d = 2^n$-dimensional state vector using a short program and then view it on the Bloch sphere. Suppose we take $n=2$ (two qubits) and generate a random state $\ket{\psi}$. It can be done as given below. Running this different times will move the arrow (on the Bloch sphere) corresponding to different random state. The reader is encouraged to check that they are appropriately normalized, i.e., $\langle \psi \vert \psi \rangle = 1 $.  The output on the Bloch sphere for a random state looks like that given in Fig.~\ref{fig:Bloch00}. Note that orthogonal states (such as $\ket{0}$ and $\ket{1}$) are antipodal on the 
Bloch sphere. 

\begin{mdframed}[backgroundcolor=celadon!6] 
	\begin{lstlisting}[language=Python]
from qiskit.quantum_info import random_statevector, Statevector
from qiskit.visualization import plot_bloch_multivector

rand_sv  = random_statevector(4).data
print(rand_sv) 
plot_bloch_multivector(rand_sv)
	\end{lstlisting}
\end{mdframed}
Now, since we have a bipartite random state, we can compute EOF as discussed in Sec.~\ref{sec:ent_mstate}.  
For example, if we compute this for Bell state mentioned above, it will be 1 (maximally entangled). We give the code below:
\begin{mdframed}[backgroundcolor=celadon!6] 
	\begin{lstlisting}[language=Python]
from qiskit import *
from qiskit.quantum_info import *

rand_sv  = random_statevector(4).data
print("Entanglement of Formation = ", entanglement_of_formation(rand_sv))		
	\end{lstlisting}
\end{mdframed}
\begin{mybox}
	\label{mybox:Q999}
	\textsc{$\blacktriangleright$ Question 21:}  Set up a quantum circuit using \QIS~to create the state given by Eq.~\ref{eq:Q0} and show that it is not entangled.
\end{mybox}
Let us now show the program to implement $\rm{CSWAP}$ gate as discussed in the text using \QIS.
Note that if we take $q_{0} = \ket{+}$, $q_{1} = \ket{0}$, and $q_{2} = \ket{1}$, then the action of 
$\rm{CSWAP}$ looks like:

\begin{equation}
	\ket{1} \otimes \ket{0} \otimes (1/\sqrt{2}) (\ket{0} + \ket{1}) = \frac{1}{\sqrt{2}} \Big(\ket{100} + \ket{101} \Big)
	 \xrightarrow{\rm{CSWAP}} \frac{1}{\sqrt{2}} \Big(\ket{100} + \ket{011} \Big)
\end{equation}

\begin{mdframed}[backgroundcolor=celadon!6] 
	\begin{lstlisting}[language=Python]
from qiskit import * 
from qiskit.quantum_info import *
from qiskit.visualization import *

qc = QuantumCircuit(3)
qc.h(0); 
qc.x(2); 
st = Statevector.from_instruction(qc)
display(array_to_latex(st, prefix="\\text{In state} = "))
qc.cswap(0,1,2)
st = Statevector.from_instruction(qc)
display(array_to_latex(st, prefix="\\text{ Out state} = "))
qc.draw()			
	\end{lstlisting}
\end{mdframed}
In the program above, we have used  ${\texttt{qc.cswap(0,1,2)}}$ to implement
$\rm{CSWAP}$. Suppose we do not have this three qubit gate and only have access to $\rm{CNOT}$ gate. 
We encourage the reader to write the equivalent program. We can also entangle $n$ qubits (using GHZ state construction) as follows:

\begin{mdframed}[backgroundcolor=celadon!6] 
	\begin{lstlisting}[language=Python]
from qiskit import *
from qiskit.visualization import *
from qiskit.quantum_info import * 
		
# We will now construct GHZ state 
qn = 5 # five qubit GHZ

circ = QuantumCircuit(qn)
circ.h(0)
circ.cx(0, 1)
circ.cx(0, 2)
circ.cx(0, 3)
circ.cx(0, 4)
circ.draw()
ghz = Statevector.from_instruction(circ)
display(array_to_latex(ghz, prefix="\\text{Statevector} = "))
	\end{lstlisting}
\end{mdframed}
Suppose we prepare some initial state and someone applies a quantum gate in our absence, is there a way in \QIS~to see what this transformation was? The way to do it, this is (apart from usual stuff):
\begin{mdframed}[backgroundcolor=celadon!6] 
	\begin{lstlisting}[language=Python]
job = execute(qc, Aer.get_backend('unitary_simulator'))
result = job.result()
print(result.get_unitary(qc, decimals=3))
	\end{lstlisting}
\end{mdframed}
We now turn to the computation of fidelity of two states which was discussed in the main text. Consider two-states given by:
\[ \ket{\psi} = \frac{1}{2} \Big( \ket{001} + \ket{101}  - \ket{011} - \ket{111} \Big), \]
and 
\[ \frac{1}{\sqrt{2}} \Big( \ket{000} + \ket{111}  \Big). \]
Compute the fidelity between them using \QIS. 
\begin{mdframed}[backgroundcolor=celadon!6] 
	\begin{lstlisting}[language=Python]	
from qiskit import * 
from qiskit.quantum_info import *
from qiskit.visualization import *
from numpy import * 

qc1 = QuantumCircuit(3)
qc1.h(0);
qc1.h(1); 
qc1.x(2);  
qc2 = QuantumCircuit(3)
qc2.h(0)
qc2.cx(0,1); 
qc2.cx(0,2); 

backend = Aer.get_backend('statevector_simulator')
sv1 = execute(qc1, backend).result().get_statevector(qc1)
sv2 = execute(qc2, backend).result().get_statevector(qc2)
print(state_fidelity(sv1, sv2))		
	\end{lstlisting}
\end{mdframed}

\subsection{Quantum full-adder} 

Similar to the classical logic gates where one can
implement full-adder, we can do similar operation using 
quantum gates. The code is given below. 

\begin{mdframed}[backgroundcolor=celadon!6]
\begin{lstlisting}[language=Python]
# Quantum full-adder program 		

from qiskit import *
from qiskit.visualization import *

qc = QuantumCircuit(4,2)

# Preparing inputs
# We do: A and B to a Bell state and CarryIn to a superposition state
qc.h(0)
qc.cx(0,1)
# To Bell State: Hadamard followed by CNOT 
qc.h(2) # CarryIn to superposition by Hadamard 

# Adder starts 
qc.ccx(0,1,3) # Toffoli gate [three qubit] 
qc.cx(0,1) 
qc.ccx(1,2,3) 
qc.cx(1,2) 
qc.cx(0,1) 

qc.measure(2,0) # Sum
qc.measure(3,1) # Carry-out 
counts = execute(qc, Aer.get_backend('qasm_simulator'), shots=1024).result().get_counts()
print(counts) 
plot_histogram(counts)
\end{lstlisting}
\end{mdframed}

The circuit for quantum full-adder is given by (we have left out the part where qubits are initialized, see code for details):
\begin{equation}
	\Qcircuit @C=1.2em @R=1.4em @! {
		\lstick{a} & \ctrl{3} & \ctrl{1} & \qw & \qw & \ctrl{1} & \qw & \rstick {p_1} \\
		\lstick{b} & \ctrl{2} & \targ  & \ctrl{2} & \ctrl{1} & \targ & \qw & \rstick {p_2}\\ 
		\lstick{q_2} & \qw & \qw & \ctrl{1} & \targ & \qw & \qw & \rstick {\text{carry-in}}\\
		\lstick{q_3} & \targ & \qw & \targ & \qw & \qw & \qw & \rstick {\text{sum}} 
	}
\end{equation}


\subsection{Grover's algorithm with general diffuser routine}
We give the \QIS~code for Grover's algorithm for the interested reader. 
We implement two cases of oracle for finding the target state $\ket{11}$ and $\ket{110}$
as example. The diffuser part (implementing $D$) is general for any number of qubits. 
\begin{mdframed}[backgroundcolor=celadon!6]
\begin{lstlisting}[language=Python]
from qiskit import *
import matplotlib.pyplot as plt
import numpy as np

n = 3 

def diffuser(nqubits):
    
    # D = (H)^n . (X)^n . (MCZ) . (X)^n . (H)^n
    qc = QuantumCircuit(nqubits)
    
    # Apply transformation |s> -> |0 ... 0> (H-gates)
    for qubit in range(nqubits):
        qc.h(qubit)
        
    # Apply transformation |0 ... 0> -> |1 ... 1> (X-gates)
    for qubit in range(nqubits):
        qc.x(qubit)
        
    # ****************************
    # Do multi-controlled-Z gate
    # by using Multi controlled Toffoli 
    # Recall that HXH = Z
    qc.h(nqubits-1)
    qc.mct(list(range(nqubits-1)), nqubits-1)  # Multi-controlled
    qc.h(nqubits-1)
    # ****************************
    
    # Now X to all 
    # Apply transformation |1 ... 1> -> |0 ... 0>
    for qubit in range(nqubits):
        qc.x(qubit)
        
    # Apply transformation |0 ... 0> -> |s>
    for qubit in range(nqubits):
        qc.h(qubit)
        
    # We will return the diffuser as a gate
    U_s = qc.to_gate()
    U_s.name = "Diffuser (D)"
    return U_s

oracle = QuantumCircuit(n, name = 'Oracle (O)')

if n == 2:
    oracle.cz(0,1)
if n == 3:
    oracle.cz(1, 2)
    
oracle.to_gate() 
# Oracle is CZ gate. This will change based on target state. 
# We provide two example |11> and |110> 

backend = Aer.get_backend('statevector_simulator')
gro_circ = QuantumCircuit(n,n) 
gro_circ.h(list(range(n)))
gro_circ.append(oracle,list(range(n)))

gro_circ.append(diffuser(n),list(range(n)))

backend = Aer.get_backend('qasm_simulator')
gro_circ.measure([0,1],[0,1])
job = execute(gro_circ,backend, shots=1)
result = job.result() 
result.get_counts()
\end{lstlisting}
\end{mdframed}

\subsection{Implementing QFT}

We now discuss the implementation of quantum Fourier transform in \QIS. This makes use of two quantum gates, which we have already seen a lot of. The first is the Hadamard
gate, the other is the controlled phase gate $R$. The matrix representation of these gates are given below:
\begin{equation}
	H = \frac{1}{\sqrt{2}}   \begin{bmatrix}
		1 ~&~ 1 \\
		1 ~&~ -1 
	\end{bmatrix}, 
\end{equation}
while the rotation matrices are interchangeably defined as both: 
\begin{equation}
R_{k} = \begin{bmatrix}
	1 ~&~ 0 \\
	0 ~&~ \exp\Big(\frac{2 \pi i}{2^{k}} \Big)  
\end{bmatrix},  ~~~~ R_{\theta} = \begin{bmatrix}
1 ~&~ 0 \\
0 ~&~ e^{i \theta}  
\end{bmatrix}
\end{equation}
The simplified QFT circuit for three qubits is given below, while the program in \QIS~below is for general $n$-qubit case.
\begin{align*}
	\Qcircuit @C=1em @R=.7em {
		\lstick{\ket{x_3}} & \qw    & \qw     & \ctrl{2}      & \qw    & \ctrl{1}      & \gate{H} & \rstick{\ket{y_1}} \qw \\
		\lstick{\ket{x_2}} & \qw    & \ctrl{1}    & \qw &  \gate{H} & \gate{R_{\pi/2}} & \qw      & \rstick{\ket{y_2}} \qw \\
		\lstick{\ket{x_1}} & \gate{H} & \gate{R_{\pi/2}} & \gate{R_{\pi/4}}  &  \qw & \qw  & \qw   & \rstick{\ket{y_3}} \qw
	}
\end{align*}
The code is given below:

\begin{mdframed}[backgroundcolor=celadon!6]
\begin{lstlisting}[language=Python]
# QuantumFourier program 
			
from qiskit import *
from numpy import * 
from qiskit.quantum_info import *
from qiskit.visualization import *

def qft_rotations(circuit, n):
    
    if n == 0:
        return circuit
    
    n -= 1
    circuit.h(n)
    
    for qubit in range(n):
        circuit.cp(pi/2**(n-qubit), qubit, n)
        
    qft_rotations(circuit, n)

def swap_registers(circuit, n):
    
    for qubit in range(n//2):
        circuit.swap(qubit, n-qubit-1)
    return circuit

def qft(circuit, n):
    
    qft_rotations(circuit, n) # Goes in a circle until all done
    swap_registers(circuit, n)
    return circuit

nbits = 3 # number of qubits 
N = 2**nbits 
# So for this case it will be QFT_8

qc = QuantumCircuit(nbits)

# If done nothing, encodes 000 [i.e., all qubits set to 0 i.e., |000>]
# We will try encoding 5 i.e, |101> 
qc.x(0)
qc.x(2)

sim = Aer.get_backend("aer_simulator")
qc_init = qc.copy()
qc_init.save_statevector()
statevector = sim.run(qc_init).result().get_statevector()
plot_bloch_multivector(statevector, title= "Before: |101> := |dud> ")

# Apply QFT 

qft(qc,nbits)
#qc.draw()

psi_AB = Statevector.from_instruction(qc)
plot_bloch_multivector(psi_AB, title= "After: QFT_8|101>")
\end{lstlisting}
\end{mdframed}




\subsection{Bit-flip code, Phase kickback} 

\begin{mdframed}[backgroundcolor=celadon!6]
\begin{lstlisting}[language=Python]
# A simple example of how the bit flip of q0 can be corrected
# This version can be modified to do bit flip with probability 'p' < 1

from qiskit import *
from qiskit import Aer 

backend = Aer.get_backend('qasm_simulator')
circuit = QuantumCircuit(3,1)

# Encode
circuit.cx(0,1)
circuit.cx(0,2)

# Noise 
circuit.x(0) # Impose a bit flip error by Pauli X

# Decode 
circuit.cx(0,1)
circuit.cx(0,2)
circuit.ccx(2,1,0)
circuit.measure(0,0)

job = execute(circuit, backend, shots=1024)
counts = job.result().get_counts()
print(counts)
\end{lstlisting}
\end{mdframed}
Now we do phase kickback to reproduce
the state mentioned in (\ref{eq:phase_kb1})
with $\phi = \pi/4$. 

\begin{mdframed}[backgroundcolor=celadon!6]
\begin{lstlisting}[language=Python]
# Phase kickback 

from qiskit import * 
from numpy import * 
from qiskit.quantum_info import Statevector
from qiskit.visualization import plot_bloch_multivector, plot_histogram, array_to_latex

def phase_kickback():
    qreg = QuantumRegister(2, 'q')
    qc1   = QuantumCircuit(qreg)
    qc1.x(qreg[1])
    qc1.h(qreg[0])
    qc1.cp(pi/4, 0, 1) # Controlled T 
    return qc1

qc = phase_kickback()
qc.draw()
co = Statevector.from_instruction(qc)
plot_bloch_multivector(co)
# We can see the qubit 0 has been rotated by pi/4 around 
# the Z-axis of the Bloch sphere as expected.	
\end{lstlisting}
\end{mdframed}

\subsection{VQE solution to anharmonic quantum oscillator for three qubits} 
In this section, we provide code which accompanies Sec.~\ref{subsec:AHO}. 
We first decompose the Hamiltonian in terms of Paulis operators and then use it as input to 
VQE solver by defining some appropriate ansatz.
\begin{mdframed}[backgroundcolor=celadon!6]
\begin{lstlisting}[language=Python]
#!pip install qiskit ipywidgets
import numpy as np
import math 
import time
import warnings
import itertools
from qiskit import Aer
from qiskit.algorithms import VQE
from qiskit.opflow import MatrixOp
from qiskit.opflow import X, Y, Z, I
from qiskit.algorithms.optimizers import SLSQP
from qiskit.algorithms.optimizers import COBYLA
from qiskit.circuit.library import EfficientSU2
from qiskit.utils import algorithm_globals, QuantumInstance
from qiskit.visualization.array import array_to_latex
 
def decompose(H):
    
    def dagger(a):
        return np.transpose(a).conj()

    def HS(M1, M2):
        # Hilbert-Schmidt prod. of two matrices 
        return np.trace(dagger(M1) @ M2)  

    def c2s(c):
        if c == 0.0:
            return "0"
        if c.imag == 0:
            return "%g" % c.real
        elif c.real == 0:
            return "%gj" % c.imag
        else:
            return "%g+%gj" % (c.real, c.imag)
    
    sx = np.array([[0, 1],  [ 1, 0]], dtype=np.complex128)
    sy = np.array([[0, -1j],[1j, 0]], dtype=np.complex128)
    sz = np.array([[1, 0],  [0, -1]], dtype=np.complex128)
    id = np.array([[1, 0],  [ 0, 1]], dtype=np.complex128)
    S = [id, sx, sy, sz]
    labels = ['I', "X", "Y", "Z"]
    final = ""

    if nbits == 2:
        for i, j in itertools.product(range(4),range(4)):
            a = (1/2**nbits) * HS(np.kron(S[i], S[j]), H)
            # Formula: (1/2^2) (Si otimes Sj) dot H
            if a != 0.0:
                term = c2s(a),'*',labels[i],'^',labels[j] 
                t2 = str(term).replace(',', '')
                t2 = str(t2).replace("'", "")
                
                if final == "":
                    final = "".join((final, t2))
                else: 
                    final = "+".join((final, t2))
    
    if nbits  == 3:
        for i, j, k in itertools.product(range(4),range(4),range(4)):
            a = (1/2**nbits) * HS(np.kron(np.kron(S[i], S[j]),S[k]), H)
            
            if a != 0.0:
                term = c2s(a),'*',labels[i],'^',labels[j],'^',labels[k]
                t2 = str(term).replace(',', '')
                t2 = str(t2).replace("'", "")
                
                if final == "":
                    final = "".join((final, t2))
                else: 
                    final = "+".join((final, t2))
    
    return final

# Alternative way 

def numberToBase(n, b, n_qubits):
    if n == 0:
        return np.zeros(n_qubits,dtype=int)
    digits = np.zeros(n_qubits,dtype=int)
    counter = 0
    while n:
        digits[counter] = int(n % b)
        n //= b
        counter += 1
    return digits[::-1]

def decomposePauli(H):
    
    sx = np.array([[0, 1],  [ 1, 0]], dtype=np.complex128)
    sy = np.array([[0, -1j],[1j, 0]], dtype=np.complex128)
    sz = np.array([[1, 0],  [0, -1]], dtype=np.complex128)
    id = np.array([[1, 0],  [ 0, 1]], dtype=np.complex128)
    S = [id, sx, sy, sz]
    dim_matrix=np.shape(H)[0]
    n_qubits=int(np.log2(dim_matrix))
    if(dim_matrix != 2**n_qubits):
        raise NameError("Matrix is not power of 2!")
    hilbertspace = 2**n_qubits
    n_paulis = 4**n_qubits
    pauli_list = np.zeros([n_paulis,n_qubits],dtype=int)
    for k in range(n_paulis):
        pauli_list[k,:] = numberToBase(k,4,n_qubits)
    weights = np.zeros(n_paulis,dtype=np.complex128)
    for k in range(n_paulis):
        pauli=S[pauli_list[k][0]]

        for n in range(1,n_qubits):
            pauli = np.kron(pauli,S[pauli_list[k][n]])
        weights[k] = 1/hilbertspace* np.dot(pauli,H).trace()

    nnz = [i for i, e in enumerate(weights) if e != 0]
    for i in nnz:
        print (weights[i],pauli_list[i])

# Yet another way
def decompose_ham_to_pauli(H):

    n = int(np.log2(len(H)))
    N = 2 ** n
    # Basic checks!
    
    if H.shape != (N, N):
        raise ValueError("The Hamiltonian should have shape (2**n, 2**n), for any qubit number n>=1")
        
    if not np.allclose(H, H.conj().T):
        raise ValueError("The Hamiltonian is not Hermitian")

    sI = np.eye(2, 2, dtype=complex)
    sX = np.array([[0, 1], [1, 0]], dtype=complex)
    sZ = np.array([[1, 0], [0,-1]], dtype=complex)
    sY = complex(0,-1)*np.matmul(sZ,sX)
    paulis = [sI, sX, sY, sZ]
    paulis_label = ['I', 'X', 'Y', 'Z']
    obs = []
    coeffs = []
    matrix = []

    for term in itertools.product(paulis, repeat=n):
        matrices = [pauli for pauli in term]
        coeff = np.trace(ft.reduce(np.kron, matrices) @ H) / N 
        coeff = np.real_if_close(coeff).item()

        if not np.allclose(coeff, 0): 
            coeffs.append(coeff)
            obs.append(''.join([paulis_label[[i for i, x in enumerate(paulis) 
            if np.all(x == t)][0]]+str(idx) for idx, t in enumerate(reversed(term))]))
            matrix.append(ft.reduce(np.kron, matrices))

    return obs, coeffs , matrix

g = 0.02 # Coupling used to create ham_HO.txt as input to this code 
H = np.loadtxt("ham_HO.txt", dtype='f', delimiter='\t')
nbits = math.log2(np.shape(H)[0])

# Two options until nbits=3. 

qubitOp = MatrixOp(primitive=H)
Hps = decompose(H)
print (Hps) 

Hps = (4 * I ^ I ^ I)+(-0.152955 * I ^ I ^ X)+(-0.5 * I ^ I ^ Z)+(-0.12289 * I ^ X ^ X)+(-0.0629948 * I ^ Y ^ Y)+(-1 * I ^ Z ^ I)+(0.0237627 * I ^ Z ^ X)+(-0.0280252 * X ^ I ^ X)+(-0.0561195 * X ^ X ^ X)+(0.0287333 * X ^ Y ^ Y)+(0.0107047 * X ^ Z ^ X)+(-0.0280252 * Y ^ I ^ Y)+(-0.0287333 * Y ^ X ^ Y)+(-0.0561195 * Y ^ Y ^ X)+(0.0107047 * Y ^ Z ^ Y)+(-2 * Z ^ I ^ I)+(0.0872346 * Z ^ I ^ X)+(0.0842295 * Z ^ X ^ X)+(0.041655 * Z ^ Y ^ Y)+(0.0207442 * Z ^ Z ^ X)
# Nice formatting using LaTeX
# H_matrix = H.to_matrix()
# array_to_latex(H_matrix)    

start_time = time.time()
var_form = EfficientSU2(qubitOp.num_qubits, su2_gates=['ry'], entanglement="full", reps=1)

rngseed = 5
warnings.filterwarnings("ignore")
backend = Aer.get_backend("statevector_simulator", max_parallel_threads=1, max_parallel_experiments=0)
q_instance = QuantumInstance(backend, seed_transpiler=rngseed, seed_simulator=rngseed)
#optimizer = SLSQP(maxiter=600)
optimizer = COBYLA(maxiter=600)
# COBYLA works little better!

# Run the VQE
vqe = VQE(ansatz=var_form,optimizer=optimizer,quantum_instance=q_instance)
ret = vqe.compute_minimum_eigenvalue(Hps) 
vqe_result = np.real(ret.eigenvalue)
print("VQE Result:", vqe_result)
exact = 0.5 - (11/8)*g*g -(465/32)*g*g*g*g
print ("Exact result for cubic oscillator upto O(g^4) is", exact)
print ("Error is", np.round(abs((exact-vqe_result)/exact)*100,10), "percent")
end_time = time.time()
runtime = end_time-start_time
print('Program runtime: ',runtime, "seconds")

# WCUBIC[n_, g_] := (n + 0.5) - ( 45 g^2)/12 (n^2  + n + 11/30); 
\end{lstlisting}
\end{mdframed}
It returns an output given below: 
\begin{verbatim}  
VQE Result: 0.4994476532684504
Exact result for cubic oscillator upto O(g^4) is 0.499447675
Error is 4.3511e-06 percent
Program runtime:  2.118819236755371 seconds
\end{verbatim}  

\subsection{Phase estimation}

We provide in this subsection the code to estimate the phase
as discussed in the main text. 
\begin{mdframed}[backgroundcolor=celadon!6]
\begin{lstlisting}[language=Python]
from qiskit import * 
from qiskit.quantum_info import *
from qiskit.visualization import *
from numpy import * 
import sys

qbits=4
cbits=3
theta= 2*pi*(1/4) # Actually \theta is without 2*pi but abuse notation!
qc = QuantumCircuit(qbits, cbits)

def qft_dagger(qc, n):

    # Swap is crucial. Remove this to see change in result 
    for qubit in range(n//2):
        qc.swap(qubit, n-qubit-1)
        for j in range(n):
            for m in range(j):
                qc.cp(-pi/float(2**(j-m)), m, j)
            qc.h(j)

for qubit in range(3):
    qc.h(qubit)

qc.x(qbits-1)  # Initialize |psi> to |1> 

sheets = 1
for qubit_first_reg in range(qbits-1):
    for i in range(sheets):
        qc.cp(theta, qubit_first_reg, qbits-1); # Target is last qubit (i.e. first qubit of second register)
        sheets *= 2

# Apply inverse QFT
qft_dagger(qc, qbits-1)
# Measure
qc.barrier()
for n in range(3):
    qc.measure(n,n)

# Get results and plot 

backend = Aer.get_backend('aer_simulator')
job = execute(qc,backend,shots=2048)
result = job.result() 
count = result.get_counts()
plot_histogram(count)
\end{lstlisting}
\end{mdframed}

\subsection{Decomposing any unitary in single-qubit and CNOT gates}

In this subsection, we provide the code which reproduces the 
gate count of the shown in Fig.~\ref{fig:qsd}. 
\begin{mdframed}[backgroundcolor=celadon!6]
\begin{lstlisting}[language=Python]
import numpy as np
from qiskit import QuantumCircuit, QuantumRegister
from qiskit.quantum_info.synthesis import two_qubit_decompose
from qiskit.compiler import transpile

def random_unitary_matrix(n):
    # Generate a random complex matrix
    mat = np.random.randn(n, n) + 1j * np.random.randn(n, n)
    # Compute the QR decomposition
    q, r = np.linalg.qr(mat)
    # Return the unitary matrix
    return q

def quantum_shannon_decomposition(circuit, unitary_matrix, qubits):
    # Create a temporary circuit with the desired number of qubits
    temp_qubits = QuantumRegister(len(qubits))
    temp_circuit = QuantumCircuit(temp_qubits)

    # Apply the unitary operator to the temporary circuit
    temp_circuit.unitary(unitary_matrix, qubits)

    # Decompose each two-qubit gate in the temporary circuit
    decomposed_circuit = transpile(temp_circuit, basis_gates=['ry', 'rx', 'cx'], optimization_level = 2)
    # 3 is highest optimization in QISKIT or ibm_q

    # Apply the decomposed gates to the original circuit
    circuit.compose(decomposed_circuit, qubits, inplace=True)

# Example usage
qbits=3
qubits = [i for i in range(0, qbits)]

# Create a quantum circuit
circuit = QuantumCircuit(qbits)
# Apply the Quantum Shannon Decomposition
unitary_matrix = random_unitary_matrix(2**qbits)
quantum_shannon_decomposition(circuit, unitary_matrix, qubits)

print(dict(circuit.count_ops()))
\end{lstlisting}
\end{mdframed}

\subsection{Sample \MAT~code for to compute vN entropy}
We need to install the appropriate package (as discussed in Sec.~\ref{sec:ent_mstate}) and then execute the commands below. 
\begin{mdframed}[backgroundcolor=celadon!6] 
	\begin{lstlisting}[language=Mathematica]	
ent = Sqrt[1/2]*(KetV[0,2] \[CircleTimes] KetV[0,2]+ KetV[1,2] \[CircleTimes] KetV[1,2])
rho12 = DM[ent];
rhonew = {{5/12, 1/6, 1/6}, {1/6, 1/6, 1/6}, {1/6, 1/6, 5/12}};
vNEntropy[rhonew, {1}, {3}]*Log[2] // N
rho1 = PickRandomRho[4];
rho2 = PickRandomRho[4];
Chop[RelEnt[rho1, rho2]]
rstate = PickRandomRho[4]
vNEntropy[rstate, {1}, {4}] * Log[2]
	\end{lstlisting}
\end{mdframed}

%% file: qcomp_notes_v3.bbl
\providecommand{\href}[2]{#2}\begingroup\raggedright\begin{thebibliography}{100}

\bibitem{Landauer_1991}
R.~Landauer, ``Information is physical,''
  \href{http://dx.doi.org/10.1063/1.881299}{{\em Physics Today} {\bfseries 44}
  no.~5, (May, 1991) 23--29}. \url{https://doi.org/10.1063%2F1.881299}.

\bibitem{Landauer_1996}
R.~Landauer, ``The physical nature of information,''
  \href{http://dx.doi.org/10.1016/0375-9601(96)00453-7}{{\em Physics Letters A}
  {\bfseries 217} no.~4-5, (Jul, 1996) 188--193}.
  \url{https://doi.org/10.1016%2F0375-9601%2896%2900453-7}.

\bibitem{Preskill:1999he}
J.~Preskill, ``{Quantum information and physics: some future directions},''
  \href{http://dx.doi.org/10.1080/09500340008244031}{{\em J. Mod. Opt.}
  {\bfseries 47} (2000) 127--137},
  \href{http://arxiv.org/abs/quant-ph/9904022}{{\ttfamily
  arXiv:quant-ph/9904022}}.

\bibitem{Landauer_1961}
R.~Landauer, ``Irreversibility and heat generation in the computing process,''
  \href{http://dx.doi.org/10.1147/rd.53.0183}{{\em {IBM} Journal of Research
  and Development} {\bfseries 5} no.~3, (Jul, 1961) 183--191}.
  \url{https://doi.org/10.1147%2Frd.53.0183}.

\bibitem{Bennett_1982}
C.~H. Bennett, ``The thermodynamics of computation{\textemdash}a review,''
  \href{http://dx.doi.org/10.1007/bf02084158}{{\em International Journal of
  Theoretical Physics} {\bfseries 21} no.~12, (Dec, 1982) 905--940}.
  \url{https://doi.org/10.1007%2Fbf02084158}.

\bibitem{Benioff_1980}
P.~Benioff, ``The computer as a physical system: A microscopic quantum
  mechanical hamiltonian model of computers as represented by turing
  machines,'' \href{http://dx.doi.org/10.1007/bf01011339}{{\em Journal of
  Statistical Physics} {\bfseries 22} no.~5, (May, 1980) 563--591}.
  \url{https://doi.org/10.1007%2Fbf01011339}.

\bibitem{Benioff_1982}
P.~{Benioff}, ``{Quantum mechanical hamiltonian models of turing machines},''
  \href{http://dx.doi.org/10.1007/BF01342185}{{\em Journal of Statistical
  Physics} {\bfseries 29} no.~3, (Nov., 1982) 515--546}.

\bibitem{Bennett_1973}
C.~H. Bennett, ``Logical reversibility of computation,''
  \href{http://dx.doi.org/10.1147/rd.176.0525}{{\em {IBM} Journal of Research
  and Development} {\bfseries 17} no.~6, (Nov, 1973) 525--532}.
  \url{https://doi.org/10.1147%2Frd.176.0525}.

\bibitem{Manin:1980abc}
Y.~Manin, ``{Computable and Uncomputable (in Russian)},''
\newblock 1980.

\bibitem{Feynman1982}
R.~P. Feynman, ``Simulating physics with computers,''
  \href{http://dx.doi.org/10.1007/BF02650179}{{\em International Journal of
  Theoretical Physics} {\bfseries 21} no.~6, (1982) 467--488}.

\bibitem{Qinfobook}
M.~A. Nielsen and I.~L. Chuang, {\em Quantum Computation and Quantum
  Information: 10th Anniversary Edition}.
\newblock Cambridge University Press, USA, 10th~ed., 2011.

\bibitem{Shor2000}
P.~W. {Shor}, ``{Introduction to Quantum Algorithms},'' {\em arXiv e-prints}
  (Apr., 2000) quant--ph/0005003,
  \href{http://arxiv.org/abs/quant-ph/0005003}{{\ttfamily
  arXiv:quant-ph/0005003 [quant-ph]}}.

\bibitem{DiVincenzo_2000}
D.~P. DiVincenzo, ``The physical implementation of quantum computation,''
  \href{http://dx.doi.org/10.1002/1521-3978(200009)48:9/11<771::aid-prop771>3.0.co;2-e}{{\em
  Fortschritte der Physik} {\bfseries 48} no.~9-11, (Sep, 2000) 771--783}.
  \url{https://doi.org/10.1002%2F1521-3978%28200009%2948%3A9%2F11%3C771%3A%3Aaid-prop771%3E3.0.co%3B2-e}.

\bibitem{Jordan:2017lea}
S.~P. Jordan, H.~Krovi, K.~S.~M. Lee, and J.~Preskill, ``{BQP-completeness of
  Scattering in Scalar Quantum Field Theory},''
  \href{http://dx.doi.org/10.22331/q-2018-01-08-44}{{\em Quantum} {\bfseries 2}
  (2018) 44}, \href{http://arxiv.org/abs/1703.00454}{{\ttfamily
  arXiv:1703.00454 [quant-ph]}}.

\bibitem{SA2016}
S.~Aaronson and L.~Chen, ``Complexity-theoretic foundations of quantum
  supremacy experiments,'' 2016.
\newblock \url{https://arxiv.org/abs/1612.05903}.

\bibitem{Google2019}
 \url{https://ai.googleblog.com/2019/10/quantum-supremacy-using-programmable.html}.

\bibitem{Dirac_1939}
P.~A.~M. Dirac, ``A new notation for quantum mechanics,''
  \href{http://dx.doi.org/10.1017/s0305004100021162}{{\em Mathematical
  Proceedings of the Cambridge Philosophical Society} {\bfseries 35} no.~3,
  (Jul, 1939) 416--418}. \url{https://doi.org/10.1017%2Fs0305004100021162}.

\bibitem{shi2002both}
Y.~Shi, ``Both toffoli and controlled-not need little help to do universal
  quantum computation,'' {\em arXiv preprint quant-ph/0205115} (2002) .

\bibitem{aharonov2003simple}
D.~Aharonov, ``A simple proof that toffoli and hadamard are quantum
  universal,'' {\em arXiv preprint quant-ph/0301040} (2003) .

\bibitem{2008arXiv0803.2316S}
V.~V. {Shende} and I.~L. {Markov}, ``{On the CNOT-cost of TOFFOLI gates},''
  \href{http://dx.doi.org/10.48550/arXiv.0803.2316}{{\em arXiv e-prints} (Mar.,
  2008) arXiv:0803.2316}, \href{http://arxiv.org/abs/0803.2316}{{\ttfamily
  arXiv:0803.2316 [quant-ph]}}.

\bibitem{PhysRevLett.62.2124}
G.~J. Milburn, ``Quantum optical fredkin gate,''
  \href{http://dx.doi.org/10.1103/PhysRevLett.62.2124}{{\em Phys. Rev. Lett.}
  {\bfseries 62} (May, 1989) 2124--2127}.
  \url{https://link.aps.org/doi/10.1103/PhysRevLett.62.2124}.

\bibitem{Feynman1986-FEYQMC}
R.~P. Feynman, ``Quantum mechanical computers,''
  \href{http://dx.doi.org/10.1007/BF01886518}{{\em Foundations of Physics}
  {\bfseries 16} no.~6, (1986) 507--531}.

\bibitem{D_r_2000}
W.~D{\"u}r, G.~Vidal, and J.~I. Cirac, ``Three qubits can be entangled in two
  inequivalent ways,'' \href{http://dx.doi.org/10.1103/physreva.62.062314}{{\em
  Physical Review A} {\bfseries 62} no.~6, (Nov, 2000) }.
  \url{https://doi.org/10.1103%2Fphysreva.62.062314}.

\bibitem{Boykin_2000}
P.~Boykin, T.~Mor, M.~Pulver, V.~Roychowdhury, and F.~Vatan, ``A new universal
  and fault-tolerant quantum basis,''
  \href{http://dx.doi.org/10.1016/s0020-0190(00)00084-3}{{\em Information
  Processing Letters} {\bfseries 75} no.~3, (Aug, 2000) 101--107}.
  \url{https://doi.org/10.1016%2Fs0020-0190%2800%2900084-3}.

\bibitem{Kitaev_1997}
A.~Y. Kitaev, ``Quantum computations: algorithms and error correction,''
  \href{http://dx.doi.org/10.1070/rm1997v052n06abeh002155}{{\em Russian
  Mathematical Surveys} {\bfseries 52} no.~6, (Dec, 1997) 1191--1249}.
  \url{https://doi.org/10.1070%2Frm1997v052n06abeh002155}.

\bibitem{Barenco95}
A.~Barenco, C.~H. Bennett, R.~Cleve, D.~P. DiVincenzo, N.~Margolus, P.~Shor,
  T.~Sleator, J.~A. Smolin, and H.~Weinfurter, ``Elementary gates for quantum
  computation,'' \href{http://dx.doi.org/10.1103/PhysRevA.52.3457}{{\em Phys.
  Rev. A} {\bfseries 52} (Nov, 1995) 3457--3467}.
  \url{https://link.aps.org/doi/10.1103/PhysRevA.52.3457}.

\bibitem{dawson2005}
C.~M. Dawson and M.~A. Nielsen, ``The solovay-kitaev algorithm,'' 2005.
\newblock \url{https://arxiv.org/abs/quant-ph/0505030}.

\bibitem{1994cond.mat..9111D}
D.~P. {DiVincenzo} and J.~{Smolin}, ``{Results on two-bit gate design for
  quantum computers},'' {\em arXiv e-prints} (Sept., 1994) cond--mat/9409111,
  \href{http://arxiv.org/abs/cond-mat/9409111}{{\ttfamily
  arXiv:cond-mat/9409111 [cond-mat]}}.

\bibitem{Barenco:1995dx}
A.~Barenco, ``{A Universal two bit gate for quantum computation},''
  \href{http://dx.doi.org/10.1098/rspa.1995.0066}{{\em Proc. Roy. Soc. Lond. A}
  {\bfseries 449} (1995) 679},
  \href{http://arxiv.org/abs/quant-ph/9505016}{{\ttfamily
  arXiv:quant-ph/9505016}}.

\bibitem{2002quant.ph..5115S}
Y.~{Shi}, ``{Both Toffoli and Controlled-NOT need little help to do universal
  quantum computation},'' {\em arXiv e-prints} (May, 2002) quant--ph/0205115,
  \href{http://arxiv.org/abs/quant-ph/0205115}{{\ttfamily
  arXiv:quant-ph/0205115 [quant-ph]}}.

\bibitem{2003quant.ph..1040A}
D.~{Aharonov}, ``{A Simple Proof that Toffoli and Hadamard are Quantum
  Universal},'' {\em arXiv e-prints} (Jan., 2003) quant--ph/0301040,
  \href{http://arxiv.org/abs/quant-ph/0301040}{{\ttfamily
  arXiv:quant-ph/0301040 [quant-ph]}}.

\bibitem{Shende2006}
V.~Shende, S.~Bullock, and I.~Markov, ``Synthesis of quantum-logic circuits,''
  \href{http://dx.doi.org/10.1109/tcad.2005.855930}{{\em {IEEE} Transactions on
  Computer-Aided Design of Integrated Circuits and Systems} {\bfseries 25}
  no.~6, (June, 2006) 1000--1010}.
  \url{https://doi.org/10.1109/tcad.2005.855930}.

\bibitem{Hill_1997}
S.~Hill and W.~K. Wootters, ``Entanglement of a pair of quantum bits,''
  \href{http://dx.doi.org/10.1103/physrevlett.78.5022}{{\em Physical Review
  Letters} {\bfseries 78} no.~26, (Jun, 1997) 5022--5025}.
  \url{https://doi.org/10.1103%2Fphysrevlett.78.5022}.

\bibitem{Wootters_1998}
W.~K. Wootters, ``Entanglement of formation of an arbitrary state of two
  qubits,'' \href{http://dx.doi.org/10.1103/physrevlett.80.2245}{{\em Physical
  Review Letters} {\bfseries 80} no.~10, (Mar, 1998) 2245--2248}.
  \url{https://doi.org/10.1103%2Fphysrevlett.80.2245}.

\bibitem{Wooters2001lk}
W.~K. Wootters, ``Entanglement of formation and concurrence,'' {\em Quantum
  Info. Comput.} {\bfseries 1} no.~1, (Jan, 2001) 27--44.

\bibitem{DJ1994}
D.~Coppersmith, ``An approximate fourier transform useful in quantum
  factoring,''. \url{https://arxiv.org/abs/quant-ph/0201067}.

\bibitem{Wootters_1982}
W.~K. Wootters and W.~H. Zurek, ``A single quantum cannot be cloned,''
  \href{http://dx.doi.org/10.1038/299802a0}{{\em Nature} {\bfseries 299}
  no.~5886, (Oct, 1982) 802--803}. \url{https://doi.org/10.1038%2F299802a0}.

\bibitem{Dieks_1982}
D.~Dieks, ``Communication by {EPR} devices,'' {\em Physics Letters A}
  {\bfseries 92} no.~6, (Nov, 1982) 271--272.

\bibitem{Kumar_Pati_2000}
A.~K. Pati and S.~L. Braunstein \href{http://dx.doi.org/10.1038/35004532}{{\em
  Nature} {\bfseries 404} no.~6774, (Mar, 2000) 164--165}.
  \url{https://doi.org/10.1038%2F35004532}.

\bibitem{ML_DE}
D.~Leach and A.~Malvino, ``Digital principles and applications,''.

\bibitem{1992DJ}
D.~Deutsch and R.~Jozsa, ``Rapid solution of problems by quantum computation,''
  \href{http://dx.doi.org/10.1098/rspa.1992.0167}{{\em Proceedings of the Royal
  Society of London. Series A: Mathematical and Physical Sciences} {\bfseries
  439} no.~1907, (Dec, 1992) 553--558}.
  \url{https://doi.org/10.1098%2Frspa.1992.0167}.

\bibitem{Cleve_1998}
R.~Cleve, A.~Ekert, C.~Macchiavello, and M.~Mosca, ``Quantum algorithms
  revisited,'' \href{http://dx.doi.org/10.1098/rspa.1998.0164}{{\em Proceedings
  of the Royal Society of London. Series A: Mathematical, Physical and
  Engineering Sciences} {\bfseries 454} no.~1969, (Jan, 1998) 339--354}.
  \url{https://doi.org/10.1098%2Frspa.1998.0164}.

\bibitem{Grover96}
L.~K. {Grover}, ``{A fast quantum mechanical algorithm for database search},''
  {\em arXiv e-prints} (May, 1996) quant--ph/9605043,
  \href{http://arxiv.org/abs/quant-ph/9605043}{{\ttfamily
  arXiv:quant-ph/9605043 [quant-ph]}}.

\bibitem{Grover_1997}
L.~K. Grover, ``Quantum mechanics helps in searching for a needle in a
  haystack,'' \href{http://dx.doi.org/10.1103/physrevlett.79.325}{{\em Physical
  Review Letters} {\bfseries 79} no.~2, (Jul, 1997) 325--328}.
  \url{https://doi.org/10.1103%2Fphysrevlett.79.325}.

\bibitem{Shor:1994jg}
P.~W. Shor, ``{Polynomial time algorithms for prime factorization and discrete
  logarithms on a quantum computer},''
  \href{http://dx.doi.org/10.1137/S0097539795293172}{{\em SIAM J. Sci. Statist.
  Comput.} {\bfseries 26} (1997) 1484},
  \href{http://arxiv.org/abs/quant-ph/9508027}{{\ttfamily
  arXiv:quant-ph/9508027}}.

\bibitem{Vandersypen_2001}
L.~M.~K. Vandersypen, M.~Steffen, G.~Breyta, C.~S. Yannoni, M.~H. Sherwood, and
  I.~L. Chuang, ``Experimental realization of shor{\textquotesingle}s quantum
  factoring algorithm using nuclear magnetic resonance,''
  \href{http://dx.doi.org/10.1038/414883a}{{\em Nature} {\bfseries 414}
  no.~6866, (Dec, 2001) 883--887}. \url{https://doi.org/10.1038%2F414883a}.

\bibitem{2019arXiv190509749G}
C.~{Gidney} and M.~{Eker{\r{a}}}, ``{How to factor 2048 bit RSA integers in 8
  hours using 20 million noisy qubits},'' {\em arXiv e-prints} (May, 2019)
  arXiv:1905.09749, \href{http://arxiv.org/abs/1905.09749}{{\ttfamily
  arXiv:1905.09749 [quant-ph]}}.

\bibitem{cerezo2020variational}
M.~Cerezo, A.~Arrasmith, R.~Babbush, S.~Benjamin, S.~Endo, K.~Fujii,
  J.~McClean, K.~Mitarai, X.~Yuan, L.~Cincio, and P.~J. Coles, ``Variational
  quantum algorithms,''
  \href{http://dx.doi.org/10.1038/s42254-021-00348-9}{{\em nature reviews
  physics} (2020) }.

\bibitem{Tilly:2021jem}
J.~Tilly {\em et~al.}, ``{The Variational Quantum Eigensolver: a review of
  methods and best practices},''
  \href{http://arxiv.org/abs/2111.05176}{{\ttfamily arXiv:2111.05176
  [quant-ph]}}.

\bibitem{Peruzzo_2014}
A.~Peruzzo, J.~McClean, P.~Shadbolt, M.-H. Yung, X.-Q. Zhou, P.~J. Love,
  A.~Aspuru-Guzik, and J.~L. O'Brien, ``A variational eigenvalue solver on a
  photonic quantum processor,''
  \href{http://dx.doi.org/10.1038/ncomms5213}{{\em Nature Communications}
  {\bfseries 5} no.~1, (Jul, 2014) }.
  \url{https://doi.org/10.1038%2Fncomms5213}.

\bibitem{McClean2018}
J.~R. McClean, S.~Boixo, V.~N. Smelyanskiy, R.~Babbush, and H.~Neven, ``Barren
  plateaus in quantum neural network training landscapes,''
  \href{http://dx.doi.org/10.1038/s41467-018-07090-4}{{\em Nature
  Communications} {\bfseries 9} no.~1, (Nov., 2018) }.
  \url{https://doi.org/10.1038/s41467-018-07090-4}.

\bibitem{PRXQuantum.1.020319}
R.~Wiersema, C.~Zhou, Y.~de~Sereville, J.~F. Carrasquilla, Y.~B. Kim, and
  H.~Yuen, ``Exploring entanglement and optimization within the hamiltonian
  variational ansatz,''
  \href{http://dx.doi.org/10.1103/PRXQuantum.1.020319}{{\em PRX Quantum}
  {\bfseries 1} (Dec, 2020) 020319}.
  \url{https://link.aps.org/doi/10.1103/PRXQuantum.1.020319}.

\bibitem{Kandala}
A.~{Kandala}, A.~{Mezzacapo}, K.~{Temme}, M.~{Takita}, M.~{Brink}, J.~M.
  {Chow}, and J.~M. {Gambetta}, ``{Hardware-efficient variational quantum
  eigensolver for small molecules and quantum magnets},''
  \href{http://arxiv.org/abs/1704.05018}{{\ttfamily arXiv:1704.05018
  [quant-ph]}}.

\bibitem{Kirby2021}
W.~M. Kirby and P.~J. Love, ``Variational quantum eigensolvers for sparse
  hamiltonians,'' \href{http://dx.doi.org/10.1103/physrevlett.127.110503}{{\em
  Physical Review Letters} {\bfseries 127} no.~11, (Sept., 2021) }.
  \url{https://doi.org/10.1103/physrevlett.127.110503}.

\bibitem{QISKIT_TB}
 \url{https://qiskit.org/textbook/preface.html}.

\bibitem{Schiffer:2021xiv}
B.~F. Schiffer, J.~Tura, and J.~I. Cirac, ``{Adiabatic Spectroscopy and a
  Variational Quantum Adiabatic Algorithm},''
  \href{http://dx.doi.org/10.1103/PRXQuantum.3.020347}{{\em PRX Quantum}
  {\bfseries 3} no.~2, (2022) 020347},
  \href{http://arxiv.org/abs/2103.01226}{{\ttfamily arXiv:2103.01226
  [quant-ph]}}.

\bibitem{2021PhRvL.127k0503K}
W.~M. {Kirby} and P.~J. {Love}, ``{Variational Quantum Eigensolvers for Sparse
  Hamiltonians},'' \href{http://dx.doi.org/10.1103/PhysRevLett.127.110503}{{\em
  PRLl} {\bfseries 127} no.~11, (Sept., 2021) 110503},
  \href{http://arxiv.org/abs/2012.07171}{{\ttfamily arXiv:2012.07171
  [quant-ph]}}.

\bibitem{Cai2022}
Z.~{Cai}, R.~{Babbush}, S.~C. {Benjamin}, S.~{Endo}, W.~J. {Huggins}, Y.~{Li},
  J.~R. {McClean}, and T.~E. {O'Brien}, ``{Quantum Error Mitigation},''
  \href{http://dx.doi.org/10.48550/arXiv.2210.00921}{{\em arXiv e-prints}
  (Oct., 2022) arXiv:2210.00921},
  \href{http://arxiv.org/abs/2210.00921}{{\ttfamily arXiv:2210.00921
  [quant-ph]}}.

\bibitem{2021PRXQ....2d0330S}
A.~{Strikis}, D.~{Qin}, Y.~{Chen}, S.~C. {Benjamin}, and Y.~{Li},
  ``{Learning-Based Quantum Error Mitigation},''
  \href{http://dx.doi.org/10.1103/PRXQuantum.2.040330}{{\em PRX Quantum}
  {\bfseries 2} no.~4, (Nov., 2021) 040330},
  \href{http://arxiv.org/abs/2005.07601}{{\ttfamily arXiv:2005.07601
  [quant-ph]}}.

\bibitem{miceli2018quantum}
{Raffaele Miceli and Michael McGuigan}, ``{Quantum Computation and
  Visualization of Hamiltonians using Discrete Quantum Mechanics and IBM
  QISKIT},'' \href{http://arxiv.org/abs/quant-ph/1812.01044}{{\ttfamily
  quant-ph/1812.01044}}.

\bibitem{Higgott_2019}
O.~Higgott, D.~Wang, and S.~Brierley, ``Variational quantum computation of
  excited states,'' \href{http://dx.doi.org/10.22331/q-2019-07-01-156}{{\em
  Quantum} {\bfseries 3} (Jul, 2019) 156}.
  \url{https://doi.org/10.22331%2Fq-2019-07-01-156}.

\bibitem{Kleinert:2004ev}
H.~Kleinert, ``{Path Integrals in Quantum Mechanics, Statistics, Polymer
  Physics, and Financial Markets},''.

\bibitem{dumitrescu2018cloud}
E.~Dumitrescu, A.~McCaskey, G.~Hagen, G.~Jansen, T.~Morris, T.~Papenbrock,
  R.~Pooser, D.~Dean, and P.~Lougovski, ``Cloud quantum computing of an atomic
  nucleus,'' \href{http://dx.doi.org/10.1103/PhysRevLett.120.210501}{{\em
  physical review letters} (2018) }.

\bibitem{Araz:2022tbd}
J.~Y. Araz, S.~Schenk, and M.~Spannowsky, ``{Towards a Quantum Simulation of
  Nonlinear Sigma Models with a Topological Term},''
  \href{http://arxiv.org/abs/2210.03679}{{\ttfamily arXiv:2210.03679
  [quant-ph]}}.

\bibitem{Somma2003}
R.~Somma, G.~Ortiz, E.~Knill, and J.~Gubernatis, ``Quantum simulations of
  physics problems,''
  \href{http://arxiv.org/abs/arXiv:quant-ph/0304063}{{\ttfamily
  arXiv:quant-ph/0304063}}.

\bibitem{Sawaya2019}
N.~P.~D. Sawaya, T.~Menke, T.~H. Kyaw, S.~Johri, A.~Aspuru-Guzik, and G.~G.
  Guerreschi, ``Resource-efficient digital quantum simulation of $d$-level
  systems for photonic, vibrational, and spin-$s$ hamiltonians,''
  \href{http://arxiv.org/abs/arXiv:1909.12847}{{\ttfamily arXiv:1909.12847}}.

\bibitem{Asaduzzaman:2022bpi}
M.~Asaduzzaman, S.~Catterall, G.~C. Toga, Y.~Meurice, and R.~Sakai, ``{Quantum
  Simulation of the N flavor Gross-Neveu Model},''
  \href{http://arxiv.org/abs/2208.05906}{{\ttfamily arXiv:2208.05906
  [hep-lat]}}.

\bibitem{Shaw:2020udc}
A.~F. Shaw, P.~Lougovski, J.~R. Stryker, and N.~Wiebe, ``{Quantum Algorithms
  for Simulating the Lattice Schwinger Model},''
  \href{http://dx.doi.org/10.22331/q-2020-08-10-306}{{\em Quantum} {\bfseries
  4} (2020) 306}, \href{http://arxiv.org/abs/2002.11146}{{\ttfamily
  arXiv:2002.11146 [quant-ph]}}.

\bibitem{Yeter-Aydeniz:2021mol}
K.~Yeter-Aydeniz, E.~Moschandreou, and G.~Siopsis, ``{Quantum imaginary-time
  evolution algorithm for quantum field theories with continuous variables},''
  \href{http://dx.doi.org/10.1103/PhysRevA.105.012412}{{\em Phys. Rev. A}
  {\bfseries 105} no.~1, (2022) 012412},
  \href{http://arxiv.org/abs/2107.00791}{{\ttfamily arXiv:2107.00791
  [quant-ph]}}.

\bibitem{2019npjQI...5...75M}
S.~{McArdle}, T.~{Jones}, S.~{Endo}, Y.~{Li}, S.~C. {Benjamin}, and X.~{Yuan},
  ``{Variational ansatz-based quantum simulation of imaginary time
  evolution},'' \href{http://dx.doi.org/10.1038/s41534-019-0187-2}{{\em npj
  Quantum Information} {\bfseries 5} (Sept., 2019) 75},
  \href{http://arxiv.org/abs/1804.03023}{{\ttfamily arXiv:1804.03023
  [quant-ph]}}.

\bibitem{2020NatPh..16..205M}
M.~{Motta}, C.~{Sun}, A.~T.~K. {Tan}, M.~J. {O'Rourke}, E.~{Ye}, A.~J.
  {Minnich}, F.~G.~S.~L. {Brand{\~a}o}, and G.~K.-L. {Chan}, ``{Determining
  eigenstates and thermal states on a quantum computer using quantum imaginary
  time evolution},'' \href{http://dx.doi.org/10.1038/s41567-019-0704-4}{{\em
  Nature Physics} {\bfseries 16} no.~2, (Jan., 2020) 205--210},
  \href{http://arxiv.org/abs/1901.07653}{{\ttfamily arXiv:1901.07653
  [quant-ph]}}.

\bibitem{Knill:1996ny}
E.~Knill and R.~Laflamme, ``{A Theory of quantum error correcting codes},''
  \href{http://dx.doi.org/10.1103/PhysRevLett.84.2525}{{\em Phys. Rev. Lett.}
  {\bfseries 84} (2000) 2525--2528},
  \href{http://arxiv.org/abs/quant-ph/9604034}{{\ttfamily
  arXiv:quant-ph/9604034}}.

\bibitem{PhysRevA.52.R2493}
P.~W. Shor, ``Scheme for reducing decoherence in quantum computer memory,''
  \href{http://dx.doi.org/10.1103/PhysRevA.52.R2493}{{\em Phys. Rev. A}
  {\bfseries 52} (Oct, 1995) R2493--R2496}.
  \url{https://link.aps.org/doi/10.1103/PhysRevA.52.R2493}.

\bibitem{Laflamme:1996iw}
R.~Laflamme, C.~Miquel, J.~P. Paz, and W.~H. Zurek, ``{Perfect quantum error
  correction code},'' \href{http://arxiv.org/abs/quant-ph/9602019}{{\ttfamily
  arXiv:quant-ph/9602019}}.

\bibitem{Steane:1995vv}
A.~Steane, ``{Multiple particle interference and quantum error correction},''
  \href{http://dx.doi.org/10.1098/rspa.1996.0136}{{\em Proc. Roy. Soc. Lond. A}
  {\bfseries 452} (1996) 2551},
  \href{http://arxiv.org/abs/quant-ph/9601029}{{\ttfamily
  arXiv:quant-ph/9601029}}.

\bibitem{2009arXiv0904.2557G}
D.~{Gottesman}, ``{An Introduction to Quantum Error Correction and
  Fault-Tolerant Quantum Computation},'' {\em arXiv e-prints} (Apr., 2009)
  arXiv:0904.2557, \href{http://arxiv.org/abs/0904.2557}{{\ttfamily
  arXiv:0904.2557 [quant-ph]}}.

\bibitem{Gottesman:1998hu}
D.~Gottesman, ``{The Heisenberg representation of quantum computers},'' in {\em
  {22nd International Colloquium on Group Theoretical Methods in Physics}},
  pp.~32--43.
\newblock 7, 1998.
\newblock \href{http://arxiv.org/abs/quant-ph/9807006}{{\ttfamily
  arXiv:quant-ph/9807006}}.

\bibitem{Braunstein:2005zz}
S.~L. Braunstein and P.~van Loock, ``{Quantum information with continuous
  variables},'' \href{http://dx.doi.org/10.1103/RevModPhys.77.513}{{\em Rev.
  Mod. Phys.} {\bfseries 77} (2005) 513--577},
  \href{http://arxiv.org/abs/quant-ph/0410100}{{\ttfamily
  arXiv:quant-ph/0410100}}.

\bibitem{Seth1996}
S.~Lloyd, ``Universal quantum simulators,''
  \href{http://dx.doi.org/10.1126/science.273.5278.1073}{{\em Science}
  {\bfseries 273} no.~5278, (Aug., 1996) 1073--1078}.
  \url{https://doi.org/10.1126/science.273.5278.1073}.

\bibitem{Chandrasekharan:1996ih}
S.~Chandrasekharan and U.~J. Wiese, ``{Quantum link models: A Discrete approach
  to gauge theories},''
  \href{http://dx.doi.org/10.1016/S0550-3213(97)00006-0}{{\em Nucl. Phys. B}
  {\bfseries 492} (1997) 455--474},
  \href{http://arxiv.org/abs/hep-lat/9609042}{{\ttfamily
  arXiv:hep-lat/9609042}}.

\bibitem{Bazavov:2019qih}
A.~Bazavov, S.~Catterall, R.~G. Jha, and J.~Unmuth-Yockey, ``{Tensor
  renormalization group study of the non-Abelian Higgs model in two
  dimensions},'' \href{http://dx.doi.org/10.1103/PhysRevD.99.114507}{{\em Phys.
  Rev. D} {\bfseries 99} no.~11, (2019) 114507},
  \href{http://arxiv.org/abs/1901.11443}{{\ttfamily arXiv:1901.11443
  [hep-lat]}}.

\bibitem{Raychowdhury:2019iki}
I.~Raychowdhury and J.~R. Stryker, ``{Loop, string, and hadron dynamics in
  SU(2) Hamiltonian lattice gauge theories},''
  \href{http://dx.doi.org/10.1103/PhysRevD.101.114502}{{\em Phys. Rev. D}
  {\bfseries 101} no.~11, (2020) 114502},
  \href{http://arxiv.org/abs/1912.06133}{{\ttfamily arXiv:1912.06133
  [hep-lat]}}.

\bibitem{Alexandru:2021jpm}
A.~Alexandru, P.~F. Bedaque, R.~Brett, and H.~Lamm, ``{Spectrum of digitized
  QCD: Glueballs in a S(1080) gauge theory},''
  \href{http://dx.doi.org/10.1103/PhysRevD.105.114508}{{\em Phys. Rev. D}
  {\bfseries 105} no.~11, (2022) 114508},
  \href{http://arxiv.org/abs/2112.08482}{{\ttfamily arXiv:2112.08482
  [hep-lat]}}.

\bibitem{Jordan:2011ci}
S.~P. Jordan, K.~S.~M. Lee, and J.~Preskill, ``{Quantum Computation of
  Scattering in Scalar Quantum Field Theories},'' {\em Quant. Inf. Comput.}
  {\bfseries 14} (2014) 1014--1080,
  \href{http://arxiv.org/abs/1112.4833}{{\ttfamily arXiv:1112.4833 [hep-th]}}.

\bibitem{Klco:2018zqz}
N.~Klco and M.~J. Savage, ``{Digitization of scalar fields for quantum
  computing},'' \href{http://dx.doi.org/10.1103/PhysRevA.99.052335}{{\em Phys.
  Rev. A} {\bfseries 99} no.~5, (2019) 052335},
  \href{http://arxiv.org/abs/1808.10378}{{\ttfamily arXiv:1808.10378
  [quant-ph]}}.

\bibitem{Marshall:2015mna}
K.~Marshall, R.~Pooser, G.~Siopsis, and C.~Weedbrook, ``{Quantum simulation of
  quantum field theory using continuous variables},''
  \href{http://dx.doi.org/10.1103/PhysRevA.92.063825}{{\em Phys. Rev. A}
  {\bfseries 92} no.~6, (2015) 063825},
  \href{http://arxiv.org/abs/1503.08121}{{\ttfamily arXiv:1503.08121
  [quant-ph]}}.

\bibitem{Barata:2020jtq}
J.~Barata, N.~Mueller, A.~Tarasov, and R.~Venugopalan, ``{Single-particle
  digitization strategy for quantum computation of a $\phi^4$ scalar field
  theory},'' \href{http://dx.doi.org/10.1103/PhysRevA.103.042410}{{\em Phys.
  Rev. A} {\bfseries 103} no.~4, (2021) 042410},
  \href{http://arxiv.org/abs/2012.00020}{{\ttfamily arXiv:2012.00020
  [hep-th]}}.

\bibitem{PhysRevLett.103.150502}
A.~W. Harrow, A.~Hassidim, and S.~Lloyd, ``Quantum algorithm for linear systems
  of equations,'' \href{http://dx.doi.org/10.1103/PhysRevLett.103.150502}{{\em
  Phys. Rev. Lett.} {\bfseries 103} (Oct, 2009) 150502}.
  \url{https://link.aps.org/doi/10.1103/PhysRevLett.103.150502}.

\bibitem{2021arXiv210809004M}
J.~{Morrell}, Hector~Jose, A.~{Zaman}, and H.~Y. {Wong}, ``{Step-by-Step HHL
  Algorithm Walkthrough to Enhance the Understanding of Critical Quantum
  Computing Concepts},''
  \href{http://dx.doi.org/10.48550/arXiv.2108.09004}{{\em arXiv e-prints}
  (Aug., 2021) arXiv:2108.09004},
  \href{http://arxiv.org/abs/2108.09004}{{\ttfamily arXiv:2108.09004
  [quant-ph]}}.

\bibitem{Hestenes1952}
M.~Hestenes and E.~Stiefel, ``Methods of conjugate gradients for solving linear
  systems,'' \href{http://dx.doi.org/10.6028/jres.049.044}{{\em Journal of
  Research of the National Bureau of Standards} {\bfseries 49} no.~6, (Dec.,
  1952) 409}. \url{https://doi.org/10.6028/jres.049.044}.

\bibitem{Aaronson2015}
S.~Aaronson, ``Read the fine print,''
  \href{http://dx.doi.org/10.1038/nphys3272}{{\em Nature Physics} {\bfseries
  11} no.~4, (Apr., 2015) 291--293}. \url{https://doi.org/10.1038/nphys3272}.

\bibitem{Whitfield_2011}
J.~D. Whitfield, J.~Biamonte, and A.~Aspuru-Guzik, ``Simulation of electronic
  structure hamiltonians using quantum computers,''
  \href{http://dx.doi.org/10.1080/00268976.2011.552441}{{\em Molecular Physics}
  {\bfseries 109} no.~5, (Mar, 2011) 735--750}.
  \url{https://doi.org/10.1080%2F00268976.2011.552441}.

\bibitem{2013Berry}
D.~W. {Berry}, A.~M. {Childs}, R.~{Cleve}, R.~{Kothari}, and R.~D. {Somma},
  ``{Exponential improvement in precision for simulating sparse
  Hamiltonians},'' \href{http://dx.doi.org/10.48550/arXiv.1312.1414}{{\em arXiv
  e-prints} (Dec., 2013) arXiv:1312.1414},
  \href{http://arxiv.org/abs/1312.1414}{{\ttfamily arXiv:1312.1414
  [quant-ph]}}.

\bibitem{2015BerryTTS}
D.~W. {Berry}, A.~M. {Childs}, R.~{Cleve}, R.~{Kothari}, and R.~D. {Somma},
  ``{Simulating Hamiltonian Dynamics with a Truncated Taylor Series},''
  \href{http://dx.doi.org/10.1103/PhysRevLett.114.090502}{{\em PRL} {\bfseries
  114} no.~9, (Mar., 2015) 090502},
  \href{http://arxiv.org/abs/1412.4687}{{\ttfamily arXiv:1412.4687
  [quant-ph]}}.

\bibitem{Kempe2004}
J.~Kempe, A.~Kitaev, and O.~Regev, ``The complexity of the local hamiltonian
  problem,'' \href{http://arxiv.org/abs/arXiv:quant-ph/0406180}{{\ttfamily
  arXiv:quant-ph/0406180}}.

\bibitem{Berry2006}
D.~W. Berry, G.~Ahokas, R.~Cleve, and B.~C. Sanders, ``Efficient quantum
  algorithms for simulating sparse hamiltonians,''
  \href{http://dx.doi.org/10.1007/s00220-006-0150-x}{{\em Communications in
  Mathematical Physics} {\bfseries 270} no.~2, (Dec., 2006) 359--371}.
  \url{https://doi.org/10.1007/s00220-006-0150-x}.

\bibitem{Childs2011}
A.~M. Childs and R.~Kothari, ``Simulating sparse hamiltonians with star
  decompositions,''. \url{https://doi.org/10.1007/978-3-642-18073-6_8}.

\bibitem{2015arXiv150101715B}
D.~W. {Berry}, A.~M. {Childs}, and R.~{Kothari}, ``{Hamiltonian simulation with
  nearly optimal dependence on all parameters},''
  \href{http://dx.doi.org/10.48550/arXiv.1501.01715}{{\em arXiv e-prints}
  (Jan., 2015) arXiv:1501.01715},
  \href{http://arxiv.org/abs/1501.01715}{{\ttfamily arXiv:1501.01715
  [quant-ph]}}.

\bibitem{2016Low}
G.~{Hao Low} and I.~L. {Chuang}, ``{Hamiltonian Simulation by Qubitization},''
  \href{http://dx.doi.org/10.48550/arXiv.1610.06546}{{\em arXiv e-prints}
  (Oct., 2016) arXiv:1610.06546},
  \href{http://arxiv.org/abs/1610.06546}{{\ttfamily arXiv:1610.06546
  [quant-ph]}}.

\bibitem{2013arXiv1312.1414B}
D.~W. {Berry}, A.~M. {Childs}, R.~{Cleve}, R.~{Kothari}, and R.~D. {Somma},
  ``{Exponential improvement in precision for simulating sparse
  Hamiltonians},'' \href{http://dx.doi.org/10.48550/arXiv.1312.1414}{{\em arXiv
  e-prints} (Dec., 2013) arXiv:1312.1414},
  \href{http://arxiv.org/abs/1312.1414}{{\ttfamily arXiv:1312.1414
  [quant-ph]}}.

\bibitem{2022arXiv221107629B}
M.~E. {Beverland}, P.~{Murali}, M.~{Troyer}, K.~M. {Svore}, T.~{Hoeffler},
  V.~{Kliuchnikov}, G.~{Hao Low}, M.~{Soeken}, A.~{Sundaram}, and
  A.~{Vaschillo}, ``{Assessing requirements to scale to practical quantum
  advantage},'' {\em arXiv e-prints} (Nov., 2022) arXiv:2211.07629,
  \href{http://arxiv.org/abs/2211.07629}{{\ttfamily arXiv:2211.07629
  [quant-ph]}}.

\bibitem{eastin2004qcircuit}
B.~Eastin and S.~T. Flammia, ``Q-circuit tutorial,''
  \href{http://arxiv.org/abs/quant-ph/0406003}{{\ttfamily
  arXiv:quant-ph/0406003 [quant-ph]}}.

\bibitem{2019Sera}
A.~{Serafini}, ``{Quantum Continuous Variables: A Primer of Theoretical
  Methods},'' {\em CRC Press} (2017) .

\bibitem{2005Brau}
S.~L. {Braunstein} and P.~{van Loock}, ``{Quantum information with continuous
  variables},'' \href{http://dx.doi.org/10.1103/RevModPhys.77.513}{{\em Reviews
  of Modern Physics} {\bfseries 77} no.~2, (Apr., 2005) 513--577},
  \href{http://arxiv.org/abs/quant-ph/0410100}{{\ttfamily
  arXiv:quant-ph/0410100 [quant-ph]}}.

\bibitem{2012Weed}
C.~{Weedbrook}, S.~{Pirandola}, R.~{Garc{\'\i}a-Patr{\'o}n}, N.~J. {Cerf},
  T.~C. {Ralph}, J.~H. {Shapiro}, and S.~{Lloyd}, ``{Gaussian quantum
  information},'' \href{http://dx.doi.org/10.1103/RevModPhys.84.621}{{\em
  Reviews of Modern Physics} {\bfseries 84} no.~2, (Apr., 2012) 621--669},
  \href{http://arxiv.org/abs/1110.3234}{{\ttfamily arXiv:1110.3234
  [quant-ph]}}.

\bibitem{2001Gott}
D.~{Gottesman}, A.~{Kitaev}, and J.~{Preskill}, ``{Encoding a qubit in an
  oscillator},'' \href{http://arxiv.org/abs/quant-ph/0008040}{{\ttfamily
  arXiv:quant-ph/0008040 [quant-ph]}}.

\bibitem{2014Mirr}
M.~{Mirrahimi}, Z.~{Leghtas}, V.~V. {Albert}, S.~{Touzard}, R.~J. {Schoelkopf},
  L.~{Jiang}, and M.~H. {Devoret}, ``{Dynamically protected cat-qubits: a new
  paradigm for universal quantum computation},''
  \href{http://dx.doi.org/10.1088/1367-2630/16/4/045014}{{\em New Journal of
  Physics} {\bfseries 16} no.~4, (Apr., 2014) 045014},
  \href{http://arxiv.org/abs/1312.2017}{{\ttfamily arXiv:1312.2017
  [quant-ph]}}.

\bibitem{2019Kill}
N.~{Killoran}, T.~R. {Bromley}, J.~M. {Arrazola}, M.~{Schuld}, N.~{Quesada},
  and S.~{Lloyd}, ``{Continuous-variable quantum neural networks},''
  \href{http://dx.doi.org/10.1103/PhysRevResearch.1.033063}{{\em Physical
  Review Research} {\bfseries 1} no.~3, (Oct., 2019) 033063},
  \href{http://arxiv.org/abs/1806.06871}{{\ttfamily arXiv:1806.06871
  [quant-ph]}}.

\bibitem{Pati2000}
A.~K. Pati, S.~L. Braunstein, and S.~Lloyd, ``Quantum searching with continuous
  variables,'' {\em arXiv preprint quant-ph/0002082} (2000) .

\bibitem{Pati2003}
A.~K. Pati and S.~L. Braunstein, ``Deutsch-jozsa algorithm for continuous
  variables,'' {\em Quantum Information with Continuous Variables} (2003)
  31--36.

\bibitem{Braunstein:1997db}
S.~L. Braunstein, ``{Error correction for continuous quantum variables},''
  \href{http://dx.doi.org/10.1103/PhysRevLett.80.4084}{{\em Phys. Rev. Lett.}
  {\bfseries 80} (1998) 4084--4087},
  \href{http://arxiv.org/abs/quant-ph/9711049}{{\ttfamily
  arXiv:quant-ph/9711049}}.

\bibitem{Wang2001}
X.~Wang, ``Continuous-variable and hybrid quantum gates,''
  \href{http://dx.doi.org/10.1088/0305-4470/34/44/316}{{\em Journal of Physics
  A: Mathematical and General} {\bfseries 34} no.~44, (Oct., 2001) 9577--9584}.
  \url{https://doi.org/10.1088/0305-4470/34/44/316}.

\bibitem{Killoran2019}
N.~{Killoran}, J.~{Izaac}, N.~{Quesada}, V.~{Bergholm}, M.~{Amy}, and
  C.~{Weedbrook}, ``{Strawberry Fields: A Software Platform for Photonic
  Quantum Computing},'' \href{http://dx.doi.org/10.22331/q-2019-03-11-129}{{\em
  Quantum} {\bfseries 3} (Mar., 2019) 129},
  \href{http://arxiv.org/abs/1804.03159}{{\ttfamily arXiv:1804.03159
  [quant-ph]}}.

\bibitem{Magnus:1954zz}
W.~Magnus, ``{On the exponential solution of differential equations for a
  linear operator},'' \href{http://dx.doi.org/10.1002/cpa.3160070404}{{\em
  Commun. Pure Appl. Math.} {\bfseries 7} (1954) 649--673}.

\bibitem{Stavenger:2022wzz}
T.~J. Stavenger, E.~Crane, K.~Smith, C.~T. Kang, S.~M. Girvin, and N.~Wiebe,
  ``{Bosonic Qiskit},'' \href{http://arxiv.org/abs/2209.11153}{{\ttfamily
  arXiv:2209.11153 [quant-ph]}}.

\end{thebibliography}\endgroup
